\documentclass[draft]{agujournal2019}
\usepackage{url} 
\usepackage[inline]{trackchanges} 
\usepackage{soul}
\draftfalse

\journalname{JGR: Space Physics}

\begin{document}

%
%

\title{Atmospheric Escape Processes and Planetary Atmospheric Evolution}

%
%




\authors{ 
G. Gronoff \affil{1,2}, 
P. Arras \affil{3},
S. Baraka\affil{4},
J. M. Bell \affil{5},
G. Cessateur \affil{6},
O. Cohen \affil{7}, 
S. M. Curry \affil{11}, 
J.J. Drake \affil{8},
M. Elrod \affil{5},
J. Erwin \affil{6},
K. Garcia-Sage \affil{5},
C. Garraffo \affil{8},
A. Glocer \affil{5},
N.G. Heavens \affil{9,10},
K. Lovato \affil{9},
R. Maggiolo \affil{6}, 
C. D. Parkinson \affil{10},
C. Simon Wedlund \affil{12},
D. R. Weimer \affil{4,13}, 
W.B. Moore \affil{4,9}}

\affiliation{1}{Science directorate, Chemistry and Dynamics branch, NASA Langley Research Center, 21 Langley Blvd., Mail Stop 401B Hampton, Virginia 23681-2199 USA}
\affiliation{2}{Science Systems and Application Inc. Hampton, Virginia, USA}
\affiliation{3}{Department of Astronomy, University of Virginia, P.O. Box 400325, Charlottesville, VA 22904, USA}
\affiliation{4}{National Institute of Aerospace, Hampton, Virginia, USA.}
\affiliation{5}{Heliophysics Division, NASA Goddard Space Flight Center, 8800 Greenbelt Rd, Greenbelt, MD 20771, USA}
\affiliation{6}{The Royal Belgian Institute for Space Aeronomy (BIRA-IASB), Avenue Circulaire 3, 1180 Brussels, Belgium}
\affiliation{7}{Lowell Center for Space Science and Technology, University of Massachusetts Lowell, 600 Suffolk St., Lowell, MA 01854, USA}
\affiliation{8}{Harvard-Smithsonian Center for Astrophysics, 60 Garden St. Cambridge, MA 02138, USA.}
\affiliation{9}{Department of Atmospheric and Planetary Sciences, Hampton University, 154 William R. Harvey Way, Hampton, Virginia 23668 USA}
\affiliation{10}{Space Science Institute, 4765 Walnut St, Suite B, Boulder, Colorado, USA}
\affiliation{11}{Space Sciences Laboratory, University of California, Berkeley, 7 Gauss Way, Berkeley, CA 94720, USA.}
\affiliation{12}{Space Research Institute (IWF), Schmiedlstra{\ss}e 6, 8042 Graz, Austria}
\affiliation{13}{Center for Space Science and Engineering Research, Virginia Tech, Blacksburg, Virginia, USA.}






\correspondingauthor{Guillaume Gronoff}{Guillaume.P.Gronoff@nasa.gov}




\begin{keypoints}
\item The different escape processes at planets and exoplanets are reviewed along with their mathematical formulation. 
\item The major parameters for each escape processes are described. Some escape processes currently negligible in the Solar system may be the major source at exoplanets, or for the early Solar system. 
\item A magnetic field should not be a priori considered as a protection for the atmosphere.
\end{keypoints}

%
%

%
%


\begin{abstract}
The habitability of the surface of any planet is determined by a complex evolution of its interior, surface, and atmosphere. The electromagnetic and particle radiation of stars drive thermal, chemical and physical alteration of planetary atmospheres, including escape. Many known extrasolar planets experience vastly different stellar environments than those in our Solar system: it is crucial to understand the broad range of processes that lead to atmospheric escape and evolution under a wide range of conditions if we are to assess the habitability of worlds around other stars.

One problem encountered between the planetary and the astrophysics communities is a lack of common language for describing escape processes. Each community has customary approximations that may be questioned by the other, such as the hypothesis of H-dominated thermosphere for astrophysicists, or the Sun-like nature of the stars for planetary scientists. Since exoplanets are becoming one of the main targets for the detection of life, a common set of definitions and hypotheses are required. 

We review the different escape mechanisms proposed for the evolution of planetary and exoplanetary atmospheres. We propose a common definition for the different escape mechanisms, and we show the important parameters to take into account when evaluating the escape at a planet in time. We show that the paradigm of the magnetic field as an atmospheric shield should be changed and that recent work on the history of Xenon in Earth's atmosphere gives an elegant explanation to its enrichment in heavier isotopes: the so-called Xenon paradox.
\end{abstract}

\section*{Plain Language Summary}
 In addition to having the right surface temperature, a planet needs an atmosphere to keep surface liquid water stable. Although many planets have been found that may lie in the right temperature range,  the existence of an atmosphere is not guaranteed. In particular, for planets that are kept warm by being close to dim stars, there are a number of ways that the star may remove a planetary atmosphere.  These atmospheric escape processes depend on the behavior of the star as well as the nature of the planet, including the presence of a planetary magnetic field. Under certain conditions, a magnetic field can protect a planet's atmosphere from the loss due to the direct impact of the stellar wind; but it may actually enhance total atmospheric loss by connecting to the highly variable magnetic field of the stellar wind. These enhancements happen especially for planets close to dim stars. We review the complete range of atmospheric loss processes driven by interaction between a planet and a star to aid in the identification of planets that are both the correct temperature for liquid water and that have a chance of maintaining an atmosphere over long periods of time.
%
%

%


%
%
%
%

\section{Introduction}
The discovery of rocky exoplanets at distances from their host stars that might allow stable surface liquid water has led to a blossoming of studies of the habitability of such objects \cite{Anglada2016,Gillon2017,teegarden2019}. While the ultimate objective of this work is the discovery of life on an exoplanet, detailed investigations of such planets may also shed light on the evolution --both past and future-- of the planets in our own Solar system \cite{ArneyKane2018}, in particular, how they came to be, remain, and/or ceased to be habitable for life as we know it \cite{moore_how_2017,towers_2017,tasker_language_2017}.

The usual definition of the ``habitable-zone'' (HZ) \cite[and references therein]{KastiToonPollo:SA1988,ramirez2018more,Lammer2009}, is where a planet like the Earth would  be able to maintain 
liquid water at its surface, however it says nothing about whether the planet actually has any liquid water, or the necessary atmospheric pressure to stabilize the liquid state. This definition fails to take into account the necessary pathways to habitability: a planet forming in the habitable zone of a star will have to accrete volatiles from the protostellar nebula to be able to have an atmosphere and liquid water, and it will also have to keep them, which is not necessarily the case for the previously mentioned exoplanets  --even if we suppose they have a strong intrinsic magnetic field-- \cite{airapetian2017hospitable,GarciaSage2017,Howard2018}. The concept of the HZ is therefore distressingly incomplete, which led to the concept of Space Weather Affected Habitable Zone \cite{airapetian2017hospitable} .

\begin{figure}
 \noindent\includegraphics[width=20pc]{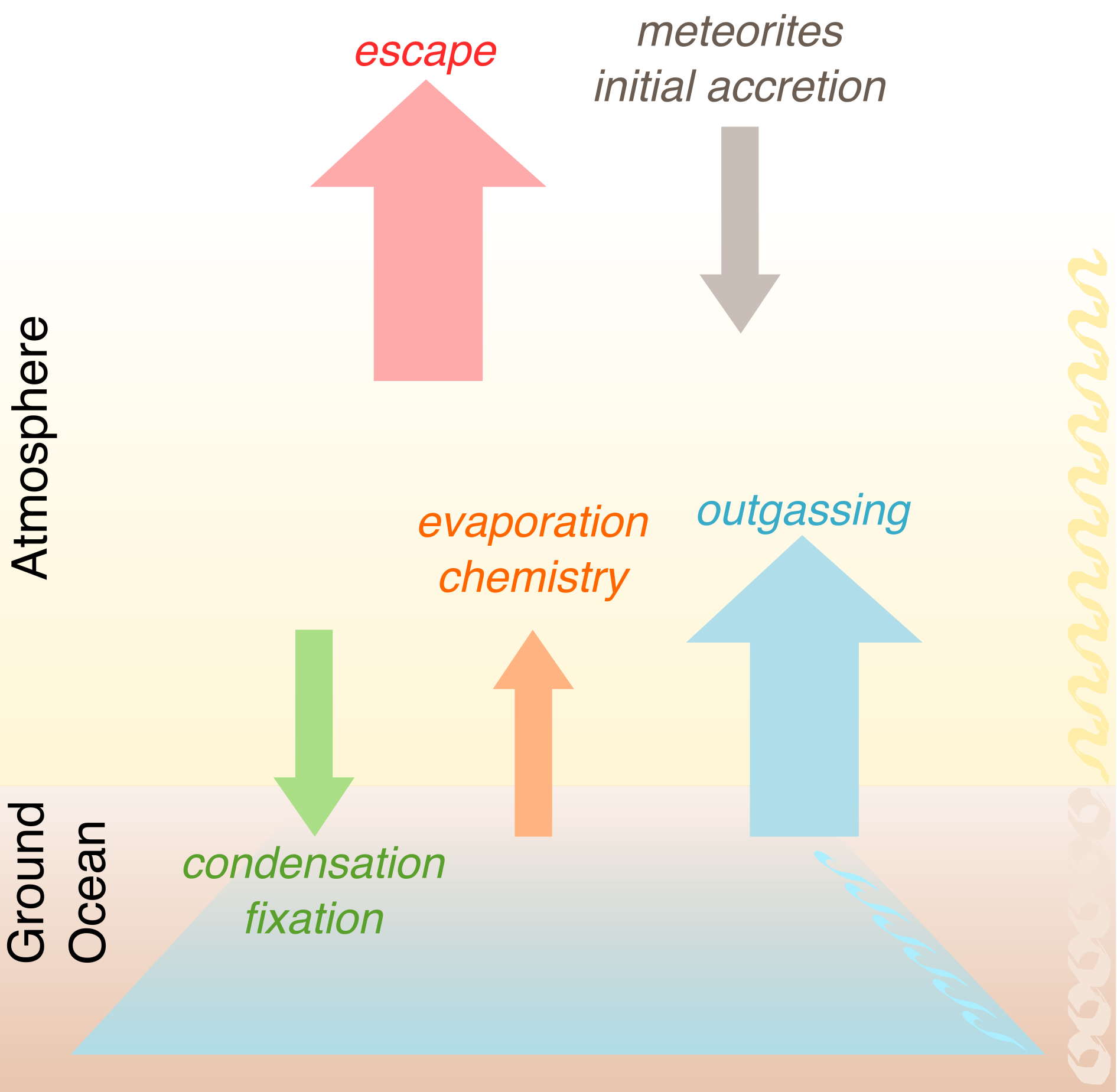}
 \caption{The processes leading to the creation and the destruction of an atmosphere. A stable balance between these processes is required for a habitable atmosphere.}
 \label{atmospherecreation}
\end{figure}

One of the best examples of the problems with this definition comes from our understanding of the early Earth: the so-called ``Faint Young Sun'' (FYS) paradox. 4 to 3~Gyr ago, the Sun was fainter by about 30\% \cite{Claire_2012}, and our models predict that surface water should have been frozen, and therefore that Earth was not in the HZ. There is however considerable evidence for an active hydrological cycle and exceptionally warm and/or clement temperatures at that period \cite{Mojzsis2001,Knauth2003,KastingOno2006,Lammer2009}. The typical solution to the FYS paradox has been to propose that the Earth's early atmosphere had a higher concentration of greenhouse gases such as CO$_2$, CH$_4$, NH$_3$, N$_2$O, etc., in a perhaps thicker atmosphere than now \cite{SaganMullen1972,Walker1981,Pavlov2000,Airapetian2016}. Greenhouse gas levels have overall implications for geological activity, cloud/aerosol formation, and atmospheric chemistry and escape that can preclude their existence, stability, or positive contribution to habitability altogether \cite{KuhnAtreya1979,Pavlov2000,Trainer2006}. Several hypotheses remain concerning the nature of the Early Earth's atmosphere; a major problem lies with the uncertainties on the nitrogen cycle in the past, and on the actual ground pressure that recent studies suggest being closer to 0.5~bar \cite{Som2016,Zerkle2017,Laneuville2018}.
For a simple example of the complexity to extend research to exoplanets, consider recent work by \citeA{Airapetian2016}, which suggests that the higher solar activity has led to chemical reactions creating N$_2$O, a very efficient greenhouse gas in the Early Earth's troposphere.

Another uncertainty comes from the magnetic activity of the host star, responsible for the space-weather conditions of close-in planets, and expected to be much stronger for lower mass stars such as the Trappist--1 system star and M dwarfs in general. Since those stars could remain as active as the young Sun throughout their lifetime \cite{airapetian2019}, it is theoretically possible that some of the planets orbiting them  are currently subject to a N$_2$O greenhouse effect while at the same time being out of the standard HZ. 

In order to produce a more useful concept of habitability, we must contend with all the processes that lead to the habitability of a planet, and how the different variables (such as the type of star, the rotation rate of the planet, etc.) affect it. The formation of a planetary atmosphere is a balance between the amount of volatiles brought during the accretion phase, and subsequently outgassed, and the subsequent escape or fixing of volatiles as the planet evolves. \cite[Figure \ref{atmospherecreation}]{Lammer2009}.

Atmospheric escape is often overlooked in this type of analysis or only approximated by an energy-limited hydrodynamic escape. Modeling based on this approximation led a fraction of the community to conclude that Pluto's atmosphere was greatly outgassing until the observations of New Horizons measured an escape rate four orders of magnitude lower than predicted \cite{ZHU2014301,gladstone_atmosphere_2016}. This leads to major questions concerning atmospheric escape that need to be solved.

\subsection{The Outstanding Questions of Atmosphere Escape}
\label{questions}
Several major questions about the evolution of planetary atmospheres have been asked \cite{Ehlmann2016}, such as: ``are [their] mass[es] and composition[s] sustainable?'',  ``how do [they] evolve with time?''. 
Recent studies of atmospheric escape have led to the following major questions, specific to escape, that are being answered through experimental studies (e.g. satellites such as Venus Express -VEX-, Mars Express --MEX--, Rosetta, Mars Atmosphere and Volatile and EvolutioN mission -MAVEN-, etc.), and theoretical work. 

\begin{enumerate}
\item \textbf{What is the current escape rate of planetary atmospheres, how does it vary with forcing parameters?} Measurements by different spacecraft enable estimates of the flux of ions and neutrals escaping a planet. However, limitations in temporal and spatial resolution render some observations very difficult; e.g., the ion plume of Mars was inferred from MEX observations, but only MAVEN could fully observe and characterize it. \cite{liemohn_mars_2014,dong_strong_2015}; ionospheric outflow at Earth is observed, but the fraction of ions coming back, and the variation of outflow with latitude, magnetic local time, and solar and geomagnetic activity, is difficult to address accurately \cite{Strangeway2000}.
\item \textbf{What was the escape rate in the past? How did it vary with the varying forcing parameters and the varying atmospheres of planets?} The isotopic composition of an atmosphere hints at changes in its composition, and can be used to evaluate the total atmospheric loss. However, if some major parameters of the composition have changed, extrapolating the current atmosphere to the past can be problematic. The Earth's atmosphere is an emblematic example of an atmosphere that has greatly changed, with the appearance of oxygen in large quantities after about 2.5~Gyr ago \cite{catling2014great}. Observations of Sun-like stars in different stages of their evolution suggest that the Sun had more sunspots and flares in the past, which, undoubtedly, changed the escape conditions of the planets in the Solar system \cite{Lammer2009}.
\item \textbf{How will escape and other atmospheric evolutionary processes shape the future of the planetary atmospheres we observe today} For example, what will the habitability of the Earth and Mars be in a billion years? Variation of the Earth's magnetic field may affect escape rates, and dramatically change the atmosphere of the Earth. At Mars, the atmospheric photochemistry may lead to H$_2$O escape with the oxidation of the crust if O is not escaping enough \cite{Lammer2003}. Recent modeling shows that CO$_2^+$ dissociative recombination is also an efficient loss channel \cite{lee_comparison_2015}. The adsorption of CO$_2$ into the crust \cite{Nakamura2002,Zent1995,hu_tracing_2015, mansfield_effect_2017} implies that future change in Mars' obliquity will increase the outgassing and therefore the surface atmospheric pressure of the planet. But what will happen if no more H$_2$O  compensates for the escape? Is it possible for all remaining Martian H$_2$O to escape? How much CO$_2$ could escape? 

\item \textbf{Does a magnetic field protect an atmosphere from escaping?}  Polar ionospheric outflow is an efficient process to accelerate ions to escape speed. Since it is driven by the energy of the solar wind, funneled by the magnetic field, the stronger the magnetic field, the more energy is available for ionospheric outflow. In that sense, a planet with a magnetic field could be more sensitive to escape \cite{gunell_why_2018}. However,  the returning component of the polar outflow is increasing, and therefore the net escape should be addressed in different conditions; there are many questions regarding how this component may evolve, and it may be so that it prevents an effective escape altogether. Do the similar escape rates measured at Earth, Venus, and Mars \cite{gunell_why_2018} mean that there is no effective shielding, or is the comparison between these planets flawed because the upper atmosphere composition, and therefore the exospheric temperature are extremely different? Is it just a coincidence that both the Earth and Titan are able to sustain a nitrogen atmosphere despite relatively large exospheric temperatures (more precisely low $\lambda_{ex}$ parameter, see Section \ref{Chiparameter}) while being immersed in a magnetosphere? Is the question of the magnetic field protection actually the relevant one?
\item \textbf{What is the escape rate from exoplanets; can we test our models against exoplanetary observations?} Some observed exoplanets are in hydrodynamic escape \cite{ehrenreich2015giant}. It is possible to observe more extreme regimes for exoplanets than for planets in the Solar system; therefore the models developed for the current Solar system conditions are likely to be inadequate for exoplanets. One of the main advantages of these tests is to be able to validate the conditions likely encountered in the early Solar system. One example of such a process that is believed to be more important in the past is sputtering, but how could we detect its efficiency at exoplanets? 
\end{enumerate}

\subsection{Analytical Approach}

The Solar system has a large variety of planetary bodies, with very different atmospheres, including Mars with a thin CO$_2$-rich atmosphere, Venus with a thick CO$_2$-rich atmosphere (both of those presenting evidence of substantial escape), or Earth with a N$_2$/O$_2$ atmosphere. The difference between these planets is, in a large part, determined by how they are losing their atmospheres. Several missions, such as MAVEN, MEX, and VEX have been giving insights on the evolution of planetary atmospheres through their escape to space, and have led to a better understanding of which important processes are active to date, and maybe in the past. In addition, work on comets, such as 67P with Rosetta, highlight some of the fundamental processes that lead to escape in slightly different regimes \cite{brain2016atmospheric}.  Unfortunately, these results cannot be simply extrapolated to exoplanets, since they may be subject to very different conditions.

To that extent, it is necessary to know: (1) what the possible mechanisms by which planets lose their atmosphere into space are, (2) how these mechanisms behave with different conditions, (3) how they produce different observables, and (4) what our current understanding of these mechanisms is. Ultimately, one would like to:
\begin{itemize}
\item Determine what the escape processes are: review all the processes that have been suggested in the literature, review what their suggested rates were, and, since definitions may vary between authors, decide for a standard definition.
\item Determine what the key parameters are for each escape process, i.e. what variations will be of importance, and how these parameters couple with each other.
\item Determine the unknown parameters that need to be addressed to answer the questions of section \ref{questions}.
\item Determine the observable for each escape process, and determine how to disentangle the observations of escape in different solar/stellar conditions to determine the relative importance of each processes.
\end{itemize}

This is why, in the present paper, we start by reviewing the different escape processes and their limitations (Section 2), what the major parameters that we need to know to calculate these escape processes and know their importance are (Section 3), before looking at how they influence the Solar system planets (Section 4) and some exoplanets (Section 5) in time. We will finally look at which measurements and models are needed to better understand the escape processes at planets and exoplanets (Section 6) before concluding.

\section{The Escape Processes}
\begin{figure}
 \noindent\includegraphics[width=40pc]{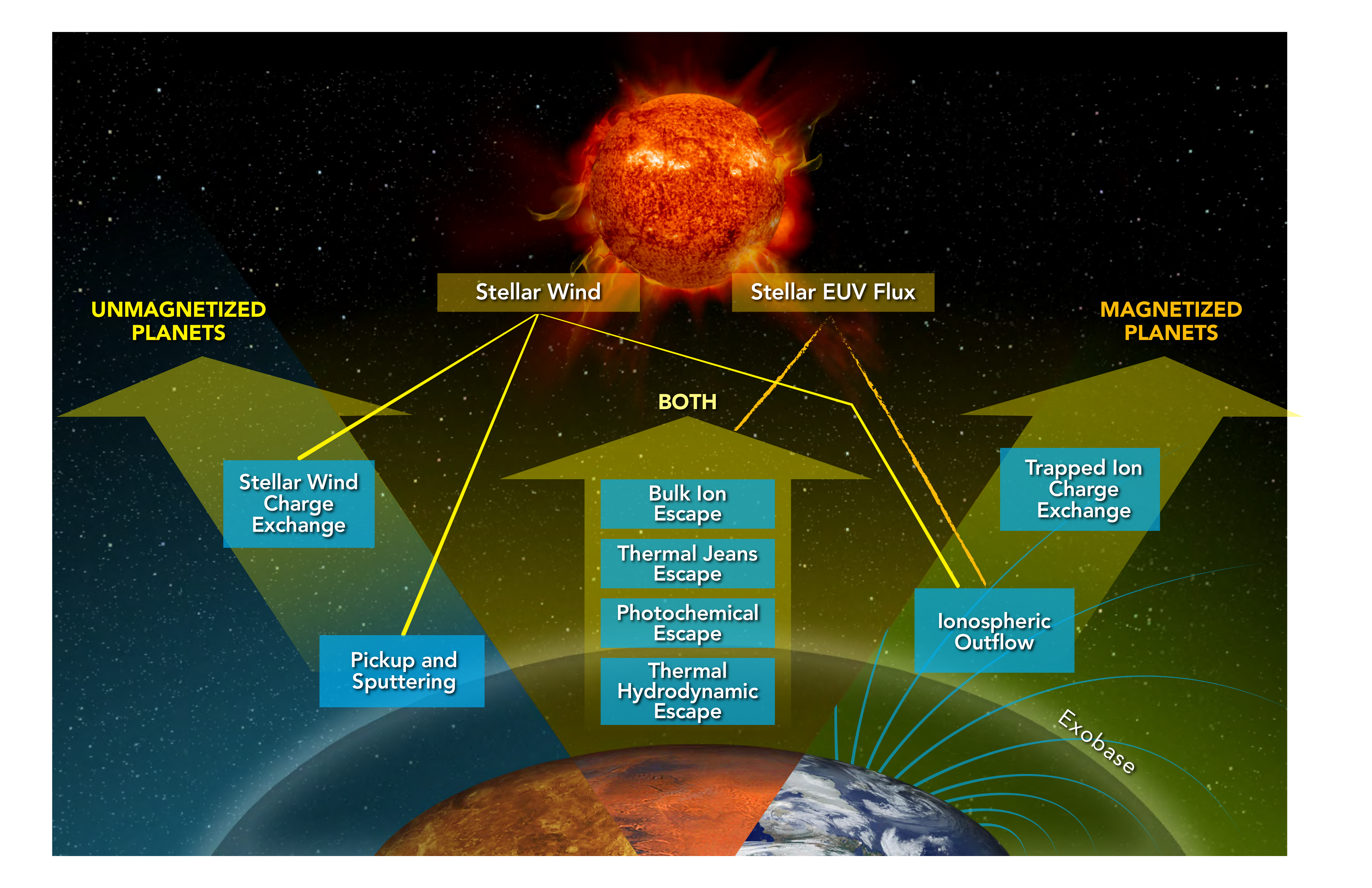}
 \caption{The main processes of atmospheric escape, along with their typical efficient altitudes domains (near the thermosphere/exobase or away from it)  and their conditions of efficiency/occurrence (magnetic field).}
 \label{escapeoverview}
\end{figure}

The escape processes are usually separated into two parts: the thermal and non-thermal processes. The thermal processes are dependent on the temperature
of the upper atmosphere, usually controlled by the host star's Extreme and X Ultraviolet (EUV-XUV) flux. The non-thermal processes are the result of more complex interactions, such as plasma interactions. Some non-thermal processes (such as sputtering) have a consistent nomenclature in the literature whereas others (such as ion outflow) have variable definitions depending on the authors. In Table~\ref{escapeprocesstable}, we summarize these escape processes and in Table~\ref{escapeparameterstable} their main parameters. Those escape processes are sketched on Figure~\ref{escapeoverview}, and an evaluation of the current escape rates can be found on Table~\ref{CurrentEscapeRatesTable}.

\begin{sidewaystable}
\tiny
\caption{The escape processes}
\begin{tabular}{|c|c|c|c|c|}
\hline
\textbf{Process} & \textbf{Origin} & \textbf{Key parameters}  \\
\hline
Jeans escape & Temperature accelerate particle above the escape velocity & Temperature, gravity, Tc: $\lambda_{ex}$ parameter $>$ 2.5 \\
\hline
Hydrodynamic escape & Thermal acceleration in a fluid way & Temperature, gravity, Tc: $\lambda_{ex}$ parameter $<$ 2.5\\
\hline
Photochemical/Ion recombination& Ion recombination releasing kinetic energy & Low gravity, molecular ion, requires ionosphere densities \\
\hline
Photochemical/Dissociations (photon, etc)& Molecular photodissociation release kinetic energy  & Requires thermosphere densities, low gravity \\
\hline
Ion Pickup& Solar wind picks up ions from ionosphere  & Requires compressed/no magnetosphere  \\
\hline
Ion Sputtering& Accelerated ions from the ionosphere translate their kinetic energy  & Requires compressed/no magnetosphere, B, U$_{sw}$ \\
\hline
Charge exchange/trapped& Fast ion trapped in magnetosphere becomes ENA through charge exchange  & Requires magnetosphere, ion density and temperature, neutral densities \\
\hline
Charge exchange/solar wind& Solar wind ion becomes ENA that can access thermosphere and increases heating & Requires large coronae, U$_{sw}$, N$_{sw}$\\
\hline
Charge exchange/particle precipitation,& Particle precipitating in thermosphere becomes ENA and translate kinetic energy  & Requires precipitaton fluxes, cross sections\\
\hline
Ionospheric outflow& creation of ion upward wind through ambipolar diffusion  &  requires fields, ionosphere\\
(often called polar wind in magnetized planets) &   &  \\
\hline
Other ion escape& Plasma instabilities leading to ions going upwards and being picked by the solar wind  & Requires fields, ion density and temperature  \\
\hline
\end{tabular}
\label{escapeprocesstable}
\end{sidewaystable}

\begin{sidewaystable}
\tiny
 \caption{Escape at Planets, parameters compiled from  \citeA{hinson_radio_2017,Young_structure_2018,johnson2013}}
 \begin{tabular}{|c|c|c|c|c|c|c|c|c|c|c|}
 \hline
 \textbf{Planet} & \textbf{Jeans $\lambda_{ex}$ parameter} & \textbf{$T_e$(K)} & \textbf{$T_c$(K)} & \textbf{g(m/s$^2$) } & \textbf{R(km)} & \textbf{H$_{exo}$(km)} &\textbf{B(Gauss-R$^3$)} & \textbf{Average Solar EUV(W/m$^2$)} & \textbf{Solar Wind Pressure(nPa) } & \textbf{$Q_c$(W)} \\
 \hline
 Mercury & 2.2 & 500 & 725 & 3.70 & 2439.7 &  & 0.002 & 9082.7 & 13.8-21.0 & 7.31 x 10$^{10}$\\
 
 Venus & 22.3 & 290 & 4307 & 8.87 & 6051.8 & 15.9 &  & 2601.3 & 1.0-12.0 & 2.64 x 10$^{12}$\\
 
 Earth & 9.4-5.0 & 800-1600 & 5020 & 9.80 & 6378.1 & 8.5 & 0.306 & 1361.0 & 1.0-6.0 & 3.51 x 10$^{12}$\\
 
 Moon & 0.8 & 226 & 400 & 1.62 & 1738.1 & &  & 1361.0 & 1.0-6.0 & 1.03 x 10$^{10}$\\
 
 Mars & 6.3-5.0 & 240-300 & 1014 & 3.71 & 3396.2 & 11.1 & & 586.2 & 0.1-1.1 & 1.68 x 10$^{11}$\\
 
 Jupiter & 311-218 & 700-1000 & 145000 & 24.79 & 71492 & 27.0 & 4.30 & 50.26 & 0.05-0.10 & 5.92 x 10$^{15}$\\
 
 Saturn & 157-98 & 500-800 & 52200 & 10.44 & 60268 & 59.5 & 0.215 & 14.82 & 0.01-0.09 & 1.06 x 10$^{15}$\\
 
 Titan & 2.3 & 180 & 280 & 1.35 & 2575 & & & 14.82 & 0.01-0.09 & 1.84 x 10$^{10}$\\
 
 for CH$_4$ & 37.3 & & 4475 & & & & & & & 3.7 x 10$^{11}$\\
 
 Uranus & 34 & 800 & 18300 & 8.87 & 25559 & 27.7 & 0.228 & 3.69 & 0.001-0.02 & 1.14 x 10$^{14}$\\
 
 Neptune & 48 & 700 & 22250 & 11.15 & 24764 & 19.1-20.3 & 0.142 & 1.508 & & 1.26 x 10$^{14}$\\
 
 Pluto & 15.1 & 68 & 408 & 0.62 & 1184 & 78 & & 0.873 & 0.006 & 8.4 x 10$^8$\\
 
 for CH$_4$ & 8.5 & & 384 & &  & 59 & & & & 1.6 x 10$^{10}$\\
\hline
 \end{tabular}
 \label{escapeparameterstable}
\end{sidewaystable}

\begin{sidewaystable}
\small
\caption{The Present Escape Values. Total escape, in s$^{-1}$, followed by the fluxes, in  cm$^{-2}$s$^{-1}$. Both are reported in the literature. While fluxes show the magnitude at a given planet, highlighting the intensity, total escape highlights the overall aspect of escape; when comparing planet to planet, none are satisfying since comparing the total escape of a Mars with e.g. Venus hides the size effects. On the other hand, comparing the fluxes from Earth with e.g Mars hides local effects like exospheric temperature. We decided to show both values. References: $^1$ - \citeA{lammer_atmospheric_2008} and references therein. $^2$ - \citeA{tian_atmospheric_2013} and references therein. $^3$ - \citeA{jakosky2018loss} and references therein. $^4$-\citeA{gunell_why_2018} and references therein. $^5$ - \citeA{Inui2019} and references therein. $^6$ At Earth, the Jeans’ flux for the solar max is basically the same flux  as for charge exchange at the solar min because H escape is diffusion limited, see Section~2.3.2.
NB: we considered the fluxes at the exobases, at Venus, a 200~km exobase has a surface of 4.9 10$^{18}$ cm$^2$; at Earth, for a 500~km exobase, 5.9 10$^{18}$~cm$^2$; at Mars, for a 200~km exobase, 1.6 10$^{18}$~cm$^2$.}
\begin{tabular}{|c|c|c|c|}
\hline
\textbf{Process} & \textbf{Venus} & \textbf{Earth} & \textbf{Mars}\\
\hline
Jeans escape & 2.5$\times$10$^{19}$ -- 5.1 $^{(1)}$ & H: 6$\times$10$^{26}$-- 10$^8$ (Solar Max)$^{(2,6)}$ & H: 1.6$\times$10$^{26}$ - 1.1$\times$10$^{27}$ -- 10$^{8}$ - 6.9$\times$10$^{8}$  $^{(3)}$\\
Charge exchange/trapped & H: 5$\times$10$^{24}$-5$\times$10$^{25}$ -- 10$^{6}$ - 10$^{7}$ $^{(2)}$ & H: 6$\times$10$^{26}$-- 10$^{8}$ (Solar Min)$^{(2,6)}$ & 10$^{22}$-10$^{23}$ -- 10$^4$-10$^5$ $^{(2)}$\\
Ion pickup & H$^+$: 10$^{25}$ --2$\times$10$^{6}$ $^{(1)}$ ; O$^+$ :1.5$\times$10$^{25}$ -- 3$\times$10$^{6}$ $^{(1)}$& Small $^{(2)}$& O$^+$: 10$^{24}$ - 10$^{6}$ $^{(2,3)}$ ; \\
           &         He$^+$:  5$\times$10$^{23}$ - 5$\times$10$^{24}$ -- 10$^{5}$ - 10$^{6}$ $^{(2)}$   &            & C$^+$: 1.6$\times$10$^{23}$--10$^5$ $^{(2)}$    \\
Sputtering & O: 5$\times$10$^{23}$-5$\times$10$^{24}$ -- 10$^{5}$-10$^{6}$ $^{(2)}$& Small $^{(2)}$&  O: 3$\times$10$^{24}$ -- 1.8$\times$10$^{6}$ $^{(3)}$ \\
          &               &       &   C: 10$^{23}$ -- 10$^{5}$ (Solar Min)  10$^{25}$ -- 10$^{7}$ (Solar Max)$^{(2)}$\\
Photochemical escape & 3.8$\times$10$^{25}$ -- 7.7$\times$10$^{6}$ $^{(1)}$& Small $^{(2)}$& O: 5$\times$10$^{25}$ - 3$\times$10$^{7}$ $^{(3)}$; C: 10$^{24}$--10$^6$ $^{(2)}$\\
Magnetized Ion Outflow & N/A & H$^+$: 8$\times$10$^{25}$ -- 1.3$\times$10$^{7}$ $^{(4)}$ & N/A\\
(Polar wind)  &  &  O$^+$: 3$\times$10$^{25}$ -- 5$\times$10$^{6}$ $^{(4)}$& (crustal escape included in outflow)\\

Unmagnetized Ion Outflow/K-H/Clouds & 5$\times$10$^{24}$-1$\times$10$^{25}$ -- 1$\times$10$^{6}$-2$\times$10$^{6}$ $^{(1)}$ & O$^+$: 6$\times$10$^{24}$ -- 1$\times$10$^{6}$ $^{(2)}$&  10$^{25}$ -  10$^{7}$ $^{(5)}$\\
\hline
\end{tabular}

\label{CurrentEscapeRatesTable}
\end{sidewaystable}

Non-thermal escape processes can be separated into Photochemical loss (Section \ref{pcloss}), Ion loss (Section \ref{ioloss}), Ionospheric outflow (Section \ref{ionosphericoutflow}), and Other losses (Section \ref{oloss}). Moreover, in order to compute the total loss of an atmosphere into space, it is necessary to take into account the problem of the ion return (Section \ref{ionreturn}). It is important to note that, while we are separating these processes, they do influence each other, and sometimes one leads to the other. For example, an ionospheric outflow process at Venus can produce fast particles involved in ion pickup and sputtering \cite{luhmann2008venus}.

\subsection{Thermal Escape}

Thermal escape is one of the most important escape processes \cite{chassefiere2004,Selsis2006}. It takes place in two regimes, Jeans escape and hydrodynamic escape, with a transition regime that is the subject of recent studies (e.g. \citeA{strobel_titans_2008, volkov_thermally_2011,volkov_kinetic_2011a, Erkaev2015}). Most of the observed isotopic fractionation in planetary atmospheres is interpreted as originating from thermal escape because of its energy efficiency at escaping large amount of gases.

 \subsubsection{Fundamental Theory}
 \paragraph{Jeans regime}

 The neutral atmospheric constituents in the upper atmosphere are in local thermodynamic equilibrium (or close to it). Therefore, their distribution  function can be approximated by a Maxwellian function \cite{Mihalas1984}:
 
\begin{eqnarray}
   f(\vec{x}, \vec{v}) &=& N \left( \frac{m}{2\pi k T} \right) ^ {3/2} e^{-\frac{m v^2}{2 k T}} \nonumber \\
                       &=& N \left ( \frac{1}{u_i\sqrt\pi} \right )^{3} e^{-v^2/u_i^2} \label{eq:maxwelldist}
\end{eqnarray}
where $u_i = \sqrt{\frac{2kT}{m_i}}$ is referred to as the thermal speed for the species $i$.

  The exobase is quantitatively defined as the level where $l_i$, the mean free path of the $i$th constituent is equal to the scale height ($H$) \cite{Hunten1973,Shizgal1996}. At the exobase, we consider that a molecule of the $i$th constituent going upwards at the escape velocity, $v_{esc} = \sqrt{2GM/r}$ will not impact another molecule, and therefore will escape.  This approximation is the equivalent of considering an atmosphere collisionless above the exobase and fully collisional below \cite{1983RvGSP..21...75F}. 
 
 By integrating the vertical flux, $v_i \cos(\theta) \times f_i$, at the exobase, for the velocities greater than the escape velocity ($v_{esc}$), and neglecting the collisions above it, we retrieve the flux of escaping molecules. 
 
  \begin{equation}\Phi_i(\textrm{escape}) = \int\limits_0^{2\pi} \int\limits_0^{\pi/2} \int\limits_{v_{esc}}^\infty v_i \cos(\theta) f(v_i) v_i^2 \sin(\theta) dv_i d\theta d\Psi \label{eq:maxwellint}  \end{equation} 
Carrying out this integration gives:
\begin{eqnarray}
  \Phi_i(\textrm{escape}) &=& N_{i} \left(\frac{k T_e}{2 \pi m_i}\right) ^ {1/2} \left(1 + \frac{m_i v_{esc} ^ 2 }{2 k T_e}\right)  e^{- \frac{m_i v_{esc}^2}{2 k T_e}} \label{eq:jeansesc1} \\
                 &=& N_{i} \left(\frac{u_i}{2 \sqrt\pi}\right) \left(1 + \frac{v_{exc}^2}{u_i^2}\right)  e^{- \frac{v_{exc}^2}{u_i^2}} \label{eq:jeansesc2} 
\end{eqnarray} 
In Eqs. \ref{eq:jeansesc1} and \ref{eq:jeansesc2}, it is important to use the values of $v_{esc}$ and $u_i$ at the exobase (using the temperature, $T_{exo}$,  and radius $r_{exo}$ at the exobase) to get a correct estimation of the escape flux. 

It is common to introduce the non-dimensional {\it Jeans parameter} to express the escape flux, and we will see later that this parameter is very useful in understanding the thermal escape process. The Jeans parameter is the ratio of gravitational energy to thermal energy, expressed at $\lambda_{ex}= \frac{(GMm_i/r)}{kT} = \frac{v_{esc}^2}{u_i^2}$. Using this, the escape flux becomes: 
\begin{equation} 
  \Phi_i(\textrm{escape}) = N_{i} \left(\frac{u_i}{2 \sqrt{\pi} }\right) \left(1 + \lambda_{ex} \right)  e^{-\lambda_{ex}} \label{eq:jeansesc3} 
\end{equation} 

Eq. \ref{eq:maxwellint} assumes that we can approximate the distribution at the exobase by a Maxwellian despite the fact that molecules faster than $v_{esc}$ are removed.
When the escape rate is high enough, a non-Maxwellian correction must be applied to consider that the high-energy tail of the Maxwellian is depleted, following \citeA{chamberlain1971}. This correction \underline{lowers} the escape rate by about 25\%. However, this correction is based on the assumption of an isothermal atmosphere below the exobase and has been evaluated for H and He escape within a O or CO$_2$ rich background atmosphere, i.e. the thermosphere of the Earth and Mars/Venus etc.

A more realistic simulation, performed by \citeA{Merryfield1994}, that considered the effect of increasing temperature with altitude, shows that the escape from the deeper layer should also be considered (i.e. it cannot be assumed that a Maxwellian is a good approximation of the atomic/molecular distribution at the exobase). In that case, the correction is an \underline{increase} of the order of 30\%. Therefore, for extremely precise determination of escape, it is important to solve the  Boltzmann equation; one of the most used techniques is the Direct Simulation Monte Carlo (DSMC) method \cite{volkov_thermally_2011,tucker_thermally_2009}, whose results show that the source of escaping particles is distributed over a wide altitude range above and below the exobase.

Overall, equation~(\ref{eq:jeansesc2}) is a good approximation for the thermal escape when the atmosphere is strongly gravitationally bound to the planet, and this formula is valid for all the constituents independently. Ideally this equation would be evaluated at or near the nominal exobase, but can be applied far below the exobase assuming no addition heating if a correction factor is applied \cite{Volkov:2015, Johnson:2016}.

It is important to note that while the Jeans parameter is the main parameter of thermal escape, the location of the exobase is extremely important. In the case of Titan or a possible early Moon atmosphere \cite{Aleinov2019} the altitude of the exobase is non-negligible compared to the radius of the planet, and while the flux per unit surface is small, it can become the most important source of loss when taking the whole exobase surface into account.

\paragraph{Hydrodynamic regime}
\label{hydrodynamicescape}
In cases where the internal energy of individual gas molecules approaches the kinetic energy required for escape ($\lambda_{ex} \approx 1$), the gas will begin to escape as a flow of continuous fluid \cite{Hunten1973,Gross1972escape, watson_dynamics_1981}.   

Qualitatively, the fundamental distinctions between the Jeans and hydrodynamic regimes can be viewed in two helpful and complementary ways. First, the Jeans regime is ``collisionless'' \cite{Shizgal1996}: it is one where collisions between molecules define an exobase as a surface (or at least a narrow region). The atmosphere is not only retained by the gravitational pull on individual molecules but also by the effective force of collision with other atmospheric molecules. In the ``collisional'' hydrodynamic regime, the molecules are so energetic that collisions are insufficient to restrict escape. Indeed, the escaping flow of lighter gases (the ones that are most likely to be escaping) is capable of exerting an effective force and dragging heavier gas molecules such as water and the noble gases \cite{pepin_origin_1991,ZahnleKasting1986}. And furthermore, hydrodynamic escape can take place far below the exobase. 

Second, the distinction between Jeans and hydrodynamic escape is analogous to that between evaporation at temperatures below the boiling point and boiling. In this analogy, the exobase is like the surface of the evaporating fluid, the force of gravity is like the atmospheric pressure, and the effective pressure of other molecules is like the effective pressure of other molecules in the liquid.   
Quantitatively, hydrodynamic escape is approached by numerical solution of an appropriate system of inviscid fluid dynamical equations \cite{watson_dynamics_1981,tian_hydrodynamic_2005}. For instance, the one-dimensional time-dependent inviscid equations for a single constituent, thermally conductive atmosphere in spherical geometry is given by \citeA{tian_hydrodynamic_2005} as: 

\begin{eqnarray} \frac{\partial(\rho r^2)}{\partial t}+\frac{\partial(\rho v r^2)}{\partial r}&=&0 \label{eq:hydrocont} \\ \frac{\partial(\rho v r^2)}{\partial t}+\frac{\partial(\rho v^2 r^2 + p r^2)}{\partial r}&=&-\rho G M + 2 p r, \label{eq:hydromom} \\ \frac{\partial(E r^2)}{\partial t}+\frac{\partial[(E+p)v r^2]}{\partial r}&=&-\rho v G M + q r^2 + \frac{\partial \kappa r^2 \frac{\partial T}{\partial r}}{\partial r}  \label{eq:hydroenergy} \end{eqnarray}

where $E = \rho(v^2/2 + e)$ (the equation for the total energy density), $e=p/[\rho (\gamma - 1)]$ (the definition of the internal energy), and $p=\rho R T$ (the ideal gas law). Here, $\rho$ is the density of the gas, $p$ is the pressure, $\gamma$ is the polytropic index of the gas, $R$ is the universal gas constant, $\kappa$ is the thermal conductivity, and $q$ is the volume heating rate. Thus, Eq. \ref{eq:hydrocont} is the continuity equation, Eq. \ref{eq:hydromom} is the momentum conservation equation, and Eq. \ref{eq:hydroenergy} is the energy conservation equation. 

If energy conservation is neglected and the temperature is held constant, steady state solutions to the above system are possible. It is convenient in that case to rewrite $p$ and $GM$ such that:
\begin{eqnarray}
p &=& \rho v_0^2 \label{eq:soundspeed} \\
GM&=&2r_0 v_0^2, \label{eq:critradius}
\end{eqnarray}
where $v_0$ is the sound speed and $r_0$ is a critical radius based on the relative gravitational potential energy and kinetic energy of a particle at the sound speed.

A single differential equation is then obtained:
\begin{eqnarray} \frac{1}{v} \frac{dv}{dr} \bigg(1-\frac{v^2}{v_0^2}\bigg) &=& \frac{2r_0}{r^2} - \frac{2}{r} \label{eq:hyddiffeq} \end{eqnarray}

Eq. \ref{eq:hyddiffeq} has an obvious critical point at $(\pm v_0, r_0)$ and thus various solutions to the differential equation can be derived by integrating from these two critical points to some other velocity and radius, assuming $v_0$ is constant. The example of interest here is the transonic outflow solution obtained by integrating Eq. \ref{eq:hyddiffeq} from the critical point to higher velocity, $v$, and radius, $r$.

\begin{equation}
\log\frac{v}{v_0}-\frac{1}{2}\bigg(\frac{v^2}{v_0^2}\bigg)+\frac{2r_0}{r}+2\log\frac{r}{r_0}=\frac{3}{2}  \label{eq:transonic}
\end{equation}

A good discussion of the various solutions and their significance can be found in \citeA{Pierrehumbert2010}. 

It is possible to derive a theoretical upper bound for hydrodynamic escape of a single constituent atmosphere at a given temperature and atmospheric level. This bound is given by Eq. \ref{eq:jeansesc2} in the limit where $\lambda_{ex} \approx 0$ \cite{Hunten1973}:

\begin{equation} \Phi_i(\textrm{escape}) = N \left(\frac{k T_e}{2 \pi m}\right) ^ {1/2} \label{eq:hydul} \end{equation} 

At this bound, escape takes the form of a one-dimensional outflow at the thermal velocity. In realistic models of atmospheres, however, hydrodynamic loss rates tend to be much lower than the theoretical limit for reasons to be discussed below.
In addition, the use of an hydrodynamic escape approach is easily abused, especially when many assumptions have to be made on the nature of the atmosphere (such as the composition and the exospheric temperature). A solution to that problem is to estimate the \textbf{critical heating rate} \cite[and Section \ref{criticalheatingrate}]{johnson2013,johnson2013_erratum}.

\paragraph{Fluid-Kinetic Models}
Applying the Jeans equation requires that the temperature and density to be known near the exobase.
As an alternative to the hydrodynamic model, one can still use the fluid equations \ref{eq:hydrocont} - \ref{eq:hydroenergy} by utilizing the Jeans escape rate and energy escape rate as upper boundary conditions.  
This Fluid-Jeans model has been adapted to hot gas giants \cite{Yelle2004}, as well as to terrestrial planets like Earth \cite{Tian2008_1}.
One advantage of these methods is the solution is valid up to the exobase, so that heating, photochemistry and diffusion can be included and their effects on the escape rate investigated.
Using this Fluid-Jeans model, \citeA{Tucker2012} and \citeA{Erwin:2013} refined the escape rate using DSMC to get a Fluid-DSMC result. 
This extends the Fluid-Jeans result to model the transition from the collisional to collisionless regimes, and demonstrates the breakdown in the fluid equations below the exobase.
These models predict that escape rates at Titan and Pluto
are roughly consistent with Jeans escape even with low gravity or with high-heating rates.

\paragraph{Limiting factors to thermal escape}
\label{enelimited}

One limit arises from the impact of thermal escape on the energetics of the upper atmosphere. Removal of the escaping material either will cool the atmosphere around the exobase or lower the altitude of the exobase. Either way, some energy source will be necessary to maintain 
escape in a steady state. That energy source is whatever stellar  
EUV $<$ 90 nm that can be absorbed near the exobase (this absorption threshold is set for H, as this is the main species evaluated in the context of energy limited escape, but is generally valid for upper atmospheric species)  which results in one form of the energy-limited escape rate \cite{watson_dynamics_1981,Erkaev2007}:

\begin{equation} \Phi_i(\textrm{limited escape}) = \epsilon F_{\odot,EUV} \left(\frac{GMm}{r}\right)^{-1} \label{eq:hydenul} \end{equation}

where $\epsilon$ is the EUV heating efficiency and $F_{\odot,EUV}$ is the solar (or stellar) EUV flux. 

When fluxes of EUV are high, energy-limited escape is defined by the balance between conductive heating of absorbed solar EUV from the exobase with adiabatic cooling of the thermosphere, as initially argued by \citeA{watson_dynamics_1981}. Thus, at increasingly higher levels of EUV flux, the thermospheric temperature profile should evolve from one in which temperature increases monotonically to the exobase to one where peak temperature is significantly below the exobase. And after a certain point, the higher the incoming solar flux, the lower the exobase temperature \cite{Tian2008_1}: a regime thought to have limited the thermal escape rate on the early Earth (the authors refer to hydrodynamic escape as the regime where the adiabatic flow is important in the upper thermosphere, even though they are using Jeans escape to define the escape rate at the upper boundary). Where increased EUV flux simultaneously reduces other cooling mechanisms (such as IR emission from CO$_2$ on Mars \cite{Tian2009}), the adiabatic cooling-driven energy limit to thermal escape is less relevant. \citeA{Erwin:2013} showed that Pluto's atmospheric escape is energy-limited even with the small EUV flux experienced at its orbit.

The final limit arises from the impact of thermal escape on the composition of the upper atmosphere. Escaping species typically cannot be supplied to the escaping region of the atmosphere at rates comparable to the various theoretical upper limits for either Jeans or hydrodynamic escape. Escape rates are then controlled by the flux of escaping species to the region of escape, a regime known as diffusion-limited escape because diffusion is the principal transport mechanism in the escape regions of the most familiar planetary atmospheres \cite{Hunten1973,KastingCatling2003}. Consider a trace gas of density $n_i$ significantly lighter than the mean molecular mass of a planetary atmosphere and present at the homopause, where eddy diffusion is too weak to mix the atmosphere thoroughly. The separate gases will unmix by molecular diffusion and segregate. Unmixing at the homopause sets the limiting diffusion rate, which is dependent on the mixing ratio at the homopause itself as well as the diffusion coefficient of the light trace gas in the heavier principal constituents $({b_{i,dom}})$ \cite{Hunten1973} (In the following equation the mass of the trace gas, is very small compared to the mass of the principal constituent, its number density is also very small, and we neglect thermal diffusion; we will explore diffusion limited escape more in Section~\ref{diffusionlimitedescape}; $n$ is the number density of the main constituent).

\begin{equation} \Phi_i(\textrm{escape}) = \bigg(\frac{b_{i,dom}}{H}\bigg) \frac{n_{i,\textrm{homopause}}}{n} \label{eq:dlesc} \end{equation}

Note that the dependence of the diffusion rate at the homopause on the concentration of the light species at the homopause makes diffusion-limited transport also dependent on all barriers to transport of the light gas lower in the atmosphere such as an atmospheric cold trap. Escape of H at Earth is a perfect example of diffusion limited escape \cite{Shizgal1996}.

Thus, thermal escape has three classes of rate limit: (1) an absolute one based on fluid dynamics at the exobase; (2) an energetic one based on the absorption of solar EUV near the exobase; and (3) a compositional one based on atmospheric vertical transport below the exobase.

 \subsubsection{Key parameters}
 \label{Chiparameter}
 The most important parameters controlling thermal escape are the atmospheric scale height in the thermosphere, H,  which depends upon the exospheric temperature  T$_e$ (Section \ref{te}) and the  mass of the atmospheric constituents, $m_i$.

 The regime of thermal escape is governed by the dimensionless Jeans parameter $\lambda_{ex}=(GMm_i/r)/(kT)$  with $r$ being taken either as the distance from the center of the planet to the surface, exobase, or the location of the molecule(s) in question. 
 There is a critical value for  $\lambda_{ex}$, \underline{below which} there is a transition between between hydrodynamic and Jeans escape. Simply equating the internal energy and the escape velocity would suggest that the critical value of $\lambda_{ex}$ is $\frac{1}{\gamma-1}$, with $\gamma$ begin the heat capacity ratio, directly linked to the degree of freedom of the molecule/atom. This would correspond to 1.5 for ideal monoatomic gases and 2.5 for ideal diatomic gases. Thus, \citeA{Selsis2006} refers to a critical value of 1.5 for simplicity. Simulations by \citeA{Erkaev2015} of an atmosphere dominated by $H_2$ show a transition in escape rates near $\lambda_{ex}= 2.5$, which implies that $\frac{1}{\gamma-1}$ is indeed a good estimate of $\lambda_{ex}$.

 Following \citeA{Selsis2006}, we can define a critical temperature $T_c$ for which $\lambda_{ex}=1.5$ for the different planets, which is valid for a H atom.

 At Titan, the eventual escape of material to space is determined by the combined effects of the deep atmosphere limiting flux and the effects of photochemical loss (for CH$_{4}$) or production (for H$_{2}$) above the homopause region \cite{Bell2014}. Each of the major species, N$_{2}$, CH$_{4}$, and H$_{2}$ possess separate critical points, but the nominal exobase is located near 1500 km, which is a significant fraction of the radius of Titan (2575 km). 

\citeA{Selsis2006} gives a table of critical temperature, $\lambda_{ex}$ (noted $\chi$ in that paper) and exospheric temperature for different objects in the Solar system. Table \ref{escapeparameterstable} is an update taking into account the recent data, e.g. from New Horizons.

In \citeA{johnson2013,johnson2013_erratum} a criterion for where the transition between Jeans escape and hydrodynamic escape should be considered, based on the heating rates, has been described.

\subsubsection{Questions and Important Points}
 \paragraph{How the transition between the thermal escape and the hydrodynamic escape is done?}
 \label{slowhydro}

 Motivated by {\it Cassini} spacecraft data for Titan, and {\it New Horizons} data for Pluto, there has been renewed interest in the physical assumptions underlying planetary escape. Following \citeA{hunten_thermal_1982}, it was assumed that if the binding parameter $\lambda_{ex} < 1$ near the exobase, an organized hydrodynamic flow would result, whereas if $\lambda_{ex} > 10$ that collisionless Jeans escape would result. Intermediate models called {\it slow hydrodynamic escape} including transport effects such as thermal conduction were also developed \cite{watson_dynamics_1981, hunten_stability_1982,strobel_titans_2008} to bridge the intermediate values of $\lambda_{ex}$ between the two limits. Recently, \citeA{volkov_thermally_2011} used DSMC to model atom/molecule motions under gravity and collisions. It was assumed that heating occurred below the base of the simulation domain, so that particles enter the domain with a Maxwell-Boltzmann distribution at a prescribed temperature. Subsequent collisions between particles then transport heat upward effectively by a heat conduction flux (although the Fourier law may be inaccurate to describe this flux). The particle density at the base of the simulation domain was parameterized through the ratio of mean free path to the scale height (the Knudsen number), which is a measure of the frequency of collisions. The surprising result of the simulations presented in \citeA{volkov_thermally_2011} was that a sharp transition occurs from the hydrodynamic to the Jeans escape limits, near $\lambda_{ex} \sim 2-3$ depending on the particle interaction law. Analytic support of these results was given in \citeA{gruzinov_rate_2011}. For $\lambda_{ex} > 3$, the bulk fluid velocity never becomes supersonic, and the escape rate is near the Jeans escape rate. Hence, given the assumptions of that study, hydrodynamic outflow is limited to small values of the binding parameter. Early in the Cassini mission to the Saturn system, \citeA{strobel_titans_2008} posited that slow hydrodynamic escape could be occurring in the upper atmosphere of Titan, due the moon's low gravity and the extended nature of its atmosphere. Further still, the combined works of \citeA{strobel_titans_2008,Strobel2012} and \citeA{Yelle2008} went a step further and suggested that hydrodynamic escape was in fact the only mechanism that could adequately reproduce the observations of methane.  However, later investigations by \citeA{Bell2011} and later in \citeA{Bell2014} demonstrated that, by self-consistently coupling dynamics, composition, and thermal structure calculations, that the in-situ measurements of methane by the Ion-Neutral Mass Spectrometer (INMS) \cite{Waite2004,Magee2009} could be explained with the atmosphere in a nearly diffusive state without the need for invoking slow hydrodynamic escape of methane. 
 
 Similar to the situation at Titan, the data obtained by the New Horizons flyby of Pluto and Charon was not consistent with a previously posited hydrodynamic escape mechanism occurring at the dwarf planet \citeA{Gladstone2015}. Prior to this observation, Pluto was suggested to be the archetype for a planetary atmosphere in a state of hydrodynamic escape. Instead, the DSMC simulation by \citeA{tucker_thermally_2009,Tucker2012}, which suggested that Pluto's atmosphere could be simulated without invoking hydrodynamic escape, seem to better match observations made by New Horizons.  Thus, despite being posited as occurring at several bodies in the Solar system, there is no clear evidence for slow hydrodynamic escape occurring in our Solar system during the current epoch.
 
\subsubsection{Observables}
\label{hydrodynamicfractionation}
When observing escape in real time, thermal escape can be viewed as principally a function of the density of the escaping species and exospheric temperature (Eq. \ref{eq:jeansesc2}). A typical technique is to infer density and temperature from airglow emission, which is also a function of density and exospheric temperature (e.g. \citeA{Chaffin:2014}). In some cases, \textit{in-situ} mass spectrometry of neutrals can enable better constraints on density (e.g. \citeA{Cui:2008}), while satellite drag can add yet another constraint jointly dependent on bulk atmospheric density and temperature (e.g. \citeA{Krauss:2012}).      

The central value of observing airglow emission for planets in the Solar system and the difficulty of obtaining additional constraints on escape from exoplanets strongly suggests that airglow emission will be the key observable for quantifying thermal escape at exoplanets, whether by Jeans or hydrodynamic escape. The expected observable for intense hydrodynamic escape is of a highly extended hydrogen corona containing relatively large amounts of heavier atoms rather than a rapid fall-off in the concentration of such atoms beyond the exobase \cite{Vidal-Madjar2003}. Airglow, however, is extremely difficult to observe at exoplanets and can be affected by particle precipitation \cite{bernard2014can}. For small/rocky planets such as a Earth-like or a Mars-like exoplanet, a technique based on CO$_2$ or O$_2$ absorption due to stellar occultation in the near UV can be used, but is extremely challenging \cite{Gronoff2014}.

The main observable for thermal escape in a planet's past is mass fractionation of the isotopic composition of the atmosphere from the stellar value. However, caution must be exercised. Isotopic composition can be affected by the outgassing of primordial materials and low-temperature chemical reactions unrelated to escape \cite{pepin_atmospheres_2006,Pope05032012}. Moreover, isotopic composition is strongly sensitive to Jeans escape but variably sensitive to hydrodynamic escape.

For Jeans escape, it can be inferred from Eq. \ref{eq:jeansesc2} that the escape rate is proportional to $m_i^{-1/2}$ for small values of $\lambda_{ex}$ and $m_i^{1/2} e^{-\lambda_{ex}}$ for large values of $\lambda_{ex}$. The former case would be hydrodynamic escape. So for Jeans escape, deuterium escapes at a rate less than atomic hydrogen. 

In the case of hydrodynamic escape, the principal escaping species drags gases lighter than the ``crossover mass'' ($m_c$) \cite{Hunten1987}. 

\begin{equation} m_c=m_{esc} + \frac{kT\Phi_{esc}}{b g X_{esc}} \label{eq:hydxovermass} \end{equation}
where $_{esc}$ refers to the principal escaping species, b is the binary diffusion coefficient (the diffusion coefficient in a 2-components gas), and $X$ is the mole fraction.  If the escape flux of the principal escaping species can be defined at a reference altitude $\Phi^{\circ}_{esc}$ and is sufficiently small, then the escape flux of the trace species at the reference altitude $\Phi^0_{trace}$ is:

\begin{equation} \Phi^{\circ}_{\textrm{trace}}=\frac{X_{\textrm{trace}}}{X_{esc}}\Phi^{\circ}_{\textrm{esc}} \bigg[\frac{m_c-m_{\textrm{trace}}}{m_c-m_{esc}}\bigg] \label{eq:twospeciesflux1} \end{equation} \cite{Hunten1987}. It is in these slower hydrodynamic escape cases that significant fractionation is possible on geological timescales. Otherwise, the larger species are carried along with the flow. And everything scales with mole fraction. \begin{equation} \Phi^{\circ}_{\textrm{trace}}=\frac{X_{\textrm{trace}}}{X_{esc}}\Phi^{\circ}_{\textrm{esc}} \bigg[1-\frac{bg^{\circ}X_{esc}}{kT F^{\circ}_{esc}}(m_{\textrm{trace}}-m_{esc})\bigg] \label{eq:twospeciesflux2} \end{equation} \cite{Hunten1987}.
In this case, fluxes are weakly dependent on mass at masses close to the mass of the principal escaping species but more strongly dependent on mass at masses much greater than that of the principal escaping species, resulting in minimal fractionation of low mass species but significant fractionation of high mass species \cite{Hunten1987, tian_atmospheric_2013}.

As noted in \citeA{pepin_origin_1991,Shizgal1996,pepin_atmospheres_2006}, the uncertainty in the hydrodynamic escape parameters, notably with the EUV output of the Young Sun, the noble gas reservoirs, the volatile outgassing (etc.), are a problem to retrieve the whole history of a planetary atmosphere. In addition, other escape processes lead to isotopic fractionation.

\subsection{Photochemical Escape}
\label{pcloss}
The dominant non-thermal loss processes vary for each planetary body. The relative significance of each process depends on planetary mass, atmospheric composition, and distance from the sun. For instance, at Mars, the current dominant non-thermal loss processes are photochemical, while at Venus is it thought to be through ionospheric escape \cite{lammer_atmospheric_2008}.

The photochemical escape of a planetary atmosphere is a non-thermal loss process due to exothermic chemical reactions in the ionosphere that provide enough kinetic energy for the escape of the neutral constituents. Photochemical escape often includes direct interactions of photons and photoelectrons with thermospheric and exospheric molecules, as well as chemical reactions of ions with neutrals and electrons. In the following, we will add the symbol $^*$ to neutral and ionized species to show that they have a non-negligible amount of kinetic energy. Such species are usually called ``hot''; and for the neutral atoms, the term ENA, for Energetic Neutral Atom, is often used.

The general method of computation for the escape of a fast atom or ion can be found in \citeA{shematovich_kinetic_1994}.
The general transport equation for any species in the atmosphere is:
\begin{eqnarray}
\frac{\partial f}{\partial t} + \vec{v}\frac{\partial f}{\partial \vec{x}}  + \frac{\vec{F}}{m}\frac{\partial f}{\partial \vec{v}} = Q + H_{h\nu} + J_{el} + J_{q} + J_{cx}
\end{eqnarray}

where $Q$ represents the productions, $H_{h\nu}$ the spontaneous transition to another state --typically by light emission--, $J_{el}$ the loss due to elastic scattering (and therefore momentum transfer) \cite{Lilensten2013}, and $J_q$ the loss due to quenching. Note the addition of an extra loss term, $J_{cx}$, for charge exchange. The transport equation should be taken into account for all the species, and they can be coupled when the loss of one species creates another one. An example of that situation is the coupled transport between H and H$^+$, where a proton undergoing a charge exchange will become a fast H, that can be re-ionized later. This equation is also valid for the excited state species, such as O($^1$S) and O($^1$D), that are  notably responsible for the green line and the red line in aurorae \cite{gronoff2012_1,gronoff2012_2}.

In the following subsections, we review the main processes creating ENA/fast ions. Charge exchange is described in a later section. While the same equation should be solved to address atmospheric escape, approximations are often used for the coupled ion/ENA equations, angular diffusion, and upper atmospheric densities \cite{rahmati_seasonal_2018}. These approximations are used for several reasons. One particularly problematic point in the simulations is the distance at which a particle is considered lost in space; some studies take a few planetary radii, other a few exospheric altitudes. Such approximations can create difficulties when comparing with observations \cite{baliukin2019}.

\subsubsection{Ion recombination}
An exothermic ion recombination (or chemical reaction) can give enough kinetic energy to one of its products so that it can escape.
Ion recombination is the most effective channel to escape O in the present Martian atmosphere. It is, in general, an efficient way to heat up an atmosphere through non-thermal process. 
It is also a process leading to the escape of heavier atoms from light planets or bodies.
The process has been largely studied in the past \cite{Shizgal1996}, and is being refined in support of the MEX and MAVEN missions \cite{cipriani_martian_2007,yagi_mars_2012,valeille_water_2010,zhao_photochemical_2015,Lillis2017}.

At Mars, the main photochemical escape process is the loss of oxygen through the reaction:
\begin{eqnarray}
\textnormal{O}_2^+ + e- &\rightarrow& \textnormal{O}(^3P) + \textnormal{O}(^3P) + 6.99 \textrm{eV} \\
&\rightarrow& \textnormal{O}(^1D) + \textnormal{O}(^3P) + 5.02 \textrm{eV}\\ 
&\rightarrow& \textnormal{O}(^1D) + \textnormal{O}(^1D) + 3.05 \textrm{eV}\\
&\rightarrow& \textnormal{O}(^1S) + \textnormal{O}(^3P) + 2.80 \textrm{eV}\\
&\rightarrow& \textnormal{O}(^1D) + \textnormal{O}(^3S) + 0.83 \textrm{eV}
\end{eqnarray}

Recent studies by MAVEN were able to show the hot oxygen corona produced by these recombination reactions \cite{Deighan2015}. A study by \citeA{Cravens2017} shows that, in the limits of the current solar conditions at Mars, a linear dependence of the escape rate to the EUV flux can be made.

Another interesting reaction is N$_2^+ + e^- \rightarrow 2 $N$^*$ which is efficient enough for the removal of $^{14}$N but not  $^{15}$N at Mars, and could explain the isotopic fractionation \cite{Shizgal1996}.

At Earth and Mars, we also have \cite{Shizgal1996,groller_hot_2014} (the channel with O($^1$D) has branching ratio close to zero):
\begin{eqnarray}
\textnormal{NO}^+ + e^- &\rightarrow& \textnormal{N}(^4S) + \textnormal{O(}^3P) + 2.78 \textrm{eV} \\
&\rightarrow& \textnormal{N}(^2D) + \textnormal{O}(^3P) + 0.39 \textrm{eV} \\
&\rightarrow& \textnormal{N}(^4S) + \textnormal{O}(^1D) + 0.81 \textrm{eV}
\end{eqnarray}

To compute the photochemical escape through these processes, it is first necessary to compute the ion density. This involves, first, computing the ion productions (via photoionization, secondary electron ionization, etc.); second, computing the resulting chemistry and transport to get the ion densities; third, computing the hot atom production, using the densities and the reaction rate; fourth, compute the actual escape by computing the transport of the hot atom. Such an escape should include collisions with other species; if the hot atom creation rate is important enough, it should be taken into account that these collisions heat up the upper atmosphere, and therefore change its profile towards more escape.

Recent work at Mars shows that the CO$_2^+$ dissociative recombination is a non-negligible source of hot oxygen \cite{lee_comparison_2015,zhao_photochemical_2015}.
In the following, the first reaction is believed to have a branching ration between 96\% and 100\%:
\begin{eqnarray}
\textnormal{CO}_2^+ + e^-&\rightarrow& \textnormal{CO}(^1\Sigma) + \textnormal{O}(^3P) + 8.27 \textrm{eV}\\
&\rightarrow& \textnormal{CO}_2(^1\Sigma_g) + 13.78 \textrm{eV} \\
&\rightarrow& \textnormal{C}(^3P) + \textnormal{O}_2(^3\Sigma_g) + 2.29 \textrm{eV}\\ 
&\rightarrow& \textnormal{C}(^3P) + 2 \textnormal{O}(^3P) - 2.87 \textrm{eV}
\end{eqnarray}

\begin{eqnarray}
\textnormal{CO}^+ + e^- &\rightarrow& \textnormal{C}(^3P) + \textnormal{O}(^3P) + 2.90 \textrm{eV} \\
&\rightarrow& \textnormal{C}(^1D) + \textnormal{O}(^3P) + 1.64 \textrm{eV}\\ 
&\rightarrow& \textnormal{C}(^3P) + \textnormal{O}(^1D) + 0.93 \textrm{eV}\\
&\rightarrow& \textnormal{C}(^1S) + \textnormal{O}(^3P) + 0.22 \textrm{eV}\\
&\rightarrow& \textnormal{C}(^1D) + \textnormal{O}(^1D) - 0.32 \textrm{eV}\\
&\rightarrow& \textnormal{C}(^3P) + \textnormal{O}(^1S) - 1.28 \textrm{eV}
\end{eqnarray}

Hot oxygen in a planetary thermosphere can also induce escape of lower mass species by sputtering \cite{shizgal_escape_1999}. 

\subsubsection{Photodissociation}

Another process leading to the creation of fast ions or atoms is the direct dissociation by photon, electron, or proton impact. 

In \citeA{shematovich_kinetic_1994} an example is given by the reaction O$_2 + h\nu \rightarrow $O$(^3$P$) + $O$(^3$P$,^1$D$,^1$S$)$; the kinetic energy given to the products is the difference between the energy of the photon and the binding energy (i.e. the threshold energy for the reaction). Similar processes can be evaluated for N$_2$, CO$_2$, etc. Photodissociation reaction are seldom considered in evaluating escape rates since the production of fast enough particle to escape is small with respect to ion recombination processes. To properly evaluate these productions, it is necessary to have an accurate set of cross sections (see Section \ref{section_crosssection}). In general, thermospheric codes consider that the kinetic energy given by these photodissociations ends up in heating, therefore one has to be careful to not count that loss of energy twice in their simulations.

\subsubsection{Di-cation dissociation}

The di-cation dissociation effect on planetary escape has been proposed by \citeA{Lilensten2013}. It is a non-thermal processes that is based on the fact that the Coulombian dissociation of a molecular doubly charged ion may give enough energy to one or both of the ions to allow their escape.

The typical example for this process is CO$_2^{2+}~\rightarrow~$CO$^+ + $O$^+$, as described in \citeA{Lilensten2013}. Other processes such as N$_2^{2+}~\rightarrow~2 $N$^+$ \cite{gronoff_2007}, or O$_2^{2+}~\rightarrow~2 $O$^+$ \cite{Simon2005, gronoff_2007} can give sufficient energies for the ion to escape. 
To account for the flux of escaping particles through that process, it is necessary to compute the transport of the fast ions from where they are created to the exobase. Since it is ions that are escaping, they are not necessarily escaping even if they reach the exobase with sufficient energy: the presence of magnetic fields could prevent their escape, and return them into the atmosphere where they could create some additional heating (the Coulomb energy being in the range of several 10~eV, such ions could not efficiently sputter, except if they are further accelerated by the solar wind). A process not accounted for in the \citeA{Lilensten2013} paper is the heating of the ionosphere and the creation of fast ENA through charge exchange of the fast ions with the atmosphere (a process similar to the one described in \citeA{Chassefiere1996h}). On the contrary, ions with energy lower than escape energy could escape due to electromagnetic forces, as will be explained in Section \ref{ionosphericoutflow} and \ref{oloss}.

The calculation of the dication escape in a non-magnetized atmosphere proceeds as follows: from $P_{i^{2+}}(z)$, the production rate of the specific dication $i^{2+}$ in function of the altitude $z$, we compute its density $n_{^{2+}} = P_{i^{2+}}(z) / L_{i^{2+}}(z) $  from the chemical loss processes $L$, neglecting the transport because of the small lifetime of the dication (for a detailed analysis of the production processes see e.g. \citeA{gronoff2012_1, gronoff2012_2}). From there, the standard transport equation of fast ion in the atmosphere can be used.
The study of \citeA{Lilensten2013}  does not take into account the loss of energy of  O$^{+*}$ impacting atmospheric O, therefore overestimating the escape (the study consider impact on CO$_2$, which has a smaller scale height). On the other hand it underestimates the escape rate by not doing a coupled equation transport and therefore not taking into account the escape of O$^*$ created by charge exchange of O$^{+*}$ with other thermospheric species.

\subsubsection{Key parameters}

Modeling photochemical loss requires the cross section for ionization by the different processes (including elastic, inelastic, and charge exchange), and the chemical reaction rates for the density/recombination (including the branching ratio and the products speed probabilities).
The ionospheric electron temperature is overall extremely important since the recombination cross section is likely to be extremely sensitive to it \cite{Sakai2016}.
For the simulation of the ion/electron composition and temperature, it is necessary to perform a 3-D modeling of the ionosphere.

\subsubsection{Questions}

The evaluation of escape rates from photochemical reactions has mainly been done for Solar system planets, especially Mars and Venus. Once we consider exoplanets or the Young Solar system, questions remains about the efficiency of each processes. The ion recombination or the usually neglected processes such as particle impact dissociation could become more important when increase in XUV or precipitating particle flux occur. This question is difficult to answer since each process affects the state of the upper atmosphere and the efficiency of each other.

\subsubsection{Observables}
\label{selfshielding}
The recombination processes create ENA at very specific energies, typically in the 5~eV range. Since collisions occurs, changing the spectral shape of the energy distribution, the direct observation of these energies peaks is extremely challenging. Indirect techniques, based on modeling the hot oxygen corona are used. At Mars, a technique to observe the product of photochemical reactions involved observing the hot oxygen geocorona \cite{Deighan2015}.  
As explained in \citeA{Shizgal1996}, photochemical escape can explain the fractionation of $^{14}$N/$^{15}$N at Mars.  A more recent work from \citeA{Mandt_2015} shows that non-thermal processes except photodissociation can explain the isotopic enrichment. The work of \citeA{Liang_2007} shows that self-shielding effects can lead to an increase in heavier isotopes (here $^{15}$N at Titan) escape from photodissociation.

\subsection{Ion Loss}
\label{ioloss}
The ion loss mechanisms begin with the interaction of the upper atmosphere and ionosphere with the solar wind. Neutral atoms can be ionized by  solar UV, charge exchange and electron impact, and can be scavenged by the solar wind. There are different processes and loss channels through which the planetary ions can escape to space, including pickup and sputtering, charge exchange, and outflow, which will have its dedicated subsection. Ion escape is believed to be one of the major sources of atmospheric escape in the current Solar system and also at exoplanets around M-dwarfs \cite{GarciaSage2017}.

\subsubsection{Pickup and sputtering escape}
\paragraph{Pick-up escape}
Pick up ion loss is due to the ionization of neutral constituents in the exosphere and upper atmosphere that sense an electric field and can be ``picked up'' and swept away. In the presence of the magnetic field, at Earth for example, the polar wind drives pick up ion escape \cite{moore1997}. At lower altitudes, this interaction can compress the magnetic field on the sunward side, forming a tail on the anti-sunward side. At high altitudes, the loss of H$^+$, He$^+$ and O$^+$ can occur when thermal plasma originating from the polar regions in the ionosphere is accelerated into the magnetosphere and escapes downtail \cite{johnson_exospheres_2008}. These processes will be detailed in Section~\ref{ionosphericoutflow} and \ref{ionreturn}.

At weakly magnetized planets, such as Mars and Venus, the lack of an intrinsic dipole magnetic field creates a scenario where the solar wind directly interacts with the upper atmosphere. In this situation, neutral constituents are ionized and picked up by the background convection electric field that is driven by the solar wind, where $\vec{E}_{\textrm{SW}}$=$-\vec{U}_{\textrm{SW}} \times \vec{B}_{\textrm{SW}}$ where $\vec{E}_{\textrm{SW}}$ is the electric field induced on an ion by the solar wind (and therefore that ion will be subject to a force $\vec{F} = q \vec{E}_{\textrm{SW}}$), $\vec{U}_{\textrm{SW}}$ is the solar wind speed and  $\vec{B}_{\textrm{SW}}$ is the interplanetary magnetic field. The main channels for ionizing planetary neutrals are photoionization, charge exchange and electron impact ionization. \citeA{curry2013a} investigates these mechanisms as a function of solar zenith angle, bulk velocity and plasma temperature, respectively, finding that the majority of pick-up ions are formed in the corona and sub-solar region of Mars. The origin of pickup ions plays a major role in their fate as escaping particles or precipitating particles \cite{fang2010b}. In the former case, the pick-up ions can accelerate to twice the solar wind speed and their gyroradii are on the order of a planetary radius, and are likely to escape. The maximum energy of a picked-up ion is $E_{max} = 2 m U^2_{\textrm{SW}} sin^2(\theta_B)$ where $\theta_B$ is the angle between the solar wind direction and the interplanetary magnetic field \cite{rahmati_maven_2015}. In the case of precipitating ions, the pick up ions will collide with neutrals in the exobase or thermosphere and transfer enough energy and momentum to the neutral that they could be able to exceed the escape velocity; a process known as sputtering. Ion precipitation also impacts the atmosphere through heating.
The sputtering process can also happen at Earth, inside the polar regions \cite{shematovich2006energetic}, but it is a small process there.

\paragraph{Pick-up equations}
If we consider $n_{\textrm{SW}}$ as the solar wind density, $n_O$ the density of oxygen where that solar wind is located, $\sigma_{CX}$ the average charge exchange cross section between the solar wind and oxygen and $\sigma_{PI}(\lambda)$ the photoionization cross section, we have an ion production of $P_I = n_O (\int \sigma_{PI}(\lambda) \Phi_{EUV}(\lambda) d\lambda + \sigma_{CX} n_{\textrm{SW}} U_{\textrm{SW}})$ (and other ionization processes can be added such as electron impact) \cite{rahmati_maven_2015,rahmati_maven_2017}, that production is balanced by the pick-up transport. If we consider $P_I(\vec{v})$ the production of ion at a speed defined by $\vec{v}$ (so that $\int P_I(\vec{v}) d\vec{v} = P_I$), $e$ the charge of the ion and $m$ its mass, then the velocity distribution function $f(\vec{x},\vec{v})$ for the picked-up ions is governed by \cite{hartle_pickup_2011}:
\begin{eqnarray}
\vec{E}_{\textrm{SW}} &=&-\vec{U}_{\textrm{SW}} \times \vec{B}_{\textrm{SW}}\\
\vec{v}.\frac{\partial f}{\partial \vec{x}} + \frac{e}{m} (\vec{E}_{\textrm{SW}} + \vec{v} \times \vec{B}_{\textrm{SW}}). \frac{\partial f}{\partial \vec{v}} &=& P_I(\vec{v}) \label{eq:pickup}\\
\Phi(\vec{x}) &=& \int v f d\vec{v}
\end{eqnarray}
with $e$ the ion charge.
 Several techniques can be used for solving Equation \ref{eq:pickup}; the complexity arises from the solar wind piling up around the planet (or the comet \cite{coates_ion_2004}, creating complex magnetic field geometries. Typically, it has been solved using test particles (Monte Carlo simulations) in fields from MHD  or self consistent hybrid codes, as by \cite{jarvinen_energization_2014}.


\paragraph{Sputtering}
The yield $Y$ of sputtered neutrals is defined by the sputtering efficiency. This yield is the ratio of the number of escaping particles and the number of incident particles, which varies inversely with the planet's gravitational energy \cite{Johnson1994,leblanc_role_2002,johnson_exospheres_2008}. Sputtering is dependent on the incident particles' energy and angle of incidence, as well as the mass of the incident particle. For lighter incident pickup ions, the direct scattering of planetary neutrals is known as ``knock-on'', which dominates at low, grazing incidence angles. For heavier incident pickup ions, the additional momentum can create a cascade of collisions at high enough energies to cause a neutral to escape, where $Y$ $\ge$ 1 \cite{leblanc_sputtering_2001,johnson_exospheres_2008}. This occurs for O$^+$ pickup ions at energies of $\sim$keV to $\sim$hundred keV. This is especially important when the pickup ion gyroradius is of the order of the planet radius, as at weakly magnetized bodies such as Mars, Venus and Titan.

\begin{figure}
	 \noindent\includegraphics[width=40pc]{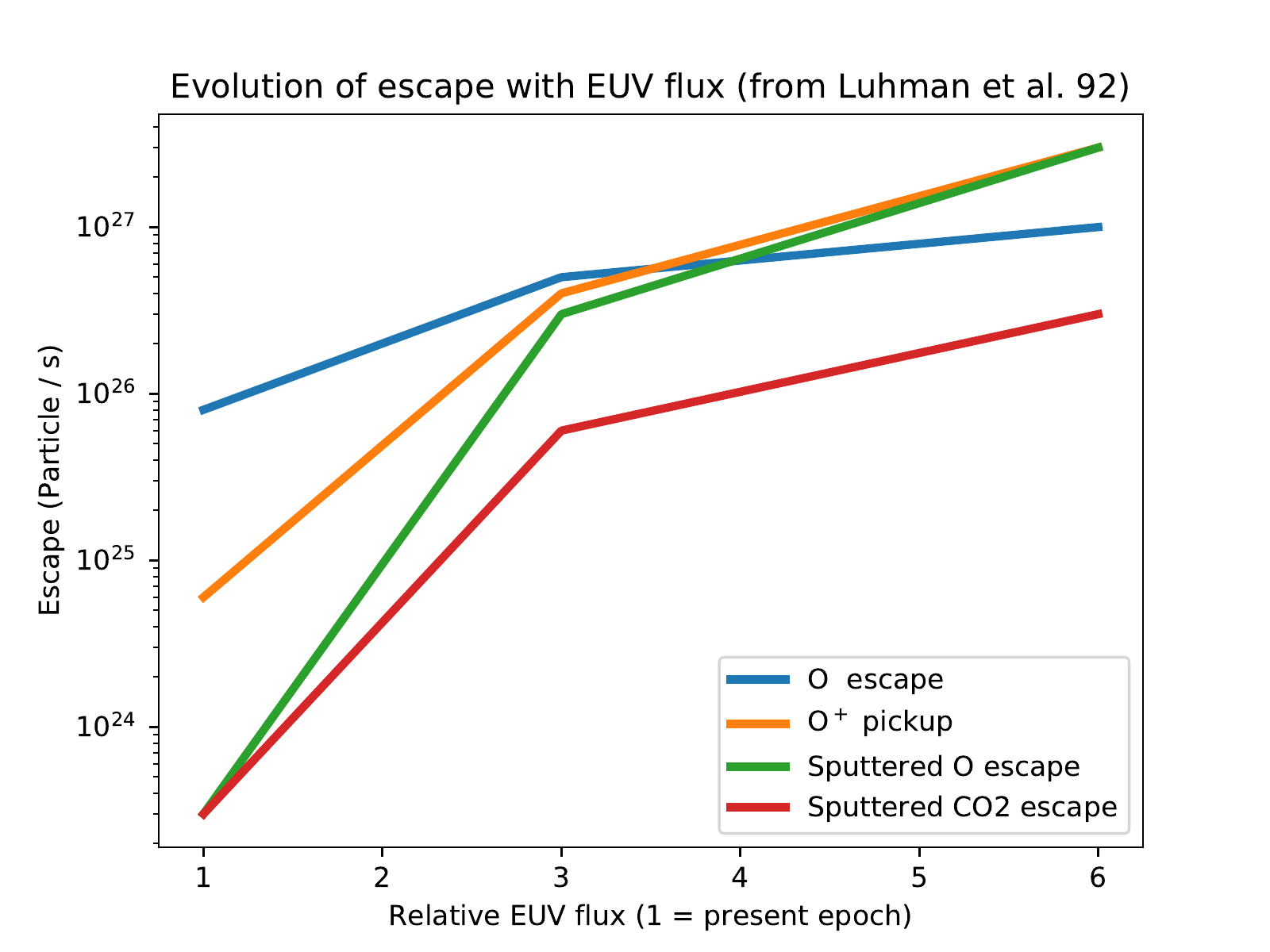}
\caption{EUV dependence of the escape process at Mars. Simulations data from \citeA{luhmann_evolutionary_1992} }
\label{euvluhman}
\end{figure}

Sputtering is widely believed to be the dominant escape process at Mars and Venus during earlier epochs of our Sun, which has major implications for exoplanetary atmospheres. \citeA{luhmann_evolutionary_1992} calculated the flux of precipitating pick-up ions and ENAs using a 1D exospheric model of the O density and a a gas-dynamic model of the solar wind and found compared to pickup ion and photochemical escape, sputtering drove the highest rates of atmospheric erosion (see Figure \ref{euvluhman}). Other studies using MHD and hybrid models have found similar results \cite{chaufray2007,wang2014}. Sputtering as a dominant driver of atmospheric escape is further supported by current isotope ratios. Specifically, Ar is an important atmospheric tracer because once Ar is in the atmosphere, the only loss process is escape to space (as opposed to volcanic outgassing from the interior, impact delivery, and mixing with the crust), which limits the exchange pathways that become complicated for most planetary volatiles \cite{jakosky2002}. Thus, measurements of the present day atmosphere reflect the importance of these exchanges over billions of years and emphasizes the need for understanding our own terrestrial planets' atmospheric evolution as a ground truth for understanding exoplanetary atmospheres.

Unfortunately, sputtering is incredibly difficult to observe as the sputtered component at Mars is indiscernible from  photochemically produced oxygen. Thus models have typically predicted what the sputtered component would be in a variety of scenarios. The passage of the Siding Spring comet close to Mars \cite{Bodewits2015} created a high flux of O$^+$ ions that impacted the atmosphere as predicted in \cite{gronoff_comet_2014}. Observations by the MAVEN Solar Energetic Particle instrument (SEP) and the Mars Odyssey-High Energy Neutron Detector (HEND)  indicate an increase in the O$^+$ pickup during the passage; however, an increase in solar activity at the same time prevents a clear conclusion on whether or not it was due to the comet \cite{sanchez2018}. \citeA{wang2016} computed the impact of these ions on the escape rate, and found that up to 10 tons of atmosphere may have escaped while 1 ton of material was added. Another formulation of the escape by sputtering can be found in \citeA{shizgal_escape_1999}. In this paper, it is the sputtering by the hot oxygen of Venus that leads to the escape of H and D. The main difference with the usual pickup-sputtering process is the origin of the hot O, from the thermosphere itself, and therefore that the forcing by an external flux and the use of yield function cannot be applied. \citeA{shizgal_escape_1999} developed a specific kinetic model for the escape.


\paragraph{Sputtering equations}
The rate of species $n_j$ escaping an atmosphere from sputtering is given by $\frac{\partial n_j}{\partial t} \approx 2\pi R_{exo}^2 <\Phi_a Y_j>$ \cite{Johnson1994} where $\Phi_a$ is the flux of the particle leading to the sputtering and $Y_j$ the sputtering efficiency for that peculiar species. 
For sputtering by an incident particle A and a target species B, and of respective masses $M_A$ and $M_B$ i.e. a thermosphere whose main constituent is B, the yield can be computed as follows.
First, for an incident particle of energy $E_A$, it is necessary to evaluate the elastic cross section $\sigma_d(E_A)$ which is related to the momentum transfer (or knock-on, elastic nuclear) stopping cross section $S_n(E_A)$ through:
\begin{eqnarray}
\gamma = \frac{4 M_A M_B}{\left(M_A + M_B\right)^2}\\
\sigma_d(E_A)= \frac{\gamma}{2} E_A S_n(E_A)
\end{eqnarray}

The overall yield is the result of single impact plus multiple impact momentum transfer at energy greater than the escape energy. 
It can be approximated by \cite{Johnson1994,johnson_sputtering_2000}:
\begin{eqnarray}
Y(\theta, E) &\approx& \frac{\alpha \beta S_n(E)}{2 U_{es} \sigma_d(\overline{E}_{es})cos^p\theta} \label{eq:CXY}\\
\overline{E}_{es} &\approx& U_{es}
\end{eqnarray}
where $\theta$ is the incident angle, $U_{es}$ is the gravitational binding energy at the exobase, $\overline{E}_{es}$ is the average energy of the escaping particle. $\alpha$, $\beta$, and $p$ are constants depending upon the impact particles, see \citeA{Johnson1994} for some numerical values in the literature. 
The sputtering yield may be enhanced by the sputtered particles that are picked-up and accelerated towards the atmosphere (equation 2 in \citeA{Johnson1994}). If the efficiency of escape for a sputtered particle is $Y_a$ and its ionization and return is $p_i$, then the effective yield is $Y_{eff}$ = $Y / \left(1-p_i\left(Y_a-1\right)\right)$.

\paragraph{Other impact processes}
The classical sputtering process involves the impact of an ion that has been accelerated by pick-up, i.e. a non-thermal processes outside of the thermosphere. \citeA{Gacesa2012} proposed a very similar mechanism where the impact of hot O from the Martian corona sputters light gases. Their computations suggest it is the main channel for HD and D$_2$ direct escape. To validate that approach, it is suggested to observe the emission of H$_2$ ro-vibrationally excited by the impact.
ENA impact on the Martian atmosphere are also a source of escape, especially when they have been created by charge exchange from the solar wind \cite{Lewkow2014}, which leads us the the other class of escape processes.


\subsubsection{Charge exchange of a magnetically trapped particle}
The basic idea of charge exchange escape is that a magnetically trapped energetic ion, such as H$^+$, exchanges its charge and becomes an energetic neutral atom (ENA) that can escape or sputter (an ion trapped in a magnetic mirror may be prevented to reach the thermosphere and therefore to efficiently sputter) \cite{Shizgal1996}. The temperature dependence is complex: at Earth it decreases with exospheric temperature for H \cite{Shizgal1996} so that the escaping flux from charge exchange plus Jeans escape is constant, reaching the diffusion-limited value. 

A simple approach adopted by \citeA{Yung1989} was to consider that the ion H$^+$ had a Maxwellian distribution at the temperature T$_{ion}$. Having exchanged its charge, the energetic neutral atom could escape, and it would have the same energy as the initial ion. The efficiency for an escaping charge-exchanged atom, with respect to the temperature of the initial ion is 
\begin{equation}
\alpha_i(R) = B_\textrm{CX} \left[1 - \frac{v_{esc}^2}{u_j(R)^2} \right] e^{-\frac{v_{esc}^2}{u_j(R)^2}}
\end{equation} 
with $u_j(R) = \sqrt{2kT_{ion}(R)/m_i}$. Considering $k_{i^+-j}$ the charge exchange rate between the ionized species $i$ and a neutral species $j$ (rate that can vary with temperature), this gives the escape flux 
\begin{equation}
    \phi = \int_{R_exo}^{R_pp} \left(\frac{R}{R_{exo}}\right)^2 \alpha_i(R) \sum_j k_{i^+-j} [i^+][j] dR.
\end{equation} 
The $B_\textrm{CX}$ factor in the definition of $\alpha_i$ is an efficiency factor, that was taken identical to the one for thermal escape in the \citeA{Yung1989} paper. The rest of the equation is similar to the thermal escape equation, except the $\frac{u_i}{2 \sqrt(\pi)}$ factor (which was taken off for considering it is hidden in the charge exchange rate). The equation in \citeA{Yung1989} paper has a negative sign that should be positive: using the equation with that negative sign leads to negative escape fluxes.  Using that equation, it happens that the charge exchange flux should increase with increasing exospheric temperature, which is not what is observed. It means that this simplified approach is not good enough for evaluating the charge exchange flux at Earth.

\citeA{Shizgal1982} developed a collisional model for computing the charge exchange induced escape. The main difference with the previous approach is that the efficiency of charge exchange with respect to the temperatures is taken into account following \citeA{1975JChPh..63..131F}. It is shown that the charge exchange is, at Earth, the most efficient mechanism to remove H from the upper atmosphere during low solar activity (low exospheric temperature) while Jeans' escape is the main mechanism during high solar activity. It is important to remember here that H escape is diffusion limited at Earth. 
In the following, A corresponds to the neutral atmosphere (O and H), $\bar{n}$ to the average density (of A, O, and H$^+$) over the region of charge exchange and $\sigma$ corresponds to the energy independent hard sphere cross section, and $a=\frac{m_A}{m_O}$
\begin{eqnarray}
 \lambda_{\textrm{\tiny{CX}}} &=& \frac{m_H v_{\textrm{esc}}^2}{2kT_{\textrm{ion}}}\\
 \hat{n} &=& \frac{\sigma_{H^+,A}}{\sigma_{H,O}}\left[\frac{\bar{n_{H^+}} \bar{n_A}}{\bar{n_O}}\right] \frac{\Gamma(a)}{1+a} \\
 \tau_{\textrm{\tiny{CX}}} &=& \frac{T_{\textrm{exo}}}{T_{H^+}} - 1\\
  \Phi_{\textrm{\tiny{CX}}}(\textrm{escape}) &=& \hat{n} \sqrt{\frac{2kT_{\textrm{exo}}}{\pi m_H}} \frac{e^{-\lambda_{\textrm{\tiny{CX}}}}}{\tau_{\textrm{\tiny{CX}}}} \nonumber\\
  &\times&\left[ (1+\tau_{\textrm{\tiny{CX}}}) - \sqrt{1+\tau_{\textrm{\tiny{CX}}}} e^{-\lambda_{\textrm{\tiny{CX}}} \tau_{\textrm{\tiny{CX}}}} \right]
\end{eqnarray}

This equation is valid for the escape of H at Earth from charge exchange. It supposes that (1) the H$^+$ density varies slowly with altitude at the location where this process is the most efficient (from the exobase to 3000~km), (2) the only species interacting are H, O, and H$^+$ , and (3) the distributions are Maxwellian, with a fixed temperature in the altitude range.


At Earth, the charge exchange is the main mechanism to remove O$^+$ from the ring current
\cite{Daglis1999}. The exchange creates ENA that can be imaged to study the ring current evolution. 

\subsubsection{Charge exchange with the solar wind}

The charge exchange between the solar wind and the upper atmospheric species can enhance the escape rate through pick-up like processes:
a species M exchanges its charge with, for example, a proton from the solar wind H$^{+*} + $M$ \rightarrow $H$^* + $M$^+$. The solar wind proton becomes an ENA, and can creates additional heating that increases the thermospheric temperature, and therefore escape \cite{Chassefiere1996h}. The created ion can escape thanks to pickup by the magnetic field.

At comets, charge transfer reactions primarily involve solar wind ions, H$^+$, He$^{2+}$ but also multiply-charged minor species such as O$^{6+}$, Si$^{10+}$, or C$^{5+}$ \cite{Cravens1997,Bodewits2007,Wedlund2016}, with water molecules continuously outgassing upon sublimation from the nucleus. As the atmosphere of a comet is in expansion, charge-transfer reactions take place over a large region of space (of the order of several $10^6$ km) and will have time to facilitate the absorption of the solar wind, converting fast ions into slow-moving ones. Charge transfer has recently been evidenced by the ESA/Rosetta ion spectrometers at comet 67P/Churyumov-Gerasimenko (67P), with the observation of H$^-$ ions \cite{Burch2015}, and He$^+$ fast ions \cite{Nilsson2015}. The latter charge-exchanged ions, originating from solar wind He$^{2+}$ ions (composing about 4\% of the bulk of the undisturbed solar wind), were present throughout the mission from a heliocentric distance ranging from 3.4 to 2~AU \cite{Wedlund2016,wedlund_solar_2019,wedlund_solar_2019-1,wedlund_solar_2019-2}. The net effect of the charge transfer of He$^{2+}$ solar wind ions with the neutral atmosphere of the comet (composed of molecules M) is the production of ENAs following the typical sequence of electron capture reactions (double charge transfer, and stripping reactions are ignored here for simplicity):
\begin{eqnarray}
	\textnormal{He}^{2+} + \textnormal{M} &\longrightarrow \textnormal{He}^{+} + \textnormal{M}^+ \label{eq:reactionHe2p}\\
   \textnormal{He}^{+} + \textnormal{M} &\longrightarrow \textnormal{He} + \textnormal{M}^+ \label{eq:reactionHep}
\end{eqnarray}
This set of reactions is equivalent to coupled differential flux continuity equations 
which can be solved analytically for the simplified case or numerically \cite{Wedlund2016,wedlund_solar_2019-1}.

Similar equations can also be derived for the coupled (H$^+$,H) system. These processes lead to the almost total conversion of the solar wind into ENAs, potentially escaping or sputtering the nucleus, by the time the solar wind impinges within a few tens of kilometres from the comet's surface, in the case of a highly outgassing nucleus (perihelion conditions). This total conversion depends on many parameters: outgassing rate, heliocentric distance, solar wind density and speed \cite{wedlund_solar_2019,wedlund_solar_2019-1,wedlund_solar_2019-2}. The effect of minor solar wind species (multiply-charged heavy ions) can be seen in the production of X-rays through charge exchange emission with the cometary atmosphere \cite{Cravens1997}. The case of comets provides a unique opportunity to study charge-exchange processes within different and varying atmospheric environments.

The observation of escape from HD 209458 has been interpreted as increased by charge exchange processes between the solar wind and the hydrogen from the upper atmosphere of the planet \cite{Holmstrom2008}. 

\subsubsection{Charge exchange with a precipitating particle}
Particles precipitating in the atmosphere of planets can give rise, through charge exchange with the ambient neutral atmosphere, to the local production of ENAs. This is particularly significant at Earth in the case of protons of solar wind origin, first accelerated in the magnetosphere and then precipitating down the magnetic field lines in the polar regions. When protons are neutralized in collisions with neutrals (mainly oxygen atoms above 200~km altitude, O$_2$ and N$_2$ below), a process referred to as \emph{electron capture}, the newly produced hydrogen ENAs, not being sensitive to the magnetic field, travel in straight trajectories, whose direction is related to the pitch angle distribution of the impinging protons, resulting in a horizontal spreading of the precipitating beam (see, for example, \citeA{Rees1989,Kozelov1994,Galand1997,Galand1998,Basu2001,Simon2007}). Hydrogen ENAs, keeping most of the kinetic energy of the incoming proton, can in turn be ionized (\emph{electron stripping}). Due to magnetic mirroring and angular re-distributions stemming from collisions between the energetic species and the atmosphere, downwelling (or precipitating) and upwelling (or backscattered) ions and ENAs will coexist at any given altitude above the $E$-region peak of the initial proton energy deposition (for a 10~keV initial proton peak will occur at 120~km altitude at Earth). The energy and angular degradation of a (H$^+$,H) beam in the atmosphere is usually formalized as a coupled system of two non-linear Boltzmann transport equations \cite{Galand1997,Galand1998}, including angular re-distributions due to the non-uniformity of the magnetic field and to collisions, for an ENA, X, and its corresponding ion, $\textnormal{X}^+$. In the following, $I$, the intensity, $\mathcal{P}$, the momentum transfer, $\mathcal{R}^\textnormal{CX}$, the charge transfer,  and, $\mathcal{Q}$, the local production,  depend upon $(\tau,E,\mu)$. 
The transport of ENAs, denoted $\textnormal{X}$, is as follows:

\begin{eqnarray}
  	\mu \frac{\partial I_{\textnormal{X}} }{\partial \tau} &=& -I_{\textnormal{X}}  
  	+ \frac{\mathcal{P}_{\textnormal{X}} + \mathcal{Q}_{\textnormal{X}} + \mathcal{R}_{\textnormal{X}^+\rightarrow \textnormal{X}}^\textnormal{CX}}{\sum_k{\sigma_{k,\textnormal{elas}}(E)\,n_k(z)}}\\
	\mu \frac{\partial I_{\textnormal{X}^+} }{\partial \tau} &=& -I_{\textnormal{X}^+}
	+ \frac{\mathcal{P}_{\textnormal{X}^+} + \mathcal{Q}_{\textnormal{X}^+} + \mathcal{R}_{\textnormal{X}\rightarrow \textnormal{X}^+}^\textnormal{CX}}{\sum_k{\sigma_{k,\textnormal{elas}}(E)\,n_k(z)}}
\end{eqnarray}

 Numerical solutions of this system have historically made use of continuous slowing-down approximations \cite{Decker1996}, DSMC techniques \cite{Basu2001,Shematovich2011}, and a semi-analytical exponential matrix solution (both with dissipative forces and angular redistributions \citeA{Galand1997,Simon2007}).

 Motivated by the Mars Express and MAVEN missions, there are an increasing number of studies of proton precipitation at Mars. \cite{Shematovich2011} have developed a DSMC model of the coupled (H$^+$,H) system in a (CO$_2$, N$_2$, O) atmosphere and applied it to Mars Express ASPERA data in solar minimum conditions. They concluded that about 20\% (10\%) of the incoming particle (energy) flux was backscattered by the atmosphere, and emphasized the role of the solar wind magnetic field pile-up region at altitudes above 100~km in increasing the backscattered flux by a factor up to 50\%.
\cite{Shematovich2017} recently studied the production of suprathermal O atoms in Mars' thermosphere via this process and concluded that a hot oxygen corona may form, creating an additional non-thermal escape flux of O that may become prevalent when extreme solar transient events, such as flares and Coronal Mass Ejections (CMEs), take place. Finally, \citeA{Halekas2017} derived the ENA flux originating from the solar wind interaction with the Martian atmosphere from the observation of protons by MAVEN/SWIA. From there, it was possible to retrieve the exospheric temperature of Mars (as well as the solar wind velocity).

At Jupiter, energetic precipitation involves protons \cite{Bisikalo1996}, but also singly or multiply-charged heavy ions such as S$^{n+}$ and O$^{n+}$ (with $n$ the charge number) \cite{Horanyi1988}, colliding with H and H$_2$ \cite{Waite:2002}. The high charged states of O at very high energies (above 200~keV/amu) are responsible for auroral X-ray emissions, as modelled in \cite{Cravens1995} and compared to X-ray observations of Jupiter. Such ion precipitation creating fast energetic atoms is also expected to play a role for satellites of Jupiter, and at Saturn, and its satellites.

\subsubsection{Charge exchange in the ionosphere}
This process is an hybrid between charge exchange and photochemical escape; it consists of having excess kinetic energy when a charge exchange is performed, such as He$^+ + $N$_2 \rightarrow $He$^* (+ 9 \textrm{eV}) + $N$_2^{+}$ \cite{Shizgal1996}. This process has been suggested to address the problem of the He budget in the Earth's thermosphere \cite{Shizgal1996,LieSvendsen1992}.

\subsubsection{Key parameters}
 The important parameters in the computation of the pick-up~/~sputtering are the solar wind parameters, that can usually be found thanks to models of the interaction of the solar wind with the planet \cite{curry2013a,Lee2017}, and the cross sections for ionization and stopping power/elastic scattering. In addition, it is necessary to have good inelastic/interaction potential \cite{Johnson1994} cross sections to be able to compute the $\alpha$, $\beta$ and $p$ parameters in Equation~\ref{eq:CXY}. 
 Finally, a particular attention should be given towards the nature of the model with respect to modifying the inputs of ion pickup models: for example, it is usually assumed that an exosphere is present in a MHD model, and mass loading will reduce the accuracy of the model. Hybrid modeling will better improve such models, as is done in a cometary environment \cite{simonwedlund2017}. A review of the comparative advantages/inconvenients of each type of solar wind models can be found in \citeA{Ledvina2008}. 

For the majority of the recent work in pick-up and sputtering, many cross sections are being used without being published, which is a major problem for the community. 
 The state-of-the-art models for sputtering are now using a DSMC approach \cite{johnson_sputtering_2000}.

For charge-exchange processes, in addition to the particle precipitation models and the solar wind models, it is important to have a good knowledge of the atmosphere composition and temperature, including the ion temperature.

\subsubsection{Questions}
How much do these processes scale up with the solar wind density, speed, and orientation? How does the creation of an induced magnetic field influence these charge exchanges processes?

\subsubsection{Observables}
\paragraph{Composition change}
The observation of the change in solar wind composition is a proof of charge exchange, for example at comets \cite{wedlund_solar_2019,wedlund_solar_2019-1,wedlund_solar_2019-2}. At Mars, the charge exchange of solar wind protons at the bow shock leads to precipitation of H that can be observed by the effects on the chemistry and by the backscatter \cite{Halekas2017}, even if the H chemistry at Mars is complex \cite{chaffin2017elevated} 
One more striking example of charge-exchange processes at Mars is the observation of heavier ions, such as O$^+$, that later lead to sputtering  \cite{Leblanc2015,Leblanc2018}.

\paragraph{Fractionation due to pickup/sputtering}

The fractionation due to pickup and sputtering is efficient because of its tendency to make the species at the top of the thermosphere escape. Since isotopes have a gravitational fractionation  at these altitudes, the overall effect is to increase the number of heavier species in the atmosphere. This is known as a Rayleigh distillation [see section \ref{gravitationalfractionation}].

\subsection{Ionospheric outflow}
\label{ionosphericoutflow}

Heating and energization of electrons and ions at a magnetized planet results in escape of ionospheric plasma, either onto open field lines where it joins the solar wind flow and is lost to interplanetary space, or onto closed or reconnecting magnetic field lines where it becomes trapped in the magnetosphere and becomes subject to magnetospheric dynamics and loss processes. The escape of ionospheric plasma is often considered in the context of magnetospheric dynamics and as a competing source of magnetospheric plasma together with the solar wind. However, it also has a vital role in the context of atmospheric escape and evolution in that a charged particle has additional plasma physics processes acting on it, as compared to a neutral particle which does not respond to the magnetic or electric field. These processes help reduce the gravitational potential barrier binding the charged particle to the planet.  

\begin{figure}
\vspace{-3cm}
 \noindent\includegraphics[width=40pc]{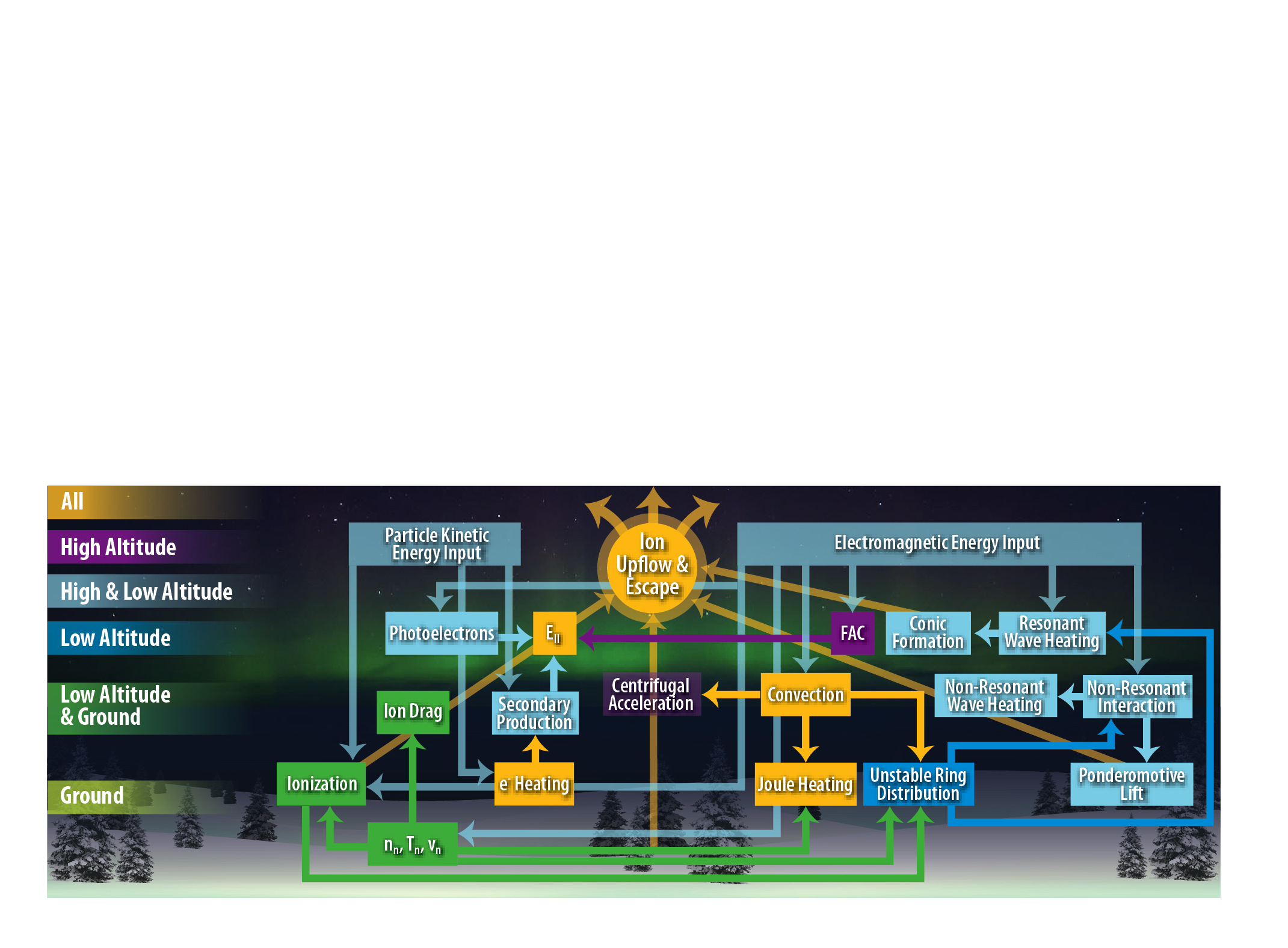}
 \caption{Processes leading to the creation of ion outflow/polar escape}
 \label{magnetizedescape}
\end{figure}

The escape of ionized particles to space has several names in the literature, ion outflow, polar wind, bulk ion escape, polar outflow, etc. This leads to some confusion as sometimes authors are generically referring to escaping plasma, but other times they are talking about outflow energized by particular processes that vary in space and time, as shown in Figure \ref{magnetizedescape}.  For instance, the ``polar wind" typically refers to the supersonic outflow of ions from the polar ionosphere accelerated by ambipolar electric fields \cite{Axford:1968,Banks:1968}. As the name implies, this polar wind is similar in concept to the solar wind, the supersonic expansion of the solar corona into space, proposed by \citeA{Parker:1958} nearly a decade before. While outflows of polar wind were initially thought to contain only light species such as protons, the first quantitative observations of O$^{+}$ in the polar wind by the Retarding Ion Mass Spectrometer on-board the Dynamics Explorer 1 (DE-1) demonstrated that heavy ions can be present as well in quite significant numbers. O$^{+}$ accelerated by wave-particle interactions in the cusp is sometimes referred to as the ``cleft ion fountain'' while the same process above the auroral region is occasionally referred to as an ``auroral wind''. The variability in location, composition, and energy of outflowing ions at Earth has led to the variety of names that describe escape along magnetic field lines. In this section, we eschew these more specific terms instead will use the term ionospheric outflow or ion outflow with the more broad meaning of any population of plasma upflowing from the planet at high altitude.

\begin{figure}
 \noindent\includegraphics[width=40pc]{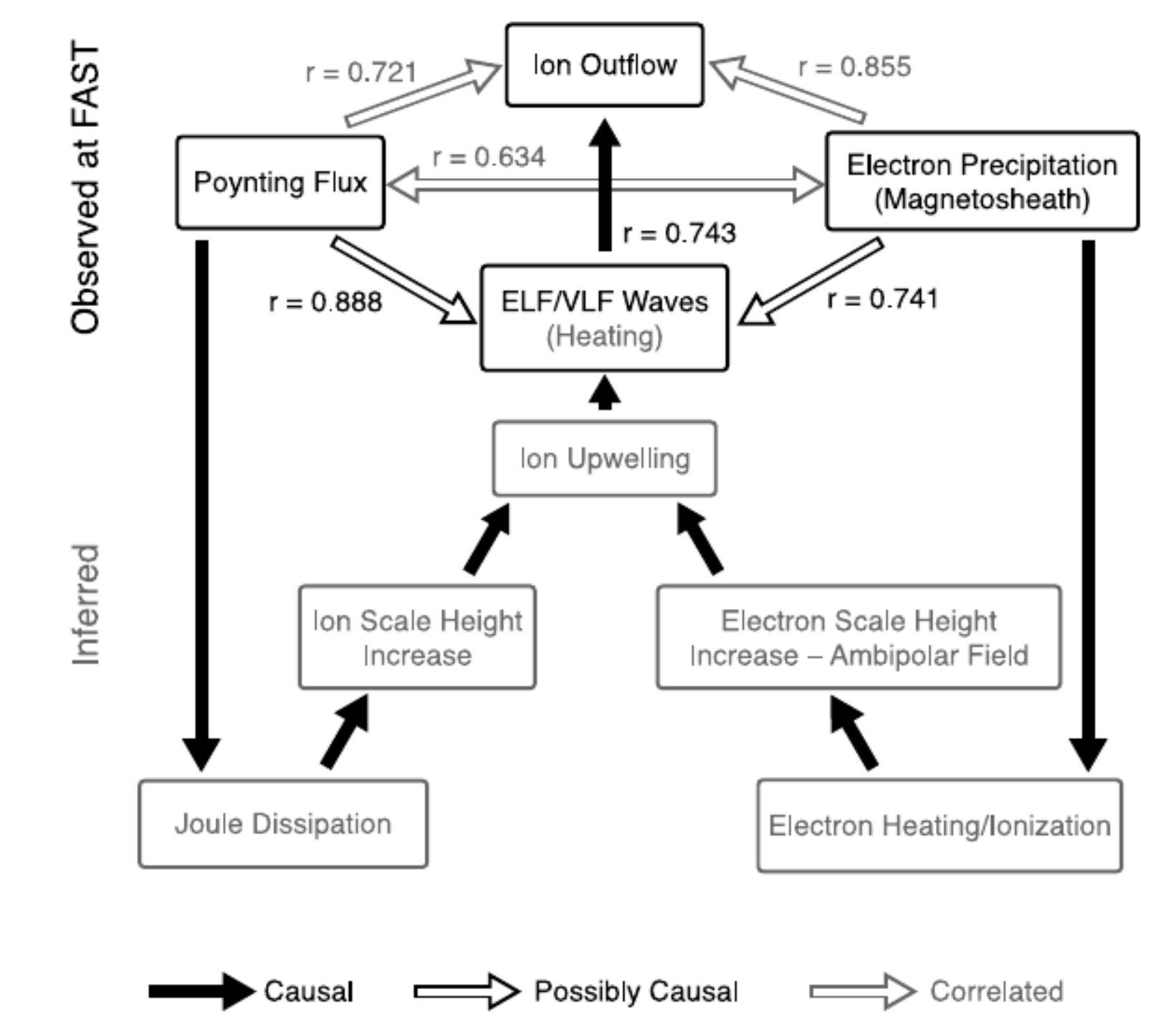}
 \caption{Correlation and causation in the ion outflow. The correlations are under current and proposed investigation to prove if they are actual causes or just coincidences/effect of similar causes. From \citeA{Strangeway2005}. }
 \label{strangeway2005}
\end{figure}

When thinking about what drives ionospheric outflows, it is instructive to consider the types of energy input. These break down into two broad categories as outlined in \cite{Strangeway2005} (see Figure \ref{strangeway2005}): (1) particle and (2) electromagnetic energy input from the magnetosphere. Both downward Poynting flux and soft electron precipitation from the magnetosphere were shown to correlate very well with outflow of ions observed by the Fast Auroral Snapshot (FAST) Explorer spacecraft. While correlation is not the same as causation, it so happens that there are a number of causal mechanisms associated with each type of energy input:
\begin{enumerate}
    \item Particle: Suprathermal electrons (Photoelectrons, auroral electrons, secondary electrons,...) enhancing the ambipolar electric field and depositing energy to the thermal electron population.
    \item Electrodynamic: Transverse heating of ions as a result of wave-particle interactions, ponderomotive forcing from Alfv\'en waves, field-aligned currents driving E$\parallel$, low altitude frictional heating driving upwelling, centrifugal force due to field line convection and curvature change and/or magnetic field co-rotation with the planet.
\end{enumerate}
The varied timescales and spatial regions over which these processes act result in dynamic outflow that varies spatially. At lower altitudes, the influence of different drivers separates the upflowing plasma into what has been called Type 1 and Type 2 outflow \cite{Wahlund:1992}, where \textbf{Type 1 involves strong electric fields and Joule heating, and Type 2 involves particle precipitation and enhanced electron temperatures}. At high altitudes, the escaping plasma also exhibits temporal and spatial variability. The polar region at Earth typically contains lower energy polar wind outflow, whereas additional energization, particularly from wave particle interactions, results in an energetic ion outflow and preferential acceleration of heavy ions in the auroral and cusp regions. Figure \ref{strangeway2005} shows the different pathways to ion outflow and some of the unknown.

\subsubsection{Suprathermal electron effect}
Suprathermal electrons refer to electrons whose mean energy is much greater than the thermal energy. The source of these electrons are either from XUV light shining on the atmosphere creating photoelectrons, precipitating electrons of magnetospheric origin (auroral electrons), or secondary electrons formed by impact ionization of the neutral atmosphere. This population is known to alter the ion outflow solution though two main processes:
\begin{enumerate}
\item Formation of the self-consistent ambipolar electric field.
\item Coulomb collisions between the superthermal and thermal electrons raising T$_e$.
\end{enumerate}
Relative to ions, suprathermal electrons are unbound by gravity and in absence of any other process would escape. However, this would lead to a net charge in the plasma violating the quasi-neutrality condition. Therefore, an electric field forms that retards the electrons and accelerates the ions, reducing the gravitational potential barrier.
Another pathway through which these electrons influence the outflow is through the deposition of energy to the thermal electrons raising the electron temperature and eventually the ion temperature. 

Photoelectrons, formed from ionization of the atmosphere by solar/stellar radiation, have been particularly well studied in the context of ionospheric outflows. 
There have a large number of theoretical studies \cite{Lemaire:1972b,Tam:1995,Tam:1998,Khazanov:1997,Wilson:1997,Su:1998} and observational studies \cite{Lee:1980, Peterson:2008,Kitamura:2011} showing that this population is critical to setting up the quiet time outflow solution.

\subsubsection{Electrodynamic Energy Input}

Waves also play an important role in the acceleration of plasma in the high-latitude, high-altitude polar region. They do this primarily through two mechanisms: the ponderomotive forces of Alfv\'en waves \cite{Li:1993a,Guglielmi:1996,Khazanov:1998a,Khazanov:2000a, Khazanov:2004a} and wave heating \cite{Retterer:1987,Crew:1990,Barghouthi:1997,Bouhram:2003,Waara:2011}.

Ponderomotive forcing due to low frequency electromagnetic waves allows electromagnetic energy from the magnetosphere to transfer energy to the ionospheric plasma. 
It arises from a non-resonant interaction between the particle encountering different portions of the wave during different parts of the particle gyration. The ponderomotive forcing depends on the wave mode, propagation direction, frequency, and background fields. While there are several types of pondermotive force derived in the literature, a useful description of the total field-aligned force $F\parallel$ from Alfv\'en waves as given by \citeA{Lundin:2006} is:
\begin{equation}
F_\parallel=-\frac{mc^2}{2B^2}\left[\frac{E^2}{B}\frac{\partial B}{\partial z} - \frac{1}{2}\frac{\partial E^2}{\partial z}\pm\frac{1}{c_A}\left(\frac{\partial}{\partial t} + \nu\right)E^2\right]
\end{equation}
Where m is the mass, $c_A$ the Alfv\'en speed, and $\nu$ the collision frequency. $E$ the electric field and $B$ the magnetic field.

Although the upward ponderomotive acceleration of ions is not species-dependent, it is countered by a downward force on electrons, resulting in a downward ambipolar field and a resulting species-dependent reduction to the acceleration \cite{Miller:1995}.

In contrast, wave-heating arises from the resonant interaction of particles with the portion of the turbulent wave spectrum that corresponds to the cyclotron motion of the particle. This preferentially heats the ions perpendicularly to the magnetic field. The mirror force converts this excess perpendicular energy into organized parallel motion. 
When modeling this interaction, the wave-heating is often represented as a diffusion term on the right hand side of the Boltzmann equation having a form like \cite{Crew:1985}:
\begin{eqnarray}
\frac{1}{v_{\perp}}\frac{\partial}{\partial v_{\perp}}\left(v_{\perp}D_{\perp}\frac{\partial f}{\partial v_{\perp}}\right)
\end{eqnarray}
where f is the velocity space distribution function, $v_{\perp}$ is the perpendicular velocity, and $D_{\perp}$ is a diffusion coefficient. The diffusion coefficient can be written approximately as \cite{Crew:1990}:
\begin{eqnarray}
D_{\perp} = (\pi q^{2}/2m^{2})|E_{L}|^{2}(\Omega(l))
\end{eqnarray}
Where $|E_{L}|^{2}$ is the electric field spectral density of left hand polarized waves,  and $\Omega(l)$ is the gyrofrequency of an ion of mass $m$ and charge $q$ at position `$l$' along a field line. Clearly this term acts to add energy to the ions transverse motion around the field increasing the first adiabatic invariant and enhancing the mirror force which accelerates the ion. 

\begin{figure}
 \noindent\includegraphics[width=40pc]{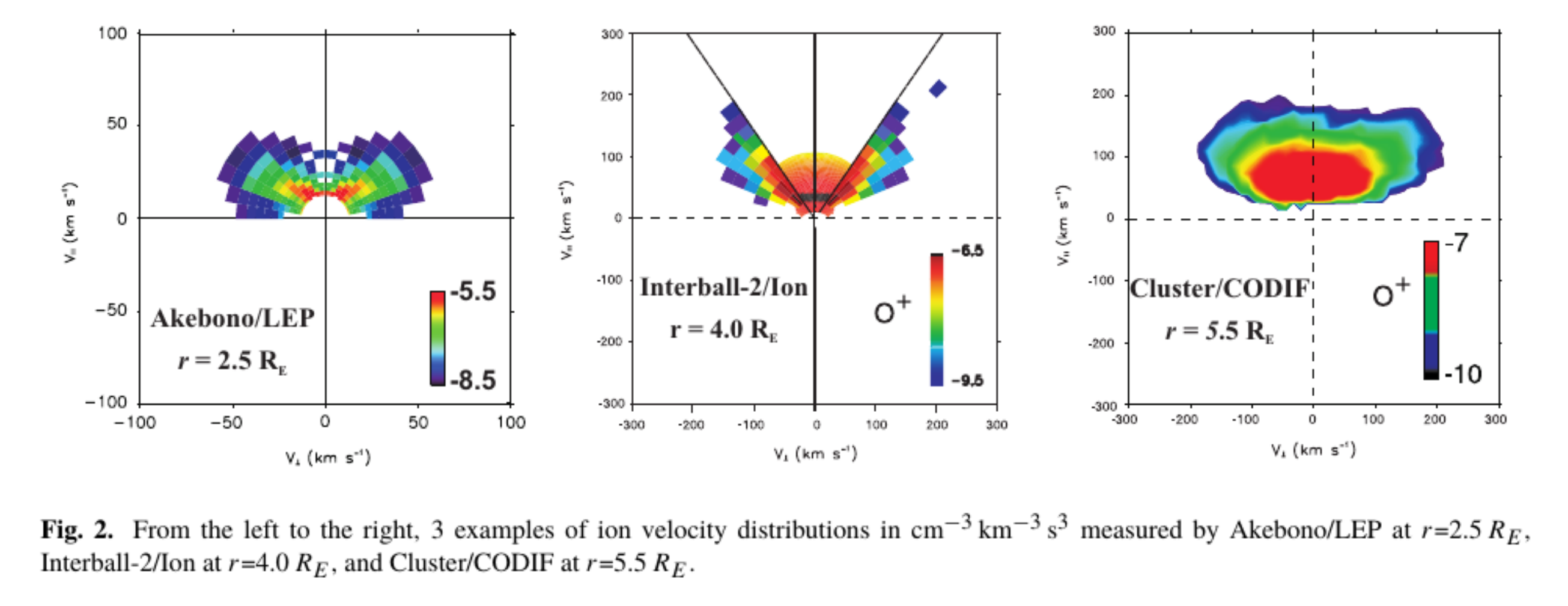}
 \caption{The conic distribution function of ions as observed by different satellites. These observations are a signature of wave excitation typical of polar outflow. From \citeA{Bouhram2004}, Creative Commons.}
 \label{conic}
\end{figure}

Resonant wave-heating has a clear signature in the shape of the ion distribution function. When the wave heating is active, the distribution function becomes increasingly perpendicular and pancake shaped. The mirror force, which acts more strongly on particles with higher perpendicular velocity, causes the distribution to ``fold'' upward into a characteristic ``V'' shape. In three-dimensional velocity space this looks like a cone, and hence the name ``conic'' distribution. The observation of a conic distribution is a clear signature of the presence of resonant wave-particle interactions. Figure \ref{conic}, from \citeA{Bouhram2004}, shows three examples of this feature observed by different satellites.  

\subsubsection{Field-Aligned Currents}
Field-aligned currents (FAC, also known as ``Birkeland currents''), driven by a planet's magnetic interaction with the stellar wind, are another form of electromagnetic energy input that contributes to ionospheric escape. This process was looked at by \citeA{Gombosi:1989}, who found that including a field aligned current causes the thermal electrons to respond by possibly enhancing the ambipolar electric field. More generally, a current conservation equation can be defined as follows \cite{Glocer:2016}:
\begin{equation}
n_{e}u_{e}+n_{se}u_{se}-\sum_{i}n_{i}u_{i}=-\frac{j}{en}
\end{equation}
which states that the current density, $j$, must be equal to the difference between the  flux of electrons (thermal and suprathermal) and the flux of ions. If a large current is driven into the ionosphere, then this condition requires other populations to react. 

\subsubsection{Joule Heating}
\label{outflow:jouleheating}
Joule heating (see also Section \ref{electromagneticenergy}) refers to the frictional heating caused by the differential motion of ions being dragged through the neutral atmosphere.  In this process, the planet's magnetic field interaction with the stellar wind generates a cross polar cap potential which sets up magnetospheric convection as well as strong convective flows in the E and F regions of the polar ionosphere. This convective flow is generated by the ExB drift and is not felt directly by the neutral population. As a result there is a differential motion between the ions and the electrons. There are several presentations of Joule heating as described by \citeA{Strangeway:2012}, but fundamentally the most direct way to model this process is as a frictional heating term. This term can be presented based on Burger's fully linear approximation \cite{Burgers:1969} as:
\begin{eqnarray}
\sum_{j}\frac{\rho_{i} \nu_{ij}}{m_{i}+m_{j}}
\left[ m_{j} \left( u_{i}-u_{j} \right)^{2} \right]
\end{eqnarray}
\cite{Gombosi:1987} and \cite{Cannata:1988} examined the role of Joule heating and found that transient upflows can result from this process.

Centrifugal forces play a role both at Earth \cite{Horwitz:1994} and at Jupiter \cite{Nagy1986}, but the origin of the centrifugal forcing is different for the two planets. At Earth, the solar-wind connected field lines convect across the high altitude polar cap region, resulting in changes to the field line curvature that centrifugally accelerate the particles outward along the magnetic field. At Jupiter, solar wind-driven convection plays a less important role, but the rapid rotation of the planet results in outward acceleration at lower latitudes.

\subsubsection{Escape equations}
There are several types of methods for modeling ionospheric outflows, but they can generally be divided into two categories: hydrodynamic models and kinetic models. In the case of hydrodynamic models usually a multimoment expansion of the Boltzmann equation for each ion species is undertaken. For magnetized planets, this is taken in the low $\beta$ limit where the magnetic field is strong. In this case the gyrotropic 5 moment equations with heat flux along an expanding magnetic field are given by \cite{Gombosi:1989}:

\begin{eqnarray}
\frac{\partial}{\partial t} \left(A\rho_{i}\right) + \frac{\partial}{\partial r} \left(A\rho_{i}u_{i}\right)=AS_{i} \label{continuity}
\end{eqnarray}
\begin{eqnarray}
\lefteqn{\frac{\partial}{\partial t} \left( A\rho_{i}u_{i} \right) + \frac{\partial}{\partial r} \left(A \rho_{i} u_{i}^{2} \right) +A\frac{\partial p_{i}}{\partial r} =} \nonumber \\
& & A\rho_{i}\left(\frac{e}{m_{i}}E_{\|}-g\right) + A\frac{\delta M_{i}}{\delta t}+A u_{i}S_{i} \label{momentumeq}
\end{eqnarray}
\begin{eqnarray}
\lefteqn{\frac{\partial}{\partial t} \left(\frac{1}{2}A\rho_{i}u_{i}^{2}+ \frac{1}{\gamma_{i}-1}Ap_{i}\right)+ \frac{\partial}{\partial r}\left( \frac{1}{2}A\rho_{i}u_{i}^{3}+\frac{\gamma_{i}}{\gamma_{i}-1}Au_{i}p_{i}\right) } \nonumber \\
& & =A\rho_{i}u_{i}\left(\frac{e}{m_{i}}E_{\|}-g\right)+\frac{\partial}{\partial r}\left(A\kappa_{i}\frac{\partial T_{i}}{\partial r}\right) + A\frac{\delta E_{i}}{\delta t} \nonumber \\
& & +A u_{i} \frac{\delta M_{i}}{\delta t} + \frac{1}{2} Au_{i}^{2}S_{i}\label{energyeq}
\end{eqnarray}
In this case $m$ refers to mass, $\rho$ is the mass density, $u$ refers to the velocity, $T$ to the temperature, $p$ to the pressure, $r$ to the distance along the field line, and $e$ is the charge of an electron. The subscriptions denote ion species or electron. Other terms include the expanding cross-sectional area of the flux tube ($A$), the heat conductivity ($\kappa$), the specific heat ratio $\gamma$, and Boltzmann's constant ($k$). The electric field $(E_{\|})$ is derived as an Ohm's law from a the steady state electron momentum equation as:
\begin{eqnarray}
\lefteqn{E_{\|}=-\frac{1}{en_{e}}\left[ \frac{\partial}{\partial r} \left( p_{e} + \rho_{e}u_{e}^{2} \right) + \frac{A'}{A}\rho_{e}u_{e}^{2} \right] +} \nonumber \\ 
& & \frac{1}{en_{e}}\frac{\partial}{\partial r} \left( \sum_{i}\frac{m_{e}}{m_{i}}\left[ \left(u_{e}-u_{i}\right)S_{i}-\frac{\delta M_{i}}{\delta t}\right]+\frac{\delta M_{e}}{\partial t}\right)\label{efield}
\end{eqnarray}
The source on the right hand side of equations \ref{continuity}-\ref{energyeq} represent the source due to ion production and loss ($S_{i}$), the source due to momentum transfer ($\frac{\delta M_{i}}{\delta t}$), and the source due to energy transfer ($\frac{\delta E_{i}}{\delta t}$), which includes Joule heating effects. There are several derivations of the collision based source terms for momentum and energy transfer, but a common choice are determined by using Burgers' fully linear approximation \cite{Burgers:1969}. Specific expressions for these terms can be found in the textbook by \cite{Schunk:2000book}. The equations outlined above are used in the Polar Wind Outflow Model \cite{Glocer:2012,Glocer:2009}. However, other codes use different hydrodynamics expansions including higher moment approximations. For example, the model presented by \citeA{Varney:2014} uses the 8 moment approximation, while \citeA{Barakat:1982} uses the 16 moment approximation. We do not elaborate further on these approach here but refer the interested reader to those papers. 

Kinetic solutions to ionospheric outflow typically solve the Boltzmann equation in some approximation. In the steady state this equation is given by \citeA{Khazanov:1997, khazanov2010kinetic}:
\begin{eqnarray*}
\mu \frac{\partial f_{\alpha}}{\partial t} + \mu v \frac{\partial f_{\alpha}}{\partial r} - \frac{1-\mu^{2}}{2B}\frac{\partial B}{\partial r}v\frac{\partial f_{\alpha}}{\partial \mu}=
\end{eqnarray*}
\begin{eqnarray}
-\left(\frac{e}{m_{e}}E_{\|}-g\right)\left(\mu\frac{\partial f_{\alpha}}{\partial v}+\frac{1-\mu^{2}}{v}\frac{\partial f_{\alpha}}{\partial \mu}\right)
\end{eqnarray}

Solving this equation usually takes one of two forms. \citeA{Khazanov:1997} use a direct solution of the Vlasov equation along a field line, whereas \citeA{Barakat:2006} use a 3D macroscopic particle-in-cell Monte Carlo technique. (Depending on the study, the time dependent or the static case are solved).

\subsubsection{Questions}

The PWOM model \cite{Glocer:2009} was applied to exoplanets \cite{airapetian2017hospitable} with the study of the atmospheric escape of a Earth-like planet at Proxima-b \cite{GarciaSage2017}. These studies were able to show that large escape rates happen even in the presence of magnetic fields. This model has the advantage of taking into account the diffusion of ions, assuring that this escape is not limited by ionospheric modeling. The questions remaining for this process is how it evolves in the absence of a permanent magnetic field, i.e. when there is only an induced magnetic field. Studies by \citeA{Collinson2019} tends to indicate that ambipolar diffusion helps the ion escape at Venus, but it remains to be modeled exhaustively. The long-lived doubly charged ions, observed/predicted in several ionospheres \cite{Simon2005,lilensten2005prediction,gronoff_2007,thissen2011doubly}, are easier to lift and therefore to escape through these processes. Are the O$^{2+}$ observed by ISEE \cite{1981JGR....86.9225H} due to these processes, and are they a significant source of escape? The dependence upon the EUV flux in certain situations \cite{Young1982} is consistent with such an hypothesis.

\subsubsection{Observables}
As explained, the polar wind is directly observed by plasma instruments. The dependence of the escape efficiency upon the q/m ratio means that lighter isotopes are easier to lift, therefore enhancing the gravitational distillation of the ions (which can be affected by the self-shielding effect Section \ref{selfshielding}).

\subsection{Other ion escape}
\label{oloss}
While the main sources of losses could be linked to the previously cited ones, other ion escape mechanisms, have been reported in the literature. They mainly come from the observation of ``bulk'' ion escape at Mars or Venus, during specific solar conditions \cite{Halekas2016}. This general denomination groups together escape that could come from very different processes while leading to loss of ions in an organized way. Overall, it is transfer of momentum from the solar wind to the ionosphere that makes these plasma escape, and it could be considered as ion pickup in a first approximation, but with more complex MHD effects. As explained in \citeA{Terada2002}, the main problem is that the production of ions above the ionopause is less than the escaping flux, which means that different processes have to diffuse these ions from the ionosphere to above the ionopause where they would be picked up.  

This is this second class of processes, that comes from complex interactions between the ionosphere and the interplanetary magnetic field to help transfer ions to the top of the thermosphere, that are studied here. These ions are usually not energetic enough to escape, but their presence makes it easier for pickup and ``bulk'' escape. It is to be noted that, when conditions are extreme, it is possible to reach levels where the production/diffusion of ions is limiting the escape. Numerous models of ion pickup do not take that situation into account, leading up to unrealistic ion escape fluxes \cite{Egan2019}.

In magnetospheres, similar problems arise: the different ion energizing processes lead to the creation of a plasmasphere, i.e. ions trapped inside the magnetosphere, and these ions are removed either by falling back to the planet or leaving through different processes \cite{Seki2015,jackman2014large}. 


\subsubsection{Fluid processes: Kelvin-Helmholtz and other instabilities}
Kelvin-Helmholtz instabilities (KHI) \cite{Johnson2014} have been observed at Mercury \cite{Sundberg2010}, Venus (\citeA{Pope2009, Lammer2006,Terada2002} and references therein: the observations of flux rope and detached plasma clouds are linked to KHI through modeling), Earth \cite{Johnson2014} and Mars \cite{Ruhunusiri2016}. They occur at the interface between two fluids or plasmas having a velocity shear and lead to upward pressure gradients and the formation of a vortex. KHI have effects in a variety of planetary processes. In the case that interest us, i.e. the development of KHI at the interface between an ionosphere and the solar wind, it leads to the transfer of solar wind momentum to the ion and, ultimately, to their acceleration into space. \citeA{penz2004} computed, for Mars, O$^+$ escape values of the order of 2~10$^{23}$ -- 3~10$^{24}$ ions/s.
Rayleigh-Taylor instabilities, ion-ion instabilities, electron-ion instabilities, have also been proposed as mean of momentum exchange leading to escape \cite{dubinin_ion_2011}.

\subsubsection{Pick-up processes: the ion plume}
The ion plume of Mars was inferred from Mars Express observations, and fully characterized by  MAVEN  \cite{liemohn_mars_2014,dong_strong_2015}. It originates from the interaction between the solar wind and the ionophere of Mars, which creates an upward electric fields through $\vec{E} = -\vec{U}_{SW} \times \vec{B}$. This process can be looked as a special case of ion-pickup since it is observed in such-models. 
At Mars, this plume escape for O$^+$ is estimated to be 30\% of the tailward escape, equivalent to 23\% of the total ion escape \cite{dong_strong_2015}.


\subsubsection{Ambipolar fluxes / outflow anomalies / snowplow}
The ``cold'' ions, i.e. ions not energized at suprathermal temperature and coming from the ionosphere, are prominent in the plasmasphere  \cite{Li2017}. Several processes leads to the filling of that plasmasphere. Ion upwelling \cite{Strangeway2005}, which can be linked to the ambipolar electric field, but at levels that do not lead to escape, is one of these processes at Earth. For unmagnetized planets, ions are transported in the upper layers of the ionosphere by ambipolar electric field \cite{Collinson2019,Akbari2019}. The draping of the interplanetary magnetic field (IMF) could lead to additional induced field with respect to the processes described in Section \ref{ionosphericoutflow}, this process is called the ``snowplow'' \cite{Halekas2016}.  From there, the transfer of momentum from the solar wind to the ionosphere creates detached plasma clouds that are escaping. It is important to note that the $\vec{E}\times\vec{B}$ drift can help these ions escape as observed at Mars, where different escape rates are observed in the +E and -E hemispheres \cite{Inui2019}. 


\subsubsection{Plasmaspheric ion losses / substorm losses / plasmoids / flux ropes }
As previously shown, the charge exchange is a major ion loss process in magnetospheres. A fraction of ions can be lost from escape from the tail, but, as shown from the observations at Jupiter, this is a low percentage of the escape \cite{jackman2014large} (At Jupiter, other processes have to be taken into account to balance the input of plasma from Io.). This escape  from the tail works through the creation of a plasmoid: the pressure of the IMF elongates the planetary magnetic field, leading to the reconnection of that magnetic field. This reconnection means that a part of the inital magnetosphere is no more linked to the magnetic field of the planet and is ejected into space, along with the plasma it contains. Ejection of plasmoids is also associated to a return of plasma towards the planet. (This is different for situations when magnetic fields are weaker, such as Mars). 
The equivalent of these plasmoids have been observed at Mars near the magnetospheres created by the crustal magnetic field, they are usually named ``flux ropes'' \cite{hara2017maven}, are linked with coronal mass ejection disturbances and other processes \cite{Hara2017}. They could be responsible for up to 10\% of the present day ion escape at Mars \cite{brain2010}. Finally, plasmoid escape has also been observed at Venus, in the induced magnetic field, and looks like the Earth's or Jupiter's plasmoids \cite{Zhang567}.


\subsubsection{Questions}
These ion escape processes are actively studied with missions such as MEX, VEX, MAVEN, etc., as well as numerical models. Most of the questions are linked to the actual amount of ions escaping due to these processes and how these evolve with the solar/stellar activity. From an observation point of view, it may be difficult to distinguish between processes from the a single point observation of the amount and location of plasma escaping \cite{Inui2019}. In addition, some processes can be seen as generalization of other processes (e.g. the ``snowplowing'' is a generalization of the ion outflow observed in magnetospheres). These points led to the above definitions and organization of these ion escape processes.  

\subsection{Ion Return and Net Escape Rates}
\label{ionreturn}
While the ionospheric outflow processes detailed above determine the escape of plasma from the ionosphere, a significant fraction of this plasma becomes trapped in Earth's magnetosphere. Magnetospheric ions mostly consist in a mixture of H$^{+}$ and O$^{+}$ ions. Contrary to H$^{+}$ ions, which can either originate from the solar wind or the ionosphere, O$^{+}$ ions almost exclusively originate from the ionosphere and are used as tracers of ionospheric material in the magnetosphere. They have been observed by several spacecraft, including GOES 1 and 2 \cite{Young1982}, ISEE \cite{LennartssonShelley1986,Lennartsson1989}, Van Allen Probes\cite{Fernandes2017}, GEOTAIL \cite{Nose2009,Ohtani2011}, and Cluster \cite{MaggioloKistler2014,KistlerMouikis2016}. All these observations show an increase the amount of O$^{+}$ ions in the magnetosphere, and thus of ionospheric material, with increasing solar EUV/UV flux and geomagnetic activity, i.e. with the amount of energy deposited din the ionosphere. Once in the magnetosphere, ionospheric material enters magnetospheric circulation patterns, which may ultimately result in loss to interplanetary space or return to the ionosphere. \citeA{Seki2001} estimate the fraction of Earth's oxygen lost to interplanetary space at about $1/10$ of the ionospheric oxygen outflow during periods of low solar activity, based on the estimate of the O$^{+}$ loss due to the main four escape routes for terrestrial ions: the escape of cold detached plasmaspheric particles through the magnetopause, of high-energy ring current/dayside plasma sheet particles through the magnetopause, of plasmasheet ions through antisunward flow in the nightside plasma sheet, and of terrestrial ion beams through the lobe/mantle. Note that the charge exchange loss of ring current ions was not considered by \citeA{Seki2001}. The outflow and loss rates are enhanced during high solar and geomagnetic activity but may not account for all magnetospheric loss mechanisms, particularly for low energy ions that are difficult to observe. This estimate, then, should be considered a lower bound on escape, but the important point here is that not all outflowing ions escape from the magnetosphere-ionosphere system.\\
However, recent observations above the polar ionosphere in the magnetospheric lobes by the Cluster spacecraft, provide evidence for a higher loss rate for ionospheric ions flowing through the lobe and mantle region. \citeA{Slapak2017} showed that energetic ions (in the range of a few hundred to several thousand of eV) escaping from the cusp region through the magnetospheric lobes/mantle have a high probability of being loss to interplanetary space rather than being returned to the ionosphere. They even claim that over geological times a quantity of oxygen lost by the Earth's atmosphere could be roughly equal to the amount of the present atmospheric oxygen content if the young Sun was actually more active than nowadays. 
Furthermore, the flux of precipitating ions as estimated by the the DMSP satellites is only of the order of 10$^{24}$ \cite{Newell:2010}: 1 to 2 orders of magnitude lower than the estimated flux of outflowing ionospheric ions. These new observations provide strong evidence against a high return rate of ionospheric ions and rather suggest that a significant fraction of ionospheric ions escaping from the ionosphere may actually be definitively lost into the interplanetary space.

\section{Major Parameters and Concepts}

In order to address the escape rate of an atmosphere and to retrieve its evolution with time, it has been demonstrated that several processes are in action. To evaluate whether or not they are negligible at a certain period in time, or to approximate the calculations, several concepts have been proposed, such as the energy limited escape or the critical heating rate for hydrodynamic escape. The two major parameters in the different models are the energetic inputs, from the EUV-XUV fluxes to the electron precipitations, and the atmospheric structure and composition.
Finally, it is very important to take into account the evolution with time, from the time dependence on small scales (typically sensitive to the solar/stellar activity) to the evolution of the atmospheric escape through eons, leading to isotopic fractionation, which is the main probe for the history of our Solar system's atmospheres (in the absence of better in-situ measurements, e.g. trapped gases in rocks \cite{Jakosky1991}).

\subsection{Limiting parameters}

\subsubsection{Critical Heating Rate}
\label{criticalheatingrate}

Present theory (Section \ref{hydrodynamicescape}) incompletely describes transition from Jeans escape to hydrodynamic escape. 
Transonic models \cite{Murray-Clay2009} have been used to describe rapid escape from exoplanets and from Pluto \cite{strobel_n2_2008}. However, \citeA{johnson2013,johnson2013_erratum} have recently discovered that this model for Pluto gave an incorrect upper atmospheric structure \cite{Tucker2012}. This erroneous prediction of the upper atmospheric structure results from applying the Jeans expressions at the exobase \cite{chamberlain_theory_1987} for uncertain boundary conditions at infinity \cite{tian_hydrodynamic_2008-1} when simulating rapid escape using continuum gas dynamics.  In this context, a sonic point is assumed to occur at an altitude $r_*$, above which the density and temperature dependence can be simply characterized \cite{parker_dynamical_1964,parker_dynamical_1964b}. The hydrodynamic, energy-limited (see Section \ref{limitedenergy}) escape rate, applied to exoplanet atmospheres \cite{Lammer2009}, is often assumed to imply that sonic boundary conditions are applicable \cite{erkaev_xuv_2013}. \citeA{johnson2013,johnson2013_erratum} used molecular kinetic simulations to show that this is not the case.
Ignoring viscosity, \cite{parker_dynamical_1964,parker_dynamical_1964b} used the momentum and energy equations to describe escape when the dominant heat source is internal. This same model was applied to planetary atmospheres primarily heated at an altitude $r_{a}$. For a Jeans parameters at $r_{0}$ (the lower altitude considered) as large as $\lambda_{0}$ $\sim$ 40, such models were assumed to produce a transonic expansion, which is often referred to as a slow hydrodynamic escape \cite{strobel_n2_2008}. However, rapid escape can occur for large Jeans parameters only when the Knudsen number is low, i.e. when the collisional approximation cannot be assumed. Therefore, the gas does not go sonic in the collision-dominated region and the escape rate computed in the ``slow hydrodynamic escape'' paradigm is a few times larger than the Jeans rate \cite{volkov_kinetic_2011a,volkov_thermally_2011}. As we explained in Section~\ref{slowhydro}, the observations that led to the ``slow hydrodynamic escape'' hypothesis could be explained by alternative processes based on chemistry.

From there \citeA{johnson2013,johnson2013_erratum} have developed a criterion to check if a transonic solution will exists, i.e. if we can approximate the escape by a hydrodynamic model. Assuming that $r_0 < r_* <r_x$, which should be the case in hydrodynamic escape, it was found that the net heating rate $Q_{net}$  should follow equation \ref{eq:criticheating}:
\begin{eqnarray}
Q_{net} > Q_{c} &\approx& 4\pi r_{*}\frac{\gamma}{c_{c}\sigma_{c}Kn_{m}}\sqrt{\frac{2U(r_{*})}{m}}U(r_{0}) \label{eq:criticheating}\\
U(r) &=& \frac{GmM}{r}
\end{eqnarray}

K is the Knudsen number and $c_{c}$ is determined by the energy dependence of the total collision cross section, $\sigma_{c}$.

If heat is primarily absorbed over a broad range of r below $r_{x}$, we can use $Kn_{m}\sim$1 as an approximation. Here, it can be seen that $Q_{c}$ does not explicitly depend on $T_{0}$, but on the sonic point only where a lower bound can be obtained by replacing $r_{*}$ with $r_{a}$, the mean absorption depth. This mean absorption depth is estimated from $ \sigma_{a}$ , the absorption cross section. At threshold, the sonic point will approach $r_{x}$, such that $r_{*} \sim r_{a}[1 + (\frac{\sigma_{a}}{c_{c}\sigma_{c}})\lambda_{ave}]$ where $\lambda_{ave}  \sim (\lambda_{a} + 2\gamma)/2$ which slightly increases $Q_{c}$. Using Pluto as an example, UV/EUV absorption at $r_{a}$ $ \sim$ 1.5 times Pluto’s radius, $Kn_{m}$ $ \sim$ 10$^{-3}$, and  $r_{*}$ $ \sim$ $r_{a}$ $ \sim$ $r_{0}$, Equation (\ref{eq:criticheating}) gives $Q_{c}$ $ \sim$ 4.5 x 10$^{10}$ W for Pluto, which is well above the largest heating rate, and shows that hydrodynamic escape should not be applied for the dwarf planet. We compiled the values of $Q_c$ in Table~\ref{escapeparameterstable}.

\subsubsection{Energy limited escape -- Radiation / Recombination - limited escape }
\label{limitedenergy}
The estimation of mass loss rate of exoplanets often assume an energy-limited escape (Section \ref{enelimited}). The basis of that assumption is that an exoplanet thermosphere is mainly composed of H, heated by ionization of H. From there, it is supposed that a large quantity of that heat is transformed into hydrodynamic escape. Therefore, one uses an efficiency coefficient $\epsilon$ (sometimes $\eta$) for transforming EUV-XUV energy into escape. This led to equation \ref{eq:hydenul}, with the standard efficiency coefficients found in the literature. \citeA{Erkaev2007} shows that this equation can be slightly modified to account for stellar gravity effects that affect close-in planets. 

For giant planets close to very active stars, the radiation-recombination limited escape is often used as a harsher limit to the energy limited escape, because the H$^+$ can recombine, reducing some of the energy in the system \cite{luger2017evolution,linsky_host_2019}, this leads to an escape proportional to $\sqrt{F_{XUV}}$ instead of $F_{XUV}$. In the case of H atmospheres where the heating is only supposed to come from ionization, there is also a case where the escape is limited by the number of ionizing photons.

This approach has been developed to study close-in giant planets \cite{salz2016energy}, and led to energy diagrams \cite{ehrenreich_mass-loss_2011} to evaluate the mass loss from giant exoplanets. Unfortunately, it notably neglects the radiative cooling processes in the upper atmosphere of the planet, i.e. it neglects the problem of the upper atmosphere temperature (Note: \citeA{Lopez2017} includes radiative cooling in an energy-limited diffusion approach). 
The main problem of the energy limited escape approximation is that it is too often applied for rocky exoplanets while concealing these major limitations:
\begin{enumerate}
\item \textbf{The escape regime is not evaluated:} if the exoplanet is not in hydrodynamic escape, the energy-limited escape equation will give an overestimate of the escape.
\item \textbf{The atmospheric profile is not evaluated:} what is exactly the profile of the atmosphere, and therefore the efficiency of the escape if it is truly in hydrodynamic regime?
\item \textbf{The atmospheric composition is not taken into account, H is assumed to be the only species:} this is related to the other problems; the presence of cooling species such as CO$_2$ may totally change the escape regime; diffusion-limited processes may prevent H to be present in large quantities in the thermosphere, etc.
\item \textbf{Only photo-ionization heating is taken into account:} Joule heating or particle precipitations can be large sources of heating for close-in exoplanets.
\item \textbf{Non-thermal processes are not addressed:} those can dramatically change the profile of the escaping species.
\end{enumerate}
Energy-limited escape models can be interesting for studying H-rich rocky planets early in their histories, for which the escape of H may not have been diffusion-limited but energy-limited \cite{tian2005hydrogen}, however energy-limited escape is less relevant to more comprehensive habitability studies.  



\subsubsection{The diffusion-limited escape}
\label{diffusionlimitedescape}

Some escape processes can be very efficient, and limited by the amount of particle available for the escape, the bottleneck for the escape of these particles will then be the diffusion from the lower layers of the atmosphere to the upper atmosphere. Typically, the escape of H at Earth is diffusion limited. It follows the Equation \ref{eq:dlesc}.
At Titan, like Earth, H$_{2}$ escape is determined by the limiting flux through the homopause deep in the lower thermosphere \cite{Cui:2008, Strobel2012, Bell2014}. However, there is currently a discrepancy between the densities of H$_{2}$ measured in-situ by INMS and those produced by modeling studies \cite{Magee2009,Cui:2008, Strobel:2002, Bell2014}. Despite this discrepancy, all modeling studies to date have indicated that the H$_{2}$ upwelling into the lower thermosphere, combined with additional H$_{2}$ produced in the thermosphere, sets the eventual planetary escape flux of H$_{2}$.
A more complete theory of diffusion-limited escape, including the cases where the diffusing species has a non-negligible mass with respect to the main species can be found in \citeA{Hunten1973}.

\subsection{Energetic inputs}

 \subsubsection{The EUV/XUV flux}
The EUV-XUV flux modifies the temperature of the exosphere and the exobase altitude. It therefore changes the concentration of particles above the exobase. It is also responsible for the creation of hot atoms through photochemical processes. At the Earth, the EUV-XUV flux varies substantially as a function of solar activity. When the variability of the solar irradiance is rather low for the visible and the IR, with less than 0.1\% and 1\% from minimum to maximum respectively, the solar irradiance variability in the XUV /EUV can be more than doubled with a direct impact on the upper atmosphere \cite{2007LRSP.H}. This variability is of two different origins: one depends on sporadic explosive events such as flares with time scales from minutes to hours, while the second one is linked to the full Sun disk activity with longer cycles, from days to years. The latter one is then related to the appearance and disappearance of active regions on the solar disk, which causes then the variability on a 27-day solar rotation scale, associated with a 13.5-day modulation from the center-to-limb variation. The long-term monitoring of the solar EUV flux, however, is a difficult task, mainly because of the heavy degradation experienced by the solar instruments that are in orbit \cite{2013SoPh389B}. Before 2002 with the launch of the TIMED satellite \cite{2005JGRA..11001312W}, measurements of the solar EUV flux variability were rather scarce. This has led to the development of several empirical approaches for reconstructing the solar XUV/EUV part of the spectrum. 

A common approach lies with using solar proxies such as the radio measurements at 10.7 cm (F10.7) \cite{1990SoPh..127..321T} and the MgII core-to-wing index \cite{1986JGR....91.8672H}. Many models are then using a linear combinations involving these proxies and their 81-days running means, or even non linear combinations \cite{1981AdSpR...1...39H,2003JGRA..108.1059L,2006AdSpR..37..315R}. However, no single index can properly reconstruct the solar XUV/EUV irradiance at all time scales \cite{2009GeoRL..3610107D}. Moreover, some widely used proxies, such as F10.7, are not really suited for the XUV/EUV lines reconstruction, whose originated from the solar corona. The F10.7 index is, however, used as the solely index to estimate the solar variability within thermospheric and ionospheric models. For the solar minimum in 2008, when the thermospheric density dropped by 28\%, the F10.7 only decreased by 4\% \cite{2010GeoRL..3712102E}, outlying then the limitations of the F10.7 index for ionospheric studies \cite{2010GeoRL..3716103S}. More appropriate solar proxies has been recently suggested such as the radio measurements at 3 cm and 30 cm which are directly linked to chromospheric and corona emissions \cite{2017JSWSC...7A...9D}. 

A different approach considers that the solar spectrum is a linear combination of reference spectra that coming from different regions of the solar disk. Those regions are attributed to the quiet Sun, coronal holes and active regions and can be disentangled using solar images or solar magnetograms. Their respective contrast can be obtained by an empirical approach \cite{1998ApJ...496..998W} or using the differential emission measure \cite{2004A&A...419..345K}. A few terms is normally needed to reconstruct the solar irradiance in the XUV/EUV spectral range \cite{2008A&A...487L..13A}. This strongly outlines that the spectral variability is highly coherent through the spectrum, but this only for time scales that exceed the dynamic time of solar flares, since the solar atmosphere is strongly structured by the magnetic field. The solar spectrum in the XUV/EUV can then be reconstructed from measurements of a few correctly chosen passbands \cite{2011A&A...528A..68C,2012_Cessateur_SWSC}. For the short term spectral variability, a specific model has been developed, the Flare Irradiance Spectrum Model (FISM) \cite{2008SpWea...605001C}, based on TIMED/SEE and SDO data.

The effects of the solar XUV/EUV variability on Earth's upper atmosphere have been quantified with empirical models \cite{Bowman:2008}, that specify the exospheric temperatures as a function of indices of EUV radiation at different wavelengths \cite{Tobiska:2008}. At Mars, \citeA{luhmann_evolutionary_1992} computed the influence of the EUV flux on the escape processes. It is complicated by the fact that the solar wind pressure is also included in the calculations: the EUV flux increases, therefore the density of hot oxygen above the exobase increases (and the altitude of the exobase increases). Therefore the escape of hot oxygen increases, the density of pickup ions increases as well, and so the sputtering and the sputtered atoms. These non-linear effects lead to the large variations in the escape rates as computed in  Figure~\ref{euvluhman}. More recent modeling and data shows that the actual increase is less important than that previous simulations \cite{lillis_characterizing_2015}. The correlation of Mars Express' observations of ion escape at Mars with the EUV flux show that it is difficult to draw a direct relation between the two \cite{Ramstad2015} in the 7 year span these observations took place. However, the non-linearity of the dependence, and the fact that negligible escape processes can become very important for extreme EUV-XUV flux, such as in the conditions in the beginning of the solar system, is still valid.

\subsubsection{The electron flux}

\paragraph{The auroral-like electron flux}

The energetic electron flux at high latitudes is produced as a result of the interaction of the solar wind and interplanetary magnetic field with the magnetic field and magnetosphere of the planet, which in turn drives ionospheric electric fields and currents.  Upward currents may contain a significant, downward energy flux from electrons \cite{Fuller-Rowell:1987}.  The energy flux at the Earth typically ranges from under 1 GW up to 20 GW \cite{Newell:2010}.

As Mars only has a limited magnetosphere, there is no significant energy deposited by the aurora as discovered by Mars Express in 2005 \cite{Bertaux2005}. This conclusion can be challenged by the observations of global aurora during solar events \cite{Schneider2018}. Diffuse electron \cite{2017Icar..293..132T} and proton aurora \cite{Deighan2018} observed by MAVEN may carry significant amounts of energy, but the total flux still needs to be estimated. 

On the other hand, both Jupiter and Saturn do have large internal magnetic fields and correspondingly large magnetospheres, so there is considerable power in their aurorae. As there are no direct measurements available, much of what is known about the outer planets' aurorae has been obtained from UV measurements, at first on the Voyager flyby of Jupiter \cite{Broadfoot:1979}. Most recent UV observations are from the Hubble Space Telescope (HST). In a review of such observations, \citeA{Grodent:2014} indicated that the auroral emissions at Jupiter and Saturn are on the order of 1 TW and 0.1 TW respectively.  Uranus and Neptune are much weaker, at 1~GW or less, and observations are sparse. Of course, the power of the emissions is less than the kinetic energy that is deposited. The Voyager UV measurements at Jupiter has implied a power injection on the level of 12 TW \cite{Broadfoot:1981}, and \citeA{Gerard:2014_Jupiter} stated that the auroral precipitation at Jupiter has a power on the order of 10 to 50~TW.  As this level of heating is much greater than that from solar radiation, the aurora has a significant contribution to the thermal properties of the upper atmosphere.

\paragraph{The supra-thermal electrons}

Supra-thermal electrons are electrons with energy higher than the typical electron in an ionosphere: when looking at the flux of electrons in function of energy, the supra-thermal electrons are responsible for the departure of the curve from a Maxwellian at high energy.
These electrons come mainly from the precipitation of electrons from outside of the ionosphere, from local creation (typically photoionization -hence the name of photoelectrons-), but also from other ionization, including from suprathermal electron impact). 
Electric potential drops can accelerate electrons to suprathermal energies, but they occur outside the ionosphere and are responsible for some magnetospheric precipitation at Earth.
To understand the effect of the suprathermal electrons, it is necessary to compute their transport in an atmosphere. Codes such as Aeroplanets and PWOM do that.

The basis of these codes is to compute the flux of electrons by solving their transport equation. The existence of codes not based on a Monte-Carlo scheme, such as Aeroplanets, allow to fastly compute large quantities of conditions and to perform sensitivity analysis \cite{gronoff2012_1, gronoff2012_2}. We refer to these papers for the equations to solve in the ionosphere/thermosphere, and for the uncertainties encountered.

\subsubsection{The electromagnetic energy}
\label{electromagneticenergy}
The Joule heating is the heating created by the resistance of the thermosphere to the electric current due the ionospheric plasma \cite{Vasyliunas2005}. It is computed by evaluating the electric field and the conductivities.

Joule heating in the polar ionosphere has a significant effect on the exospheric temperatures, and hence the amount of outflow (Section \ref{outflow:jouleheating}).
At the Earth the total Joule heating is normally in the range of a few hundred GW, but in extreme events can range from 1~TW \cite{Lu:1998} up to 5~TW, while increasing the mean temperature of thermosphere by up to $500^{\circ}$K \cite{Weimer:2011}.  At the same time, the additional heating tends to increase the amount of nitric oxide in the thermosphere, which acts to accelerate the rate at which it cools down to the equilibrium temperature set by the solar EUV radiation \cite{Weimer:2015}.  \citeA{Wilson:2006} had found that Joule heating is most typically about 3 times the energy from precipitating particles, with the ratio varying from 2 to 7 in the different events that were studied.

At other planets there are no direct measurements of the electromagnetic energy input into their ionosphere and thermosphere, so at present it can only be estimated.
At Jupiter, \citeA{Strobel:2002} estimated the Joule and auroral particle heating to be about 1000 times larger than at the Earth for typical conditions, which would be on the order of 500 TW.

The generation of currents and electromagnetic energy at Jupiter may be dominated by processes much different from at the Earth, as the interaction of the solar wind and interplanetary magnetic field are weaker. It is thought that the planet's rotation and magnetic field provide a significant contribution to the energy sources of the heating processes \cite{Eviatar:1984, Waite:2002}.

Due to the lack of observations of the electromagnetic fields at other planets, most of what is known is derived from computer simulations, such as the Jupiter Thermospheric General Circulation Model (JTGCM), that addresses  global temperatures, three-component neutral winds, and neutral-ion species distributions \cite{Bougher:2005}.  In a case study with auroral forcing plus ion drag, \citeA{Bougher:2005} calculated exospheric temperatures at auroral latitudes ranging from 1200 to 1300~K, which match available multispectral observations.  The levels of Joule heating are in the range of 70 to 140~mW/m$^2$ in the auroral ovals, while the auroral particles produce 2 to 8~mW/m$^2$.
With different model parameters higher levels of the Joule heating can be produced and exospheric temperatures above 3000~K may be achieved.  Other numerical studies have been done, too numerous to mention here. The main point is that Joule heating can significantly modify the heat budget of the thermosphere in the Jovian gas giant, and similar processes would be expected at similar exoplanets. As there are many assumptions and approximations made in the modeling process, more work needs to be done to more accurately calculate the contribution of Joule heating to the exospheric temperatures and the resulting effects on the outflow, particularly the contributions from the solar wind dynamo.

\subsubsection{The Cross Sections and the computation of ionization}
\label{section_crosssection}
Elastic and inelastic cross sections are at the core of the computation of the energy transfer from particle precipitation to the atmosphere. To that extent cross sections for ionization, excitation, and dissociation are necessary tools for all the computations. Several efforts have been made to gather cross sections. The most comprehensive one has been recently developed with the study of upper atmospheres in mind, called AtMoCIAD. Its advantage is the inclusion  of error bars, that allows the computation of the propagation of the experimental or theoretical uncertainties \cite{gronoff2012_1,gronoff2012_2}, but also the inclusion of all kinds of particles (photons, electrons, protons, hydrogen, \ldots) colliding with atoms or molecules. 

\begin{figure}
      \noindent\includegraphics[width=40pc]{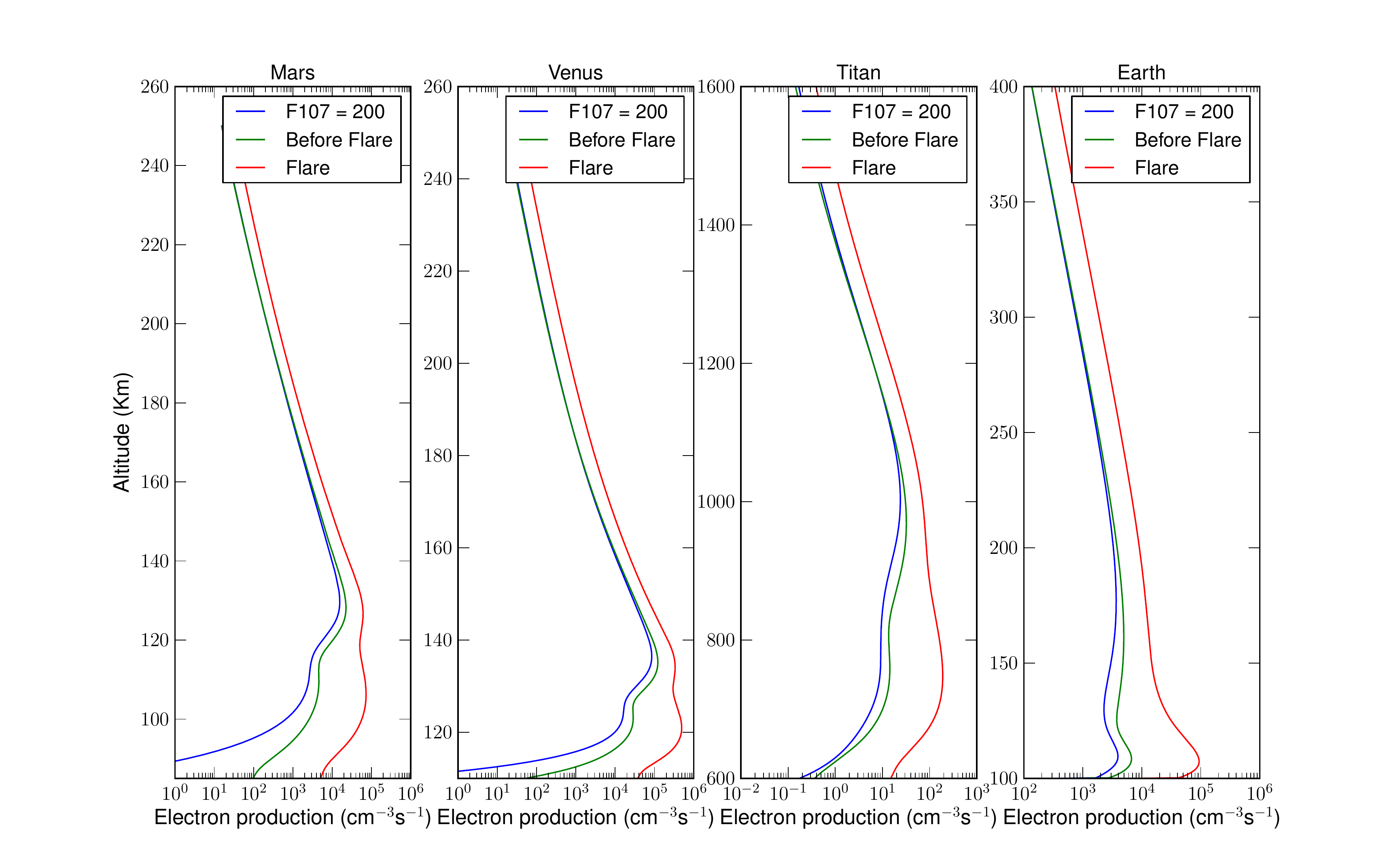}
\caption{The ionization at Mars, Venus, and Titan for similar solar conditions, including a solar flare. The neutral atmosphere is of importance in deciding at which altitude the peak is. The extent of the atmosphere, roughly determined by the scale height (since it can be function of the altitude) of the atmosphere, is the main parameter to explain the height of the peak. These ionizations were computed using the Aeroplanets model \cite{gronoff2012_1}; including both direct photoionization and secondary electron ionization.}
\label{ionizationfig}
\end{figure}

The precise knowledge of all types of cross section can improve the computation of the different conditions at different planets. A consistent set of cross sections allows to perform comparative planetology studies. An example of such a computation can be seen in Figure~\ref{ionizationfig}.

Other cross sections such as charge-exchange cross sections are of importance for escape studies.
The database maintained by the Atomic and Molecular Collisions Group of the Department of Physics and Astronomy at Rice University (Houston, USA,
http://www.ruf.rice.edu/$\sim$atmol/) is among the most populated with species of interest for space science studies \cite{lindsay2005charge}.

\subsection{Atmospheric structure}

Addressing the atmosphere structure is one of the more complex part of the study of upper atmosphere. Model should address both ionospheric problems, such as the precipitation of particles and Joule heating, as well as fluid problems like the heat transport, winds, or radiative problems such as the CO$_2$ 15$\mu$m cooling \cite[and references therein]{Johnstone2018}.

\subsubsection{Vertical Mixing and Photochemical Modeling of Atmospheres} 

Most of the planets in our Solar system have a substantial atmosphere, 
with the exception of Mercury which has a very tenuous atmosphere. Several moons in the Solar system also have atmospheres. An atmospheric gas can be made up of a variety of chemical species that were distributed unevenly at the time of the formation of the Solar system. A basic relationship between fundamental quantities governing the gas distribution of chemical species is the ideal gas law, $p = n(z)kT$, which becomes increasingly less valid for pressures greater than 1 bar, after which Van der Waals equation of state should be used \cite{Parkinson2002}.

Knowledge of the photochemical and chemical processes governing the transformation of a particular atmospheric species into another can be used to calculate the distribution of each species considered throughout the atmosphere.  Since a particular atmospheric constituent might be the source of one or more other constituents, this calculation requires the simultaneous solution of a series of coupled continuity equations, one for each atmospheric species considered, viz.,

\begin{eqnarray}
\frac{\partial n_i}{\partial t} + \nabla \cdot \phi_i =
P_i - L_i n_i 
\label{ctygeneral}
\end{eqnarray}
where $\phi_i$ is the flux of a particular species, and t is time.
The species number density is given by $n_i$, $P_i$ is the chemical production
rate and $L_i$ is the loss frequency at altitude $z$ and time $t$ \cite[see]{chamberlain_theory_1987}.
  
The solution of Equation \ref{ctygeneral} yields the distribution of the species that are being studied.  This solution is obtained by considering the various photochemical and chemical production and loss terms in addition to the effects of composition, eddy diffusion, temperature, mixing ratio and the solar flux on the various constituents distribution.  This method of solution is described for one dimension in the sections that follow.

\paragraph{1-D General Method of Solution}

The vertical distribution of a minor constituent in a planetary atmosphere is governed by the 1-dimensional 
continuity equation for each species, $i$,
\begin{eqnarray}\frac{\partial n_i}{\partial t} + \frac{\partial \phi_i}{\partial z} =P_i - L_i n_i 
\label{cty}
\end{eqnarray}
where the vertical flux, $\phi_i$, can be approximated by
\begin{equation}
\phi_i = \phi^K_i + \phi^D_i .
\end{equation}
The eddy flux, $\phi^K_i$,
\begin{equation}
\phi^K_i = -K(\frac{\partial n_i}{\partial z} + (\frac{1}{ H_{av}} + \frac{1}{T}
\frac{\partial T}{ \partial z})n_i)
\label{K}
\end{equation}
represents the vertical flux that parameterizes macroscopic
motions, such as the large scale circulation and gravity waves, and $\phi^D_i$
\begin{equation}
\phi^D_i = -D_i(\frac{\partial n_i }{ \partial z} + \frac{(1+\alpha_i)}{ T}
\frac{\partial T }{ \partial z} + \frac{n_i}{H_i})
\label{D}
\end{equation}
is the vertical flux carried by molecular diffusion. The species
number density is given by $n_i$, $P_i$ is the chemical production
rate (cm$^{-3}$ s$^{-1}$) and $L_i$ is the loss frequency (s$^{-1}$) at  
altitude $z$ and time $t$ (e.g. \citeA{chamberlain_theory_1987}). 
$D_i$ and $K$ = $K(z)$ are, respectively, the molecular and vertical eddy diffusion 
coefficients. 
The molecular diffusion coefficients, $D_i$, are taken from \citeA{MasonMarerro1970,Cravens1987} 
where applicable using the formula
$D_i = \frac{b_i}{n_{bg}} = \frac{A T^{s}}{n_{bg}}$ where $b$ 
is the binary collision parameter (expressed in terms of the coefficients A and s)
and the subscript `\textit{bg}' denotes background.
$H_i$ and $H_{av}$ are respectively the constituent and background atmospheric pressure scale heights, i.e.,
$H_i = \frac{kT}{M_i g}$ and $H_{av} = \frac{kT}{M_{av}g}$
where $M_i$ and $M_{av}$ are respectively the molecular weights of the constituent and the atmosphere.
In these calculations we have neglected the effects of the thermal diffusion factor, 
$\alpha_i$, as its inclusion contributed less than 1\% to a given species 
column in test runs.

Eddy mixing tends to homogenize the atmosphere such that, where there are no effects due to chemistry,
all species would be distributed according to the mean atmospheric pressure scale height.  Molecular diffusion
tends to separate constituents by their individual molecular weights.  The atmospheric level at which the molecular
diffusion coefficient is equal to the eddy diffusion coefficient is defined as the homopause for the
$i^{th}$ constituent.  Above this altitude, molecular diffusion dominates and the time constant for
reaching diffusive equilibrium is given by $\tau_D = \frac{H^{2}_{av}}{D_i}$ \cite{ChapmanCowling1970,Colegrove1966}.
Below the homopause, eddy diffusion dominates and the long lived species are ``mixed'',
and the mixing time constant is analogously expressed as $\tau_K = \frac{H^{2}_{av}}{K}$.

Equation~(\ref{cty}) is solved using a finite central difference
approximation for the vertical derivatives and the species densities
are solved semi-implicitly in time using a simple tridiagonal solver. For
these applications we have assumed a steady state exists and so have
driven the solution so that $\frac{1}{P}\frac{\partial n_i }{\partial t}
\rightarrow 0$.
Examples of such models and details are given in \cite{Parkinson2002,YungDemore1982} 


\paragraph{Eddy Diffusion Coefficient, $K(z)$}

One of the fundamental properties of a planetary atmosphere is the
amount of mechanical mixing forced by large scale circulation,
gravity waves and other processes.  In a one-dimensional model, this
mixing is often characterized by the eddy diffusion coefficient, 
which we will denote by $K$, $K_z$ or $K(z)$.  
The value of $K(z)$ in the vicinity of the homopause, $K_h$, is
critical in determining the onset of the importance of molecular diffusion.  
Estimates of $K_h$ for the outer planets have been obtained by various means:
e.g., analyses of the H Lyman-$\alpha$ albedo (Wallace and Hunten, 1973; 
Atreya, 1982; Ben Jaffel et al., 1993; Ben Jaffel et al., 1994),
the fall-off in hydrocarbon profiles, as measured against an H$_2$
background, using solar and stellar occultation data \cite{Atreya1981,Festou1982,ROMANI1993}, the He 584 \AA\ albedo \cite{mcconnell1981voyager,sandel1982,VERVACK1995,PARKINSON1998} and the CH$_4$ fluorescence  \cite{drossart1999fluorescence}.


\paragraph{Thermospheric-ionospheric simulations}

The modeling of a thermosphere-ionosphere is slightly different than the deeper layers of the atmosphere: a density profile has to be taken into account for each different neutral species, since they follow their own scale height. Supra-thermal species can exist, such O in the upper atmosphere of Mars, resulting from O$_2^+$ dissociation. For the ionized species, a different temperature has to be computed (and it changes with the species in the most complicated simulations). Finally electron temperatures have to be addressed.
The full description of these models is outside the scope of this paper. We refer the reader to the following studies and their included references \citeA{Johnstone2018,Bougher:2005}.

\paragraph{Importance of the 3-D modeling}

Three dimensional models (3-D) models provide a broad characterization of the whole atmosphere that couple chemistry, dynamics, and energy balance. These numerical tools, while not capable of including the details of their one-dimensional (1-D) counterparts, can capture the effects of global dynamics, diurnal chemistry, and the resulting energy balance. It has been shown that the approximations made with 1-D modeling are not able to fully reflect the reality of a planetary climate. For example, the presence of clouds, ice sheets, oceans, etc. have large effect able to change a non-habitable planet into one \cite{way2016venus,way2018climates}.
For thermospheres-ionospheres, the 3-D effects of transport and cooling lead to different results as well, which may change our view of an exoplanet. 

\subsubsection{Exospheric temperature}
\label{te}

The exospheric temperature, $T_{exo}$ is one of the most important parameters in the study of non-hydrodynamic atmospheric escape. It is the temperature at the base of the exosphere. Its effects on atmospheric escape are numerous. First, a higher $T_{exo}$ means a higher thermal escape. Secondly, with a warmer thermosphere, the exobase increases with altitude thereby increasing the exobase surface. This in turn implies a higher total escape from the planet and a greater cross section to non-thermal escape. Thirdly, a high $T_{exo}$ means that non-thermal processes can be more efficient.

The exospheric temperature depends upon (a) the UV flux (photon heating), (b) the chemical heating, (c) the electromagnetic energy (Joule heating), and (d) precipitation (auroral heating) in the atmosphere. The rate of cooling, primarily by infrared radiation, depends upon the composition and the adiabatic expansion. The equilibrium between the heating and cooling factors gives the temperature. Since wind and UV heating are important factors, major dayside-nightside exospheric temperature differences can occur. Full 3-D models such as the Global Ionosphere-Thermosphere Model (GITM) \cite{RidleyGITM:2006} are therefore necessary to obtain a correct value for the exospheric temperature. A 1-D approximation of the temperature can be made, but, in the case of the study of non-thermal escape, it may become a major problem. This is because the day-night asymmetry from the escape processes is correlated with the asymmetry from the exospheric temperature, which could lead to severe errors in the determination of the magnitude of the escape.

The exospheric temperature is determined by the equilibrium between heating and cooling. Since these processes are altitude dependent, it is often necessary to determine the structure of the thermosphere and compute the exospheric temperature from the basic equations. Empirical models exist, for example, for Earth \cite{Weimer:2011}.
In the planets of the Solar system, heating is dominated by (1) photoexcitation and cooling by (2) thermal conduction \cite{galindo2009}. The photoexcitation/photodissociation heating is due to the kinetic energy left in these processes: the difference between the threshold $E_{t,k}$ of the $k^{th}$ reaction on a species, $s$, and the energy, $E$, of the photon is transformed into heat. When the flux of photon per unit energy is $\Phi(E)$ we have:
\begin{eqnarray}
  Q_{UV,k} = \int_{E_{t,k}}^\infty (E - E_{t,k}) n_k \sigma_k(E) \Phi(E) dE
\end{eqnarray}
The thermal conduction is solved through the following equation \cite{galindo2009}:
\begin{eqnarray}
  \frac{\partial T}{\partial t} &=& \frac{1}{\rho c_p} \frac{\partial \left(k \frac{\partial T}{\partial z}  \right)}{\partial z}\\
  k &=& A T^{0.69}
\end{eqnarray}
With $\rho$ being the density (kg/m$^3$), $c_p$ the heat capacity, and A the weighted average of the thermal conductivities. 

Two major parameters are to be carefully determined when estimating the exospheric temperature, and are the most complicated to address to date: (3) the chemical heating/cooling, (4) the radiative cooling.

The chemical heating, due to the exothermic reactions, and cooling, due to the endothermic ractions, follow the ionization and dissociation by precipitating particles (including photons). Evaluating this contribution requires to carefully evaluate the chemical reactions chains and their energies. Those are atmospheric-composition dependent and can be quite complex and not well understood (e.g. Titan).

The radiative cooling is mainly due to the de-excitation of molecular species in a rotational or vibrational state. Simple approximations of that cooling can be made if the cooling species is in low quantity in the atmosphere and if it is excited only by thermal processes; i.e. if it is in a local thermodynamic equilibrium (LTE) and if the emission line (or band) is optically thin.
More complex cases exist in the atmospheres (such as non-LTE processes that are known to happen in auroral regions and optically thick cases), that require precise radiative transfer calculations \cite{mertens_new_2008,mertens_kinetic_2009}. In addition, very complex cases such as state inversion and MASER can be obtained, such as those occurring at Mars and Venus at 10 $\mu$m \cite{Mumma1993} and probably at some exoplanets \cite{cosmovici2018water}. Finally, some of the radiative species can be obtained by chemical reactions when the system is out of equilibrium, e.g. NO cooling at Earth \cite{Weimer:2015}.
For the extrapolation of Solar system planets' situation to other stellar systems, it is important to validate such approximations .

Other important parameters have to be considered depending on the cases studies: (5) NIR heating, important in the case of CO$_2$-rich planets, (6) dynamic cooling -from winds or expansion-, and (7) heating from gravity waves dissipation \cite{hargreaves1992solar}. %

\subsubsection{The exobase altitude}

The exobase is the altitude at which the scale height is equal to the mean free path of a thermalized particle (at $T_{exo}$). Above this altitude, the mean free path is greater than the scale height, and a particle with sufficient energy is likely to escape without any collision.

One can approximate the density in the thermosphere by $n(z) = n_o \times e^{\left(- \frac{z - z_o}{H}\right)}$ (nb: this is valid for a thermosphere with one constituent; if multi-constituent a H will have to be defined for each of those, but the exobase is usually defined for the main constituent). At the exobase, we have $n_{exo} = \frac{H}{\sigma}$ with $\sigma$ being the collision cross section between the main molecules. If we suppose an isothermal thermosphere, i.e. H does not vary with altitude, it is possible to easily retrieve the exobase altitude: $z_{exo} = z_0 - H ln(\frac{n_{exo}}{n_0})$.
For multi-component atmospheres and varying temperature, the evaluation becomes more complex since H and $\sigma$ (and therefore $n_{exo}$) vary with altitude.

\subsection{Time dependence and creation of observable markers}

Once the main processes leading to atmospheric escape are known, the study of their influence in time requires evaluating the evolution of the stellar forcing parameters. If possible, the study of the isotopic ratio in the planetary atmosphere will be a major input for validating the calculations and estimating the influence of other processes such as outgassing, etc.

\subsubsection{Evolution in time of the stellar forcing parameters}
\label{s:stellar_forcing}

Stellar rotation drives the magnetic activity responsible for UV to X-ray emission from Sun-like stars through a dynamo mechanism thought to be seated near the bottom of the stellar convection zone. In turn, this magnetic activity influences rotation itself through angular momentum loss to a magnetized wind that leads to a gradual slow down of the rate of spin.

Stars are born with a natural spread in their rotation periods and these initially evolve quite rapidly with time due to changes in moment of inertia as stars contract on to the main sequence.  This initial rotational evolution then involves {\em spin up}, rather than spin down. All newly-formed Sun-like stars are thought to possess a residual disk of gas, called a protoplanetary disk, within which planets form.
In the early, so called ``T Tauri", phase of evolution (named after the representative prototype) lasting a few million years, the protoplanetary disk is expected to prevent them from spin up through a mechanism known as disk-locking \cite{Rebull2002,Rebull2004}. While the detailed physics behind this is still poorly understood, the underpinning of the idea is that there is angular momentum exchange between the star and the disk modulated by magnetic fields that connect them---in essence, the disk applies a magnetic brake. After anything from a few Myr up to 10 Myr, the disk gets dispersed and stars then freely spin-up as a consequence of contraction. Once on the main-sequence, contraction has stopped and magnetic braking through the stellar wind results in an efficient spin-down process.

\begin{figure}
      \noindent\includegraphics[width=20pc]{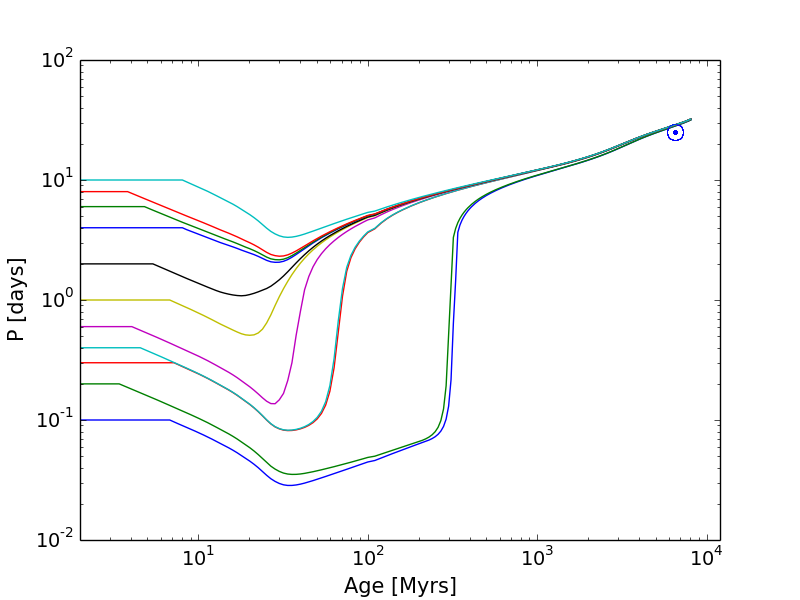}
\caption{Evolution of the rotation period of a $1 \, M_{\odot}$ star as a function of age for different initial periods during the disk-locked T~Tauri phase based on the rotation evolution model of \citeA{Garraffo.etal:18}.}
\label{rotation}
\end{figure}

Magnetic braking is determined by the magnetic fields on their surfaces \cite{WeberDavis1967, Kawaler:88}.  This self-regulating mechanism results in the rotation period evolving with time following the Skumanich law for spin-down $P_{rot} \propto t^{1/2}$ \cite{Skumanich1972}. This is the foundation of Gyrochronology \cite{Meibom.etal:15}, a very powerful tool that enables the conversion of rotation periods into stellar ages. 
Studies of the rotation periods of stars in young open clusters have revealed a bimodal distribution, recently attributed to different magnetic evolutionary paths of stars with different initial rotation periods \cite{Garraffo.etal:18}.  Stars that start off spinning faster will have smaller Rossby numbers, and this is expected to result in a more complex geometry of the surface magnetic fields.  This, in turn, has the effect of closing otherwise open field lines, preventing the stellar wind to escape removing angular momentum. As a consequence, stars with short initial rotation periods will remain rotating fast for longer than their initial slow rotators counterparts (see Figure \ref{rotation} for an illustration of the effect of different initial periods in the spin evolution of a $1\, M_{\odot}$ star). The period of time for which the initially fast rotators will remain rotating fast is larger the lower the stellar mass is. Eventually, at an age that depends on the stellar mass ($\sim 600$ Myrs for solar mass stars), initial conditions have been erased and all stars follow the Skumanich law, making Gyrochronology fairly reliable. However, the activity history of these stars can be quite different depending on their initial rotation history, and that can potentially make a difference in the survivability of their planets' atmospheres and habitability.

\begin{figure}
      \noindent\includegraphics[width=20pc]{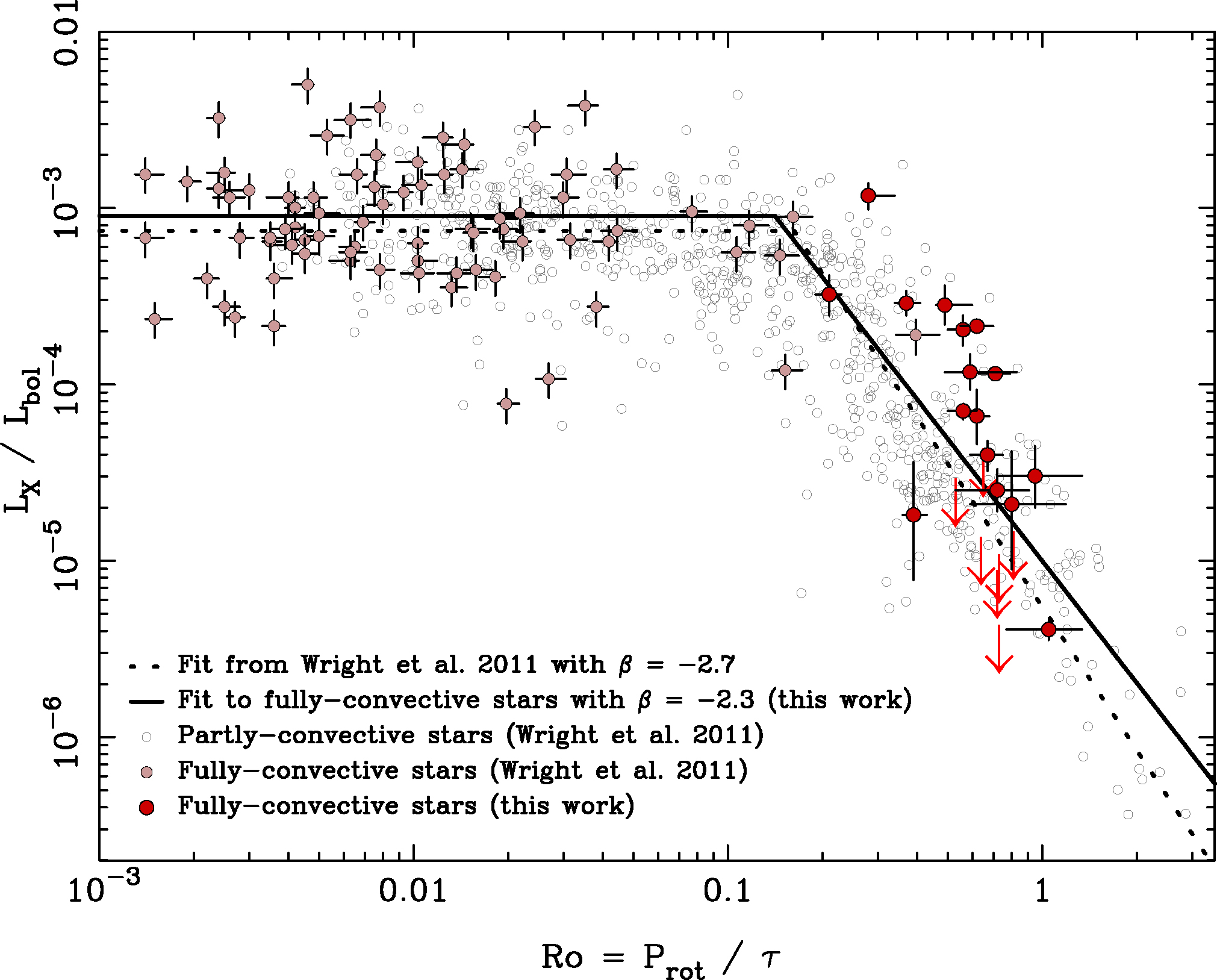}
\caption{X-ray to bolometric luminosity ratio, $L_X/L_{bol}$, as a function of the Rossby number, $R_o = P_{rot}/\tau$, for both partly convective and fully convective stars. The best-fitting activity–rotation relations found for fully convective stars by \citeA{Wright.etal:18} ($\beta = -2.3$ and $Ro_{sat} = 0.14$, solid line) and from \citeA{Wright2011} ($\beta = -2.7$ and $Ro_{sat} = 0.16$, dotted line) are shown.  From \citeA{Wright.etal:18}; see text for details.}
\label{f:lxlbolvsrossby}
\end{figure}

The establishment of rotation (or more correctly, differential rotation) as the driver for the magnetic dynamo activity that gives rise to UV, EUV and X-ray emission that drive planetary atmospheric ionization and loss processes can be traced back to the 1960s when it was noticed that Ca~II H \& K emission fluxes of stars declined linearly with stellar rotation velocity. 
The magnetic nature of stellar coronae was essentially established a decade later by the {\it Einstein} observatory and the realization that X-ray luminosity was highly correlated with stellar rotation \cite{Vaiana.etal:81,Pallavicini.etal:81,Walter.Bowyer:81}. Some fraction of the magnetic energy created within the star by dynamo action and subject to buoyant rise is dissipated at the stellar surface and converted into particle acceleration and plasma heating.
Although none of these processes are fully understood, the dependence of activity diagnostics and stellar UV and X-ray fluxes on rotation shows a very simple empirical relation in terms of a magnetic ``Rossby" number  illustrated in Figure~\ref{f:lxlbolvsrossby}. The Rossby number in this case is the ratio of the rotation period and convective turnover time near the base of the convection zone, $Ro=P_{rot}/\tau_{conv}$ (see also \citeA{Noyes.etal:84}).

Figure~\ref{f:lxlbolvsrossby} shows stellar X-ray luminosities normalized to the total stellar bolometric output, $L_X/L_{bol}$, as a function of the Rossby number for late-type stars ranging from spectral type F down to mid-M, including fully-convective M dwarfs. At slower rotation rates, $L_X/L_{bol} \propto Ro^\beta$, where \citeA{Wright.etal:18} find $\beta = -2.3$, up until a threshold at which point X-ray emission saturates, $L_X/L_{bol} \sim 10^{-3}$, close to a Rossby number $Ro=0.13$. This saturation behavior was already apparent from data obtained by the {\it Einstein} observatory \cite{Vilhu:84,Micela.etal:85}, although its origin is still debated.  It is likely that it represents saturation of the dynamo itself (see, e.g., the discussion in  \citeA{Wright2011} and \citeA{Blackman.Thomas:15}). The rotation period at which saturation sets in increases for decreasing stellar mass. For a solar mass main-sequence star,  X-ray emission saturates at a $\sim 1.25$ days period, while it can be more than $100$ days for an early M dwarf. \cite{Wright2011}.  This means that lower mass stars are expected to be saturated, and therefore comparatively more active and UV and X-ray bright, than higher mass stars for much longer.   

\begin{figure}
      \noindent\includegraphics[width=20pc]{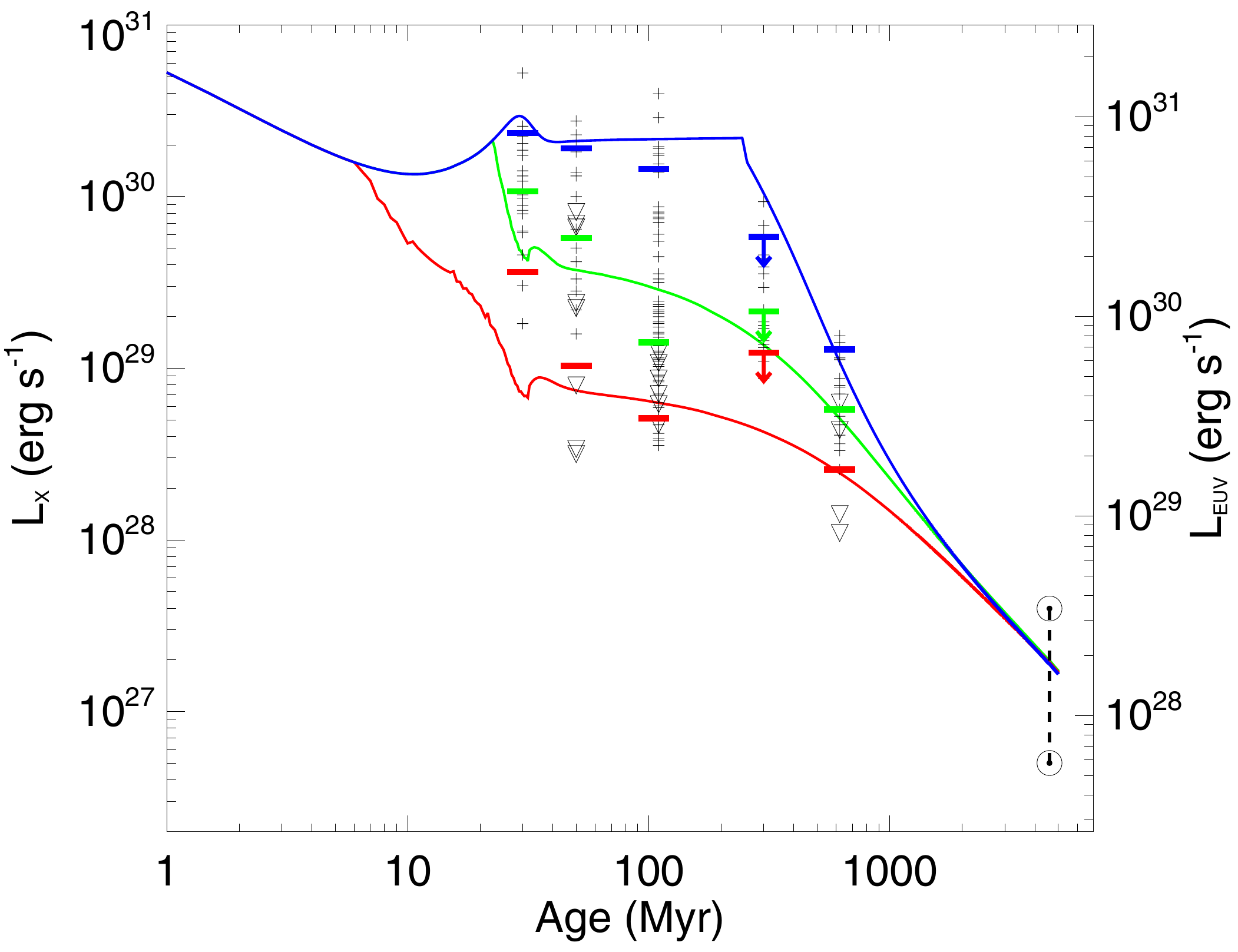}
\caption{The X-ray and EUV luminosities, $L_X$ and $L_{EUV}$, for a solar mass star as a function of time. Shown are the luminosity trajectories for three different rotation evolution tracks, together with observed
X-ray luminosities for single stars in open clusters. Upper limits are indicated by inverted triangles. Solid horizontal lines indicate 10th, 50th, and 90th percentiles of the observed distributions of $L_X$ at each age calculated by counting upper limits as detections. Two solar symbols at 4.5 Gyr
show the range in $L_X$ for the Sun over the solar cycle. From \citeA{tu2015extreme}.
}
\label{f:lxvstime}
\end{figure}

The X-ray luminosities and cumulative X-ray doses for a $1M_\odot$ star as a function of age for the different rotation histories shown in Figure~\ref{rotation} based on the rotation-activity relations of \citeA{Wright.etal:18} are illustrated in Figure~\ref{f:lxvstime}. 

\subsubsection{Isotopic Fractionations and Retrieving the History of Planetary Atmospheres}
The study of isotopes is the major tool to study the history of planetary systems. In planetary atmospheres, it allows to create an history of the escape. Unfortunately, it is usually a ill-posed problem, and hypothesis are required, such as an atmosphere with basically the same composition over eons, and it allows to retrieve the fraction lost to space, without any other data about the fraction lost to e.g. surface; which prevents one to have a good idea of the surface pressure in the past as explained in \citeA{brain2016atmospheric}; the situation is complicated once surface processes are able to perform isotopic fractionation \cite{parai2018xenon}. Once productions/outgassing are taken into account, it is possible to reach a steady state for isotopic fractionations, which complicates the interpretation \cite{mandt_isotopic_2009}.

\paragraph{Theory for hydrodynamic fractionation} we have seen in Section \ref{hydrodynamicfractionation} how to retrieve the differential flux of a species $n_b$ dragged by a species $n_a$ in hydrodynamic escape. Considering $N_b$, the total content of the species $b$ in the atmosphere, it is possible to estimate the variation of $N_b$ in function of the history of the hydrodynamic escape of the atmospheric species $a$.
Following \citeA{pepin1992origin}, we assume no replenishment and that the escaping flux of $a$ in function of time, $F_a(t)$,  follows $F_a(t) = F_a^0 f(t)$, with $f$ being a decreasing function with time. 
In that case, we find that:
\begin{eqnarray}
  \frac{dN_b}{N_b} = -\frac{F_a^0}{N_a} \left[f(t) - \frac{m_b-m_a}{m_c^0 - m_a}  \right]dt
\end{eqnarray}
Solving this equation allows to evaluate the total amount of $b$ that escaped. The most interesting conclusion that we can make from this equation, while having no knowledge of $f(t)$, it that the escape of $N_b$ stops at the time $t_2$ such that $f(t_2)=\frac{m_b-m_a}{m_c^0 - m_a}$.

\paragraph{Theory for Jeans/non-thermal fractionation} 
\label{gravitationalfractionation}
When discussing fractionation with respect to an escaping atmosphere, the Rayleigh fractionation/distillation law and its notation are often used \cite{jakosky_mars_1994,johnson_sputtering_2000,mandt_isotopic_2009}. In this nomenclature, R, the ratio of two species --often isotopes--, is the main parameter.  First, it is important to not make a confusion between the observed ratio of species b and a, $R(z) = n_b(z) / n_a(z)$, and the total ratio  $R = N_b(z) / N_a(z)\approx N_b(z_i) / N_a(z_i)|z_i <z_{homopause}$. The observed ratio is fractionated in altitude above the homopause. 
The main hypothesis of the Rayleigh distillation law is that any species $i$ is lost proportionally to its total amount: $dN_i= k_i N_i$.
Therefore we have:
\begin{eqnarray}
  \frac{dN_b}{dN_a} = \frac{k_b}{k_a}\frac{N_b}{N_a} = f \frac{N_b}{N_a}\\
  \frac{dN_b}{N_b} = f \frac{dN_a}{N_a} \Rightarrow ln\left(\frac{N_b}{N_b^0}\right) = f ln\left(\frac{N_a}{N_a^0}\right)\\
  \left(\frac{N_b/ N_a}{N_b^0/N_a^0}\right) = \left(\frac{N_a^0}{N_a}\right)^{(1-f)}\\
  \frac{N_a^0}{N_a} = \left(\frac{R}{R^0}\right)^{\frac{1}{(1-f)}}\label{rayleighdistillation}
\end{eqnarray}

The Rayleigh distillation equation \ref{rayleighdistillation} allows to evaluate the total loss of a species through escape provided the fractionation factor $f$ and the initial isotopic ratio $R^0$ and the current one $R$. If one considers an escape flux $F$ proportional to $N$, it is easy to see that $k=F/N$, and therefore that $f_e= \frac{F_b}{F_a} \times \frac{1}{R}$.
Using equation \ref{eq:jeansesc3}, it appears that the fractionation factor for Jeans escape is:
\begin{eqnarray}
  f_{Jeans} = \sqrt{\frac{m_a}{m_b}}\frac{1+\lambda_{ex,b}}{1+\lambda_{ex,a}} e^{\lambda_{ex,a}-\lambda_{ex,b}}
\end{eqnarray}


\paragraph{Outstanding problems related to fractionation}
The observation and modeling of fractionation highlighted major events in the evolution of planetary atmospheres. The D/H ratio observed at Venus, $>\sim$ 1.6 $\times$ 10$^{-2}$ \cite{DonahueEtal1982,2018SSRv..214...10M}, suggests that the loss of water at Venus may be consistent with the loss of a Earth's ocean \cite{Shizgal1996}. 

The enrichment of $^{15}$N over $^{14}$N at Mars can be explained by non-thermal processes \cite{Shizgal1996}. Since Ar does not reacts chemically, \citeA{Jakosky2017} used the $^{38}$Ar/$^{36}$Ar ratio to determine that Mars lost 66\% of its atmosphere to space. It is to be noted that the location and timing at which an isotopic ratio is measured can have an effect: \citeA{LIVENGOOD2020113387} has shown that surface adsorption at Mars performs isotopic fractionation that can be highlighted by daily variations (temperature changes the amount of gas adsorbed).

\label{xenon}
At Earth, the fractionation of noble gases have been explained by hydrodynamic escape processes \cite{Shizgal1996}, except for xenon.
Xenon is depleted by one order of magnitude relative to other noble gases and other volatile elements when normalized to the chondritic composition (e.g.\citeA{Marty2012}) and is largely enriched in its heavy isotopes relatively to Solar or chondritic xenon. 
This peculiarity of xenon compared to other noble gases is known as the ``xenon paradox''.
The specific electronic structure of xenon which makes it the most reactive element among noble gases with the lowest ionization potential (12.13 eV or 102.23 nm) and an extended photoabsorption cross section covering part of the VUV spectrum (up to about 150 nm). From this consideration and because of the difficulty to explain Xe depletion and fractionation with other mechanisms, the escape of Xe$^{+}$ driven by H$^{+}$ ion escape is considered as a plausible explanation of the xenon paradox  \cite{ZAHNLE201956}. On the other hand, \citeA{HebrardandMarty2014} proposed a scenario combining the trapping of heavy xenon isotopes in haze with an efficient escape of Xe$^{+}$ ions which is both consistent with the Xenon depletion and fractionation. 
The time history of Xe isotope fractionation has been investigated in detail by \citeA{Avice2018}, and showed that it started evolving at least 3.5 Ga ago until it reached the modern-like atmospheric Xe composition at around 2.1 Ga ago. They concluded that termination of the isotopic fractionation of Xe may coincide with the end of the hydrogen escape which has previously been suggested to explain the progressive oxygenation of the Earth's atmosphere \cite{Zahnle2013}. However, such significant escape of Xe$^{+}$ ions with no associated loss of other noble gases is challenging due to the large mass of Xe$^{+}$ and its associated large gravitational binding energy ($\sim$85~eV).
The work of \citeA{parai2018xenon} shows that xenon can be trapped in the crust from the oceans, but is basing its atmospheric fractionation on the work of \cite{pepin_origin_1991} who does not consider that xenon can be trapped back into the Earth. It may be that the xenon paradox could be solved not by escape physics, but by crustal absorption. This is in agreement with the organic haze scavenging of Xe hypothesis, investigated in \citeA{Avice2018}, that solves the paradox without requiring atmospheric escape. An interesting point is the requirement for atmospheric hazes in that hypothesis: specific atmospheric conditions, similar to the present Titan \cite{McKay2001}, are required to create those, and therefore Xe isotopic fractionation may be an indicator of the atmospheric conditions of the Archaean Earth. Such a solution has the potential to reconcile the conclusions of \citeA{parai2018xenon}, i.e. no plate tectonics before 2.5~Gyr ago or extremely dry tectonics, with the observation of ancient plate tectonics with excess water before 3.3~Gyr by \citeA{sobolev2019deep}.

\section{Escape at Solar System's Planets and Bodies}

The atmospheric escape in the Solar system shaped most of the planets and dwarf planets' atmospheres, as well as those of some satellites. Mercury may have had a proto-atmosphere above its magma ocean, like the Moon \cite{greenwood2018water}, but, unlike the Moon, no sample from the surface are available and a detailed study of the history of Mercury's atmosphere lacks too much experimental evidence. Mars, Venus, the Earth, and Titan have on the contrary a large quantity of data showing an atmosphere that has been transformed by escape. Currently H and He are the most important species escaping for these objects. We are also observing other escape processes, such as O, and we try to understand the pathways of escape of CO$_2$ at Mars, so we can understand the evolution of its atmosphere in time. Non-thermal escape processes at Jupiter and Saturn are known to fill a part of their plasmaspheres and leads to some minor escape. Overall, the giant planets are too big for efficient escape to take part and change drastically their atmospheric evolution. Uranus and Neptune are similar in that their mass prevents a lot of escape. In addition, the ice giants have only been visited by the Voyager probes and have not had a Galileo or Cassini-like mission allowing study of their atmospheres as comprehensively as Jupiter or Saturn. A recent work by \citeA{DiBraccio2019} has shown that, as for Jupiter and Saturn, plasmoids have been observed at Uranus. However, a large quantity of these plasmoids' loss can come from the satellite of the giant planets. Finally, relatively small loss are suspected to come from polar wind \cite{Glocer2007}. 

Understanding the current escape processes allows to perform some interpolation back in time, thanks to a better understanding of the conditions (especially the solar forcing in time), and of the important parameters for each escape process. Such work necessary to understand how the atmosphere could evolve into an habitable one. To have a control point to these interpolations, it is necessary to know about the isotope ratio. Unfortunately, surface processes (volcanism, adsorption) and other events such as comet falls or cosmic ray spallation processes [e.g. \citeA{pavlov2014}] can affect these results and possible concurrent models leads to the currently observed state.

\subsection{The Solar forcing in time}
\label{solarforcingintime}


The solar magnetic activity forcing changed substantially over the lifetime of the Sun, as discussed in the general stellar context in Section \ref{s:stellar_forcing}. In its early stages, the rotation of the young Sun was faster than its present rotation, with rotation periods of only few days compared to its present 27 days period.  During this time, the Sun's ratio of X-ray to bolometric luminosity declined by a factor of about 1000. Thus, in general, the most important aspect of the evolution of solar forcing is that the EUV and X-ray fluxes were much higher during the early Solar system than today, with related consequences for earlier planetary atmospheric escape rates. This time evolution is shown in Figure~\ref{f:lxvstime}.

In addition to the solar radiation, the solar wind also plays a role in planetary atmospheric escape. However, the change of the solar wind in time is much less well-defined than the solar radiation, since it is almost impossible to measure the signatures of weak winds of solar analogs. In general, the magnetic activity paradigm suggests that a more active Sun should produce a stronger solar wind. However, no clear evidence of that assumption has been discovered so far. Scaling laws have been developed based on observations of the neutral Hydrogen absorption line generated at the edge of stellar astrospheres, where the stellar wind collides with the Inter Stellar Medium (ISM) \cite{Wood2006,Wood2014}. Some modeling work was done to characterize the winds of Sun-like stars (e.g., \citeA{Cohen2010,CohenDrake2014}). In both cases, the winds of young stars were not found to be dramatically stronger than older stars, and many of the observed systems were found to deviate from the scaling low and present weaker winds than expected.  

Understanding the young Sun is important in the context of solving the Faint Young Sun Paradox (Section \ref{FYSparadox}). However, it is important to note that if the hypothesis on the atmosphere composition and pressure is off, no conclusion can be made only from the star parameters.




\subsection{Coupling with the world below}

The escaping region of an atmosphere does not exist in isolation. Once the primordial atmosphere of a telluric planet has eroded, the hydrogen and other light elements that escape ultimately come from the interior of the planet and pass through the lower/middle atmosphere to reach the region of escape. Just to understand hydrogen escape, it is crucial to understand the basic processes that control: (1) the flux of hydrogen-bearing species (e.g. water, methane, H$_2$) from their source regions at the surface throughout the atmosphere; and (2) the flux of hydrogen-bearing species to the atmosphere. 

Accurate modeling of hydrogen-bearing species are particularly important for understanding potentially habitable exoplanets. First, methane is a potential biosignature. Second, the chemistry of these species is often connected with the chemistry of O$_2$, O$_3$, CO$_2$, and CO through the HO$_x$ reactions in the atmosphere and through analogous reactions in the interior (e.g., \citeA{Kasting1993redox}). Third, sufficient hydrogen escape can modify the redox state of the atmosphere, surface and interior (e.g. \citeA{Kasting1993redox,CatlingClaire2005}).

\subsubsection{The Present: A Focus on Hydrogen-Bearing Species}

The vertical temperature structure of the atmosphere is a critical control on the ability of condensible species to move upward from the warm troposphere to where it can escape. The stratospheric water trap on present-day Earth is the classical example of such a control. Transport across such barriers may be accomplished diffusively or dynamically by means of atmospheric moist convection. The effectiveness of the cold trap for water may depend on the presence of other hydrogen-bearing species, such as methane.

\paragraph{Earth}

On present day Earth, the dominant sources of hydrogen-bearing species are evaporation from the Earth's oceans (H$_2$O) and anthropogenic sources of methane. Non-anthropogenic, biogenic sources of methane remain significant and probably greatly exceed geological sources \cite{Dlugokencky2011}, though \cite{EtiopeKlusman2002} argue that natural geological sources may be currently accounted to anthropogenic emissions in error. Atmospheric water poorly mixes into the middle atmosphere. There is a strong contrast between water vapor mixing ratios typical of the troposphere ($\approx$ 1000 ppmm) and water vapor mixing ratios near the Earth's mesopause ($\approx$ 5 ppmm), where the photodissociation of water by solar radiation at Lyman $\alpha$ wavelengths takes place
\cite{GaTechPhdThesisRoell12}. A parallel contrast exists between water vapor concentrations near the surface and in the upper troposphere ($\approx$10000 ppmm vs. $\approx$ 100 ppmm in the tropics: \cite{1993JAtS...50.1643S}). The region of rapid fall-off in water vapor mixing ratio is known as the hygropause. 

The contrast in humidity between the troposphere and the mesosphere results from the large-scale temperature structure of the atmosphere, in which the atmospheric temperature minimum is at the tropopause and lower stratosphere. Any excess water beyond the point of saturation will condense to liquid and ice, which may precipitate. Thus, moist air is freeze-dried to the equilibrium water vapor concentration at the ambient temperature. At the temperature minimum of the tropopause and the lower stratosphere, an ``atmospheric cold trap'' forms. The contrast in humidity between the surface and the upper troposphere partly arises from the same mechanism. Therefore, relatively slow vertical mixing of water vapor by large-scale processes such as the Hadley cell or synoptic-scale systems will be set by the vertical thermal structure of the atmosphere, which radiative-convective models can estimate approximately \cite{Pavlov2000}.

Mesoscale processes also have some impact on water vapor transport. Strong vertical motions in buoyant moist convection can transport ice to higher altitudes, evading ``cold trap'' effects. In some cases, moist convection can ``overshoot'' the tropopause, injecting large amounts of water ice into the stratosphere over an areally limited region. If this water ice sublimates in the stratosphere, the stratosphere is hydrated locally by the same order as the background stratospheric water vapor concentration \cite{2007ACP.....7.4977G,2010ACP....10.8267L}. If overshooting moist convection were more intense and/or more efficient at transporting water ice to the stratosphere, the contrast between tropospheric and mesospheric water vapor concentrations could be reduced. (We assume that the increase in water vapor due to overshooting convection is greater than the decrease in water vapor due to mixing resulting from downdrafts of overshooting convection.) 

Methane does not condense at atmospheric temperatures, which reduces the surface--mesosphere contrast considerably (1.8 ppm vs. 0.1 ppm) \cite{Summers1997}. It is slowly dissociated in the stratosphere and mesosphere, so its composition in the upper atmosphere will be controlled by the relative balance between chemistry and vertical transport in the stratosphere and mesosphere as well as the intensity of stratospheric-tropospheric exchange. 

\paragraph{Mars}

On present day Mars, the vertical structure of water vapor differs greatly from that of the Earth. The main source and sink of water vapor is sublimation from and condensation on the polar caps. A seasonally varying hygropause is apparent at a characteristic height of 40 km above the surface in the tropics \cite{2017Icar..293..132T,HeavensEtAl2018}. Yet detached water vapor layers are frequently observed as high as 80--90 km above the surface \cite{2013Icar..223..942M}. This structure partly reflects differences in the atmospheric temperature structure. The tropical middle atmosphere is not separated from the lower atmosphere by a strong thermal inversion analogous to the stratosphere (except perhaps in global dust storms, when the entire atmosphere is effectively inflated by the heating of dust). Thus, to first order, the depth of the Hadley cell sets the hygropause height \cite{2002Natur.416..298R}.
  Detached water vapor layers originate from mesoscale transport processes, such as injection within dust plumes in Mars's well-known dust storms \cite{2011JGRE..116.1007H,2013Icar..223..942M,2013JGRE..118..746S,GRL:GRL52939,HeavensEtAl2018,FedorovaEtAl2018} or associated with topographically-driven circulations, with or without dust storm activity \cite{2002Natur.419..697R,2006GeoRL..3316201M,GRL:GRL52939,HeavensEtAl2018}. In regional and global dust storms, convective transport of water to the middle atmosphere within dusty air can be so strong that we cannot really speak of detached water vapor layers; the mean hygropause of the planet can rise to 80 km (a change mostly caused by ascent in the tropical hygropause) \cite{FedorovaEtAl2018,HeavensEtAl2018}.

While a variety of observations suggest the presence of atmospheric methane, sufficiently little is known about it to make discussion of its surface sources and transport to the upper atmosphere entirely speculative \cite{Formisano2004,Mumma2009,Webster2015methane}.

\paragraph{Venus}

On present day Venus, the main sources of water to the atmosphere are believed to be cometary and meteoritic impacts and volcanic outgassing in uncertain proportions \cite{TaylorGrinspoon2009}. The possibility of ongoing volcanic outgassing has been bolstered by observations of temporal and spatial variability in atmospheric SO$_2$ and transient NIR emission from a prominent rift zone \cite{Marcq2013,Shalygin2015}.

Estimated lower atmospheric (5--45 km) water vapor concentrations from spectroscopic observations range from 25--50 ppmv with typical uncertainties at the 20\% level \cite{Chamberlain2013}. There is an outlier estimate of 200 ppmv in the 30--45 km altitude range \cite{Bell1991}. While this value is consistent with some \textit{in situ} measurements by entry probes, the entry probe data is mutually inconsistent and generally mistrusted \cite{MeadowsCrisp1996}. Current observations are unable to probe water vapor concentrations within 5 km of the surface, but it is speculated that water vapor might be depleted near the surface because of reactions with surface rocks \cite{Fegley2003,Chamberlain2013}.

Water vapor concentrations above the troposphere are 3--11 ppmv near the top of the sulfuric acid cloud deck at 60--70 km altitude and likely decrease to ~1 ppmv at 100 km \cite{Fedorova2008,Fedorova2016}. Water vapor at these altitudes would be vulnerable to photochemical loss processes. Water vapor in the middle atmosphere is most abundant near the Equator, a phenomenon that suggests convective transport of water vapor from the lower atmosphere into the middle atmosphere \cite{Fedorova2016}. Otherwise, the water vapor distribution in the middle atmosphere appears quite sensitive to the altitude of the cloud deck, suggesting that the sulfuric acid cloud deck is an effective hygropause for Venus due to the formation of sulfuric acid from H$_2$O and SO$_2$. 

Venus's atmosphere does not contain measurable amounts of methane at present measurement sensitivities \cite{TaylorGrinspoon2009}. Early \textit{in situ} measurements by Pioneer Venus suggested atmospheric methane concentrations were up to 6000 ppmv, but these measurements likely were contaminated by reactions within the measurement apparatus itself \cite{DonahueandHodges:1993}. Yet some methane input from meteoritic and cometary sources is possible. If the mantle of Venus has remained sufficiently reducing, a source of methane from volcanic outgassing is possible as well.

\paragraph{Titan}



The principal hydrogen-bearing species in the present day atmosphere of Titan is CH$_4$. The total amount in the atmosphere as vapor exceeds the amount present on the surface as liquid by at least a factor of 2 \cite{Lorenz2008}. Most hydrogen-bearing species in Titan's atmosphere, such as H$_2$ and various organic compounds, are likely derived from photochemical reactions involving CH$_4$ \cite{OwenNiemann2009,Strobel2012,Krasnopolsky2014titan}, though an H$_2$ source derived from serpentinization also has been proposed (e.g. \citeA{Atreya2006titanmethane}). An exception is H$_2$O, which is likely supplied to Titan by ablation of micrometeorites and/or plume material from Enceladus \cite{Lara1996titan,Coustenis1998titan,Dobrijevic2014titan}. The ultimate source of methane on Titan is believed to be episodic outgassing from Titan's deep interior \cite{Lorenz1997,Tobie2009,Wong2015}. 

CH$_4$ concentrations near the surface are $\approx$ 50000 ppmv, decrease to $\approx$ 15000 ppmv in Titan's stratosphere (above 32 km) \cite{Niemann2005titan} and remains uniformly mixed at that up to altitudes near the homopause region $\approx$ 850 - 1000 km where diffusive separation causes the relative fraction of methane to increase with altitude up to the exobase \cite{Yelle2008,Johnson2010titan, Bell2014}. The major barrier to transport is an atmospheric cold trap occasionally broken by deep convective clouds of CH$_4$ \cite{Griffith2009titan}.


\subsubsection{The Past: Coupling, Unusual Escape Regimes, and Current Evidence for Atmospheric Mass and Upper Atmospheric Composition}

\paragraph{Earth}

In the course of the Earth's early history, the upper mantle was gradually oxidized by means of coupled pathways: (1) reductants in the upper mantle were emitted into the atmosphere by volcanic processes, were transported into the upper atmosphere by processes at mesoscale to planetary scales, and escaped the Earth system by a mixture of physical and chemical processes in the upper atmosphere, resulting in an unbalanced loss of reducing power from the upper mantle; and (2) weathered (hydrated) or even oxidized crust was recycled into the mantle, where water reacted with mantle material to form hydrogen, a light, readily escaping reductant \cite{1984ceao.book.....H}. This period of mantle oxidation closed when the mantle was sufficiently
oxidized that water-hydrogen conversion in the mantle ceased \cite{1993Icar..101..108K}.

These processes are mediated by escape itself. Water that is vertically transported into the upper atmosphere photodissociates, leading to the production of H$_2$ and O$_2$ abiotically \cite{1983JGR....88.4935L}. The H$_2$ is highly vulnerable to escape, transporting reducing power out of the system, while the O$_2$ may mix down into the lower atmosphere and oxidize the crust. Simultaneously, precipitation of atmospheric water
to the surface hydrates the crust.

Planets like present day Earth exchange water with the mantle in the course of plate tectonic processes such as subduction, which are relatively efficient. If Venus' surface were wetter, some exchange would take place during putative resurfacing events \cite{Strom1994}. Similar speculations on might be made about Mars' volcanic activity. However, crustal recycling on both Mars and Venus are thought to be much weaker than the Earth and, on average, weaker in the past \cite{TaylorMcLennan2009}.

However, there are signs from the extant record of early Earth history that plate tectonics may not be the upper limit for crustal recycling rates on the Earth. Instead, the Earth may have experienced a ``heat-pipe'' phase \cite{2013Natur.501..501M}. In this phase, persistent mafic to ultramafic volcanism regularly re-surfaced the Earth. Both crustal material and surface water are cycled back into the mantle through repeated eruption and burial of older flows. The heat pipes were associated with greater eruptive volumes of volcanic material as well as faster crustal recycling than plate tectonics. Observations of the other terrestrial bodies in the Solar system are also consistent with heat pipe operation in their early phases \cite{MooreSimonWebb:EPSL2017}.

Therefore, the upper mantle of the Earth began in a far more reduced state than today and may have degassed far more intensely than today.
An important consequence of the reduced state of the Earth's early mantle (without enhanced degassing) would have been higher proportional degassing of carbon from the interior in the form of CH$_4$ as opposed to CO$_2$. In addition, formation of H$_2$ from H$_2$O in the mantle would have
resulted in significant emission of H$_2$ to the atmosphere \cite{1993Icar..101..108K}. High concentrations of atmospheric H$_2$
would have interrupted the OH radical pathway for CH$_4$ oxidation. Both gases have a demonstrable greenhouse effect \cite{Pavlov2000,2013ApJ...778..154W}, which enhances near-surface water vapor abundance. And absorption of visible/near-infrared radiation by CH$_4$ strongly heats the lower stratosphere \cite{Pavlov2000}. Based
on radiative-convective simulations with variable CH$_4$ values, \cite{Pavlov2000} argued that there would be a direct relationship between a more reduced mantle, a warmer ``cold trap'', deeper vertical mixing of water vapor, weaker contrast in water vapor concentrations across the hygropause, enhanced water vapor photolysis, and oxygen production in the upper atmosphere. 

As in the present day, CH$_4$, unlike H$_2$O, would not condense at Earth atmosphere temperatures.
Therefore, under a variety of ``cold trap'' conditions, CH$_4$ from the putative reduced mantle source would diffuse or advect beyond the hygropause to altitudes at which it will photolyze to produce H$_2$ (but not O$_2$). 

Speculations that H$_2$O and CH$_4$ would react to form an organic haze (e.g., C$_4$H$_2$ and C$_5$H$_4$), which would oppose any CH$_4$
or H$_2$ greenhouse effect, challenge this picture \cite{2001Geo....29.1003P,2008AsBio...8.1127H}. However, recent simulations of this aerosol suggest that it would be optically thin in the visible but optically thick in the ultraviolet. The haze would have little effect on greenhouse warming but shield the atmosphere below it from photolysis \cite{2010Sci...328.1266W} (and may affect the Xenon isotopic ratio, Section \ref{xenon}).

At the same time, \citeA{Pavlov2000}'s simulations greatly simplify quantitative treatment of chemistry, transport, and hydrogen escape. As is noted, ``[a figure which shows the relationship between CH$_4$ flux and atmospheric concentration] is somewhat deceptive in that it implies that atmospheric CH$_4$ concentrations can be calculated by specifying the surface CH$_4$ flux.'' One example of a complication is that higher water vapor abundance in the atmosphere will reduce the atmospheric concentration of CH$_4$ by providing an abundant source of OH radical.
The source of CH$_4$ of the Early Earth has also been suggested as being biological of origin, and a possible biosignature \cite{Arney2016}.

Another distinct type of hydrogen escape regime might occur if the entire surface were glaciated, as is speculated to have occurred during portions of Paleoproterozoic and Neoproterozoic time on the Earth. This regime has been invoked to explain mysterious rises in atmospheric p$_{\textrm{\tiny{O}}_2}$ (O$_2$ partial pressure) during the deglaciation from Snowball events.

p$_{\textrm{\tiny{O}}_2}$ climbed to 10\% or even close to present-level p$_{\textrm{\tiny{O}}_2}$ in the aftermath of the Paleoproterozoic Snowball \cite{2005E&amp;PSL.238..156B}. However, the connection between an increase in atmospheric oxygen and the Paleoproterozoic
glaciation is disputed \cite{Hoffman2013143}. It is entirely possible that the glaciations preceded the rise in oxygen by
$\approx$ 100 million years. p$_{\textrm{\tiny{O}}_2}$ then dropped to 1-3\% before rising again to 5-18\% in the
Neoproterozoic, a time when the connection between deglaciation and the oxygen rise is better established  \cite{2006Natur.444..744F,2008Natur.452..456S,2007Sci...315...92C,2008Sci...321..949C,Halverson2009351,2012Natur.489..546S}. This higher level of p$_{\textrm{\tiny{O}}_2}$ coincided with the first appearance of metazoans in the rock record at around 600 Ma \cite{2007Sci...315...92C}, and with the end of the Cryogenian era of Snowball Earth glaciations.

Simple models of atmospheric chemistry suggest that the unusually cold conditions of an entirely ice-covered Earth would favor the production of
H$_2$O$_2$ in the atmosphere \cite{2006PNAS..10318896L}.  As in present day Antarctica, this H$_2$O$_2$ would be incorporated into ice. During deglaciation, this H$_2$O$_2$ would enter an ocean rich with Fe and Mn, poisoning existing anaerobic organisms while
creating the selection pressure for the development of oxygenic photosynthesis \cite{2005PNAS..10211131K,2006PNAS..10318896L}. Oxygen limitation thereafter then would be due to oxidation and precipitation of Fe and Mn outcompeting oxygen
production by nitrogen-limited early photosynthetic autotrophs \cite{2006PNAS..10318896L}. The dependence of this mechanism on the composition and emission rate of mantle effluents (and thus geological activity/upper mantle oxidation state) is unknown. And this question does figure in interpreting and extrapolating from the Snowballs, because the Earth's mantle was likely more oxidized during Neoproterozoic time than Paleoproterozoic time.

An additional variable to consider for the past is the identity and abundance of the principal atmospheric constituent (N$_2$  at present). Models consistent with abundant liquid water that assume a Faint Young Sun either assume higher atmospheric pressure from gases such as N$_2$ (e.g. \citeA{Goldblatt2009}) or assume higher concentrations of greenhouse gases such as CH$_4$, H$_2$, and CO$_2$, whose ability to warm climate is strongly dependent on pressure broadening (e.g. \citeA{Kasting1984}). Higher atmospheric pressure also can reduce surface temperature as a result of increased molecular scattering of incoming solar radiation \cite{Goldblatt2009,Poulsen2015}. However, the sign of the net effect is unclear. Radiative-convective model simulations suggest that increased N$_2$ or O$_2$  will result in net positive radiative forcing even at p$_{\textrm{\tiny{CO}}_2}$  much less than at present \cite{Goldblatt2009,Payne2016}. Simulations with a GCM that included clouds suggest that the net radiative forcing can be negative as a result of cloud feedback effects at higher atmospheric pressure \cite{Poulsen2015}.

Data from the geological record about past atmospheric pressure has wide uncertainties but may argue against the Earth's atmosphere being much thicker in the Archean. A recent study of gas bubbles in an Archean (2.7 Ga) lava flow near paleo-sea level by \citeA{Som2016} suggests that the Earth's atmospheric pressure was no more 50\% of present and most likely $\approx$ 25\% of present at that time. Raindrop-based reconstructions also have been attempted. \citeA{Som2012} suggested an upper bound for atmospheric density of approximately twice present, but \citeA{Kavanagh2015} argued that raindrop size was more sensitive to rainfall rate than atmospheric pressure and suggested an upper bound for atmospheric density of approximately 11 times present.
 It is to be noted that the work of \citeA{Airapetian2016} and its extension (Gronoff et al. in. prep.) consider an alteration of the atmospheric chemistry by SEP events to create N$_2$O, which increase the temperature of the Early Earth even for atmospheric pressure lower than 0.5 bars.
 
Modeling suggests that if were possible to keep liquid water stable in a low pressure N$_2$ atmosphere (200 hPa), water transport by moist convection to the middle and upper atmosphere would be extremely efficient, resulting in high rates of water photolysis \cite{KleinboehlEtAl2018}. The resulting atmosphere evolves to a state in which abiotic oxygen dominates the atmosphere, unless there is a strong sink of oxygen at the surface \cite{KleinboehlEtAl2018}. Such a mechanism could explain bursts of oxygenation coincident with the formation of banded iron formations, but the model relies on a one-dimensional parameterization of moist convective adjustment \cite{Kasting1988runaway} that requires testing in a framework that more explicitly resolves the physical processes.     

\paragraph{The Moon}

The exosphere of the Moon is interesting in several ways: 1- it is easier to experiment on it: we can study the decay of artificial gases released on it by lunar lander in function of the solar activity \cite{vondrak1974creation,Vondrak1974,Vondrak1992}; 2- it has the same solar wind conditions as the one measured for space weather at Earth, and therefore studies such as the impact of CME on it are easier \cite{Killen2012}; 3- we have samples from the Moon, and we can study the possibilities of ancient atmosphere from it.

The hypothesis of a secondary atmosphere due to volcanic activity at the moon has been proposed in \cite{2017E&PSL.478..175N} based on the analysis of samples from the Apollo mission. It is possible that an an atmosphere with up to a few mb at the surface was created and stable for 1000s of years. In \citeA{Aleinov2019}, a study of the thermal escape was made, showing the limitations of the creation of such an atmosphere, as well as the climatic conditions an atmosphere would have had. These conditions are interesting since they show the transport of volatiles to the poles. it would be possible to find some clues of that atmosphere in samples from the poles.

\paragraph{Mars}

When Mars had an intrinsic magnetic field early in its history, its hydrogen-bearing species fluxes to the upper atmosphere likely occupied a phase space that could be described by the early Earth or even present day Earth phase spaces \cite{alho2015paleo}. Transition to the regime observed today may have depended on the timing of magnetic field loss. This transition has so far been modeled as a primarily CO$_2$ atmosphere condensing to form at least one permanent ice cap (\citeA{Soto2015} and references therein). The principal unknown about the last billion years or so is how fluctuations in Mars' obliquity have changed the location of surface and sub-surface ice reservoirs, which could affect the water cycle, the total atmospheric mass, and the dust cycle \cite{Fastook2008,Madeleine2009}. A lot of questions have also been asked about the effect of the magnetic field in the loss of the atmosphere. Since observation shows that similar amount of heavy ions are lost above magnetic fields at the current Mars than above non-magnetized parts \cite{Sakai2018}, it may be that it influence has been greatly exaggerated in previous studies.

Like Earth, there are some constraints on past atmospheric mass for Mars. The atmosphere filters the impact crater population by ablating the lower end of the bolide size distribution \cite{Jakosky2017}. On this basis, \cite{Kite2014} proposed that Martian paleopressure was never higher than $\approx$ 3~bar (and likely much less). A higher palopressure would have led to a collapse of the atmosphere. From meteoritic observation constraints, and considering that some isotopic reservoirs can be replenished by meteoritic/cometic falls \citeA{kurokawa2018lower} slightly modified the history presented in \citeA{Jakosky2017} and suggested a minimum paleopressure of 0.5~bar. \citeA{jakosky2018loss} suggested that Mars lost more than 0.8~bar of CO$_2$ or the equivalent of 28~m of water.

An interesting point at Mars is the observation of solar-wind H deposition in the thermosphere \cite{halekas_maven_2015}, this deposition follows a charge-exchange process, and could have led to changes in D/H ratio if large enough in the Early Solar system; however, it is probable that this deposition would have been counteracted by hydrodynamic escape.

\paragraph{Venus}

Venus, at some point during its history, likely occupied an additional phase space with respect to coupling between the surface and the exosphere: that of the runaway greenhouse \cite{Ingersoll1969}. However, this regime is somewhat analogous to the elimination of the ``cold trap'' by absorption of visible/near-infrared radiation by CH$_4$. The twist is that it is the infrared greenhouse effect of H$_2$O that breaks the cold trap.

The effect can be conceptualized semi-quantitatively. Consider a layer of the atmosphere at which vertical mixing from the surface is relatively efficient. Now raise the surface temperature by some amount by introducing a higher amount of solar insulation. To first order, the relationship between water vapor concentration and temperature should be exponential, following the Clausius-Clapeyron relation that defines the saturation curve. In the Earth's atmosphere, however, it is observed that the effects of vertical mixing and pseudo-adiabatic precipitation processes reduces the sensitivity of mean $p_{H_2O}$ to surface temperature in the lower troposphere \cite{HeldSoden2006} but may enhance it in the upper troposphere \cite{GettelmanFu2007}. Thus, water vapor concentration will increase exponentially in response to the increase in surface temperature. The layer's temperature likely will increase as well in response to the increase in surface temperature. (This is easiest to visualize at the surface itself.)

At the same time, the increase in water vapor will increase the infrared opacity of the layer, reducing outgoing longwave radiation from the layer (and below the layer). Yet the increase in the layer's temperature will result in increased outgoing longwave radiation according to Stefan-Boltzmann's Law. At low temperatures and water concentrations, it is easy to see that the principal change in outgoing longwave radiation will be due to the increase in layer temperature. However, as temperatures increase, the exponential dependence of water vapor on temperature eventually will overcome the quartic dependence of outgoing longwave radiation on temperature. Thus, for any sufficiently abundant infrared absorber condensing and evaporating, there is some critical point at which outgoing longwave radiation in the layer will decrease rather than increase with surface temperature, initiating a runaway positive feedback loop. Warming of the troposphere eventually results in its expansion and enhancement of vertical transport in the middle and upper atmosphere. For water, this runaway loop is slowed by UV hydrolysis of water in the middle and upper atmosphere and stopped by exhaustion of the surface reservoir, a process that \citeA{Ingersoll1969} argued had occurred on Venus (rather than Earth or Mars) as a result of the former's higher insulation.

This escape regime has been simulated by \citeA{kasting_loss_1983,Kumar1983,Chassefiere1996h,Chassefiere1996o}. None of these simulations challenge the basic mechanism but emphasize: (1) that hydrolysis rates will be dependent on the oxidation state of the atmosphere and buffering by chemical reactions in the crust and (2) that the EUV flux of the Sun (a major unknown early in its lifetime) is the principal control on the rate of escape. Another interesting conclusion is that the present D/H ratio in Venus' atmosphere must be a consequence of a period of reduced escape rates that closed the runaway greenhouse phase. In the ideal runaway greenhouse escape regime for Venus, D would have been stripped off as easily as H \cite{kasting_loss_1983}. The question of when that runaway escape happened is difficult as it was suggested that Venus could have been able to sustain liquid water up to a $\approx$Gyr ago \cite{way2016venus}.

\paragraph{Titan}
The large size of the atmospheric reservoir of methane in comparison with the surface reservoir of methane and methane's photochemical products \cite{Lorenz2008} strongly suggests that a methane-rich atmosphere for Titan has been a relatively unusual condition during Titan's history \cite{Lorenz1997}. Once a sufficient amount of time has passed, photochemistry will refine methane to organic compounds that will form surface deposits of liquid and solid higher order hydrocarbons. The resulting atmosphere will lose the portion of its greenhouse effect driven by pressure broadening of methane, and Titan will lose its stratosphere \cite{Wong2015}. Any hydrogen escape presumably will be restricted to photochemical loss of water derived from micrometeorite ablation, etc.

Yet the presence of CH$_4$ in Titan's atmosphere likewise implies occasional, episodic release of methane into the atmosphere by volcanism \cite{Tobie2009}. Depending on the exact nature of this volcanism, Titan could have experienced a more intense hydrogen escape regime in the past.

\section{Escape at Exoplanets}

Since their first detections around stars in the mid-late 1990s \cite{mayor1995jupiter}, a particular interest has been set to the atmospheric escape of exoplanets. In particular, the intense heating and radiation at close-in orbit planets, such as the planets orbiting M-dwarfs in the Habitable Zone (HZ), or the giants close to their host stars (the so-called ``hot-Jupiters'' that we name close-in giant in the following since the nature of, notably, their atmospheric escape cannot be considered as Jupiter-like), may lead to very high atmospheric mass-loss rate and potentially a complete evaporation of the planetary atmosphere (in addition to potential atmospheric stripping by the stellar wind) (e.g. \citeA{Lammer2003,Cohen2015ApJ...806...41C}). 
\subsection{Current observations and modeling}
\subsubsection{Close-in giants}
Observations of the close-in giant planet HD 209458 have revealed absorption in the Lyman $\alpha$ line, which associated with the existence of neutral Hydrogen (H I) at or above the estimated Rosch lobe \cite{Vidal-Madjar2003,Vidal-Madjar2004}. The fractional difference of in- and out-of-transit flux was wavelength dependent, with much of the flux decrease occurring at wavelengths more than $100\, \rm km\ s^{-1}$ from line center. Though the atom-photon cross section is much larger at line center, and one would expect much larger transit depths there, interstellar absorption and geocoronal emission contaminate wavelengths $< 50\, \rm km\ s^{-1}$ from line center, and only measurements further from line center may be trusted. Later observations have indicated the existence of heavier atoms, such as Ca II and O I at this altitude \cite{Ehrenreich2008, Linsky2010}. These observations suggest that the planet has an inflated atmosphere with a high mass-loss rate of the order of $10^{7}\;kg\;s^{-1}$, and an escape velocity of the order of $100\;km\;s^{-1}$. 

The current paradigm assumes that close-in giants lose mass from their atmospheres due to hydrodynamic escape \cite{Ben-Jaffel2007,Ben-Jaffel2008,Vidal-Madjar2008,Linsky2010}. However, it is not obvious what is the mass-loss rate and the escape speed, what is the altitude of the observations, and what is the overall structure of the inflated atmosphere. It has also been suggested that due to the fast orbital motion, the extended atmosphere may have a comet-like tail \cite{Linsky2010,Cohen2011}.

A number of models have been developed to study atmospheric escape from close-in giants \cite{Baraffe2004,Yelle2004,Garcia-Munoz2007,Lecavelier2007,Schneiter2007,Penz2008,Murray-Clay2009,Tian2009,Stone2009,Adams2011,Trammell2011,Koskinen2014}, where most of the models assumed that the intense hydrodynamic escape is due to photo-evaporation by the intense stellar radiation. The models listed above (partial list) vary in the equations they solve, their assumptions about the energy sources and distributions, their complexity, and the way they are solved. The mass-loss rate obtained by these models covers few orders of magnitude. Therefore, despite of the vast modeling effort, the nature of atmospheric escape from these close-in giants is not fully understood yet. 
The efforts by \cite{tanaka_atmospheric_2014,tanaka_atmospheric_2015} to model atmospheric escape from close-in giants have to be noticed since the model is based on MHD wave heating leading to ionospheric outflow. This is a case of the more general ionospheric outflow described in section~\ref{ionosphericoutflow}.

In the case of HD 209458b, the implies the existence of a large ``corona" or ``cloud" of atomic hydrogen. This cloud must be optically thick to Lyman-$\alpha$ at wavelengths $>100\, \rm km\ s^{-1}$ from line center out to several (optical continuum) planetary radii, approaching the planet's Hill radius, beyond which stellar tides dominate over the planet's gravity. There are two models to account for this large hydrogen density at such high altitudes. 

The first model \cite{Yelle2004} is that the absorption is due to thermal particles in the planet's upper atmosphere. Photoelectric heating from hydrogen ionization, balanced by slow adiabatic expansion, raises the temperature to $T \sim 10^4\, \rm K$. The resulting large scale height implies a slow outward decrease of the density and hence large density at high altitude. In this model, the thermal speed of the atoms is $v_{\rm th} \sim 10\, {\rm km\ s^{-1}}$ and absorption at $>100\, {\rm km\ s^{-1}}$ implies a large column of hydrogen is needed to overcome the small cross section at $>10$ Doppler widths from line center. 

The second model \cite{Holmstrom2008} relies on fast hydrogen atoms (ENA), which must move at speeds comparable to the line width. The large atomic speeds imply that vastly smaller columns are needed to attain optical depth unity. The thermal hydrogen speeds, and bulk velocity in hydrodynamic escape, are expected to be only $\sim 10\, {\rm km\ s^{-1}}$. The production of fast hydrogen atoms is through charge exchange with $v_{\rm th} \simeq v_{\rm bulk} \sim 100\, {\rm km\ s^{-1}}$ stellar wind protons. There are variants of this model in which atoms are ballistically fired outward from the planet and interact with the stellar wind \cite{Holmstrom2008} and also models in which the mean free paths of the atoms are small, and the interaction occurs in a hydrodynamic mixing layer \cite{tremblin_colliding_2013}.

These models are in a sense not independent, but rather focus on two separate aspects of the same problem, since both thermal and non-thermal hydrogen may contribute to the absorption. In particular, the density of hydrogen atoms which may interact with the stellar wind (model 2) is set by the outer limit of the upper atmosphere (model 1). It has to be noted that, even if the models are complementary, the conclusion drawn from a peculiar aspect are not totally the same: the ENA model is consistent with a much smaller escape than the thermal escape model.

\subsubsection{Rocky planets}

In the case of atmospheric escape from terrestrial/rocky planets, some modeling work has been done \cite{Tian2009,wordsworth_hydrogen-nitrogen_2013,Kislyakova2014,Cohen2015ApJ...806...41C,Gao2015,Dong2017ApJ...837L..26D}, but no reliable observations have been obtained so far, mainly due to the large size of the telescopes needed for the measurements \cite{Gronoff2014}, except in the case of extremely close-in rocky planets such as the disintegrating planet KIC 12557548b \cite{Rappaport2012}.

Major efforts went to model the planets in the HZ of Proxima Centauri B and Trappist 1.
The work of \citeA{GarciaSage2017} shows that a Earth-like planet at the location of these planets would suffer an enhanced ion escape, leading to the loss of the equivalent of the Earth's oceans over a billion years; the location of many of the planets inside the Alfv\'en surface (section \ref{alvensurf}) further prevents the existence of a sustainable atmosphere. It means that, to sustain habitability in the sense of liquid water existing at the surface, such planets would require a large amount of volatiles in their initial inventory, and that they should not lose them in the active young years of their host star. To that extent, work has been done to look at the hydrodynamic escape of planets in the habitable zone of their active stars showing that even N$_2$ would be hydrodynamic \cite{johnstone2019extreme}.
This theoretical work has been confirmed by the recent work of \citeA{kreidberg_absence_2019} that was able to show, using NASA/Spitzer observations, that the exoplanet LHS 3844b has no thick atmosphere: such an atmosphere would have been able to reduce the temperature difference between the nightside and the dayside of the planet compared to the observations.
The conclusion of that problem is that, while their are the easiest target for detecting habitable exo-atmospheres with instruments such as the James Webb Space Telescope (JWST), planets in the HZ of red-dwarfs may not be able to sustain them and therefore would be the worst target.

\subsection{The Stellar Wind and the Alfv\'en Surface}
\label{alvensurf}
The classical HZ of stars fainter than the Sun resides closer to the star. In particular, the HZ of M-dwarf stars is located at planetary orbits of less than 0.1AU. While the size of M-dwarf stars is about $0.1-0.3R_\odot$, their magnetic fields seemed to be overall stronger than the field of K, and G stars \cite{Reiners2007ApJ...656.1121R}. As a result, their Alfv\'en surface, at which the stellar wind exceeds the Alfv\'en speed and open the coronal field lines into the interplanetary space, is more extended than that of the Sun. 

\begin{figure}
 \noindent\includegraphics[width=0.8
\textwidth]{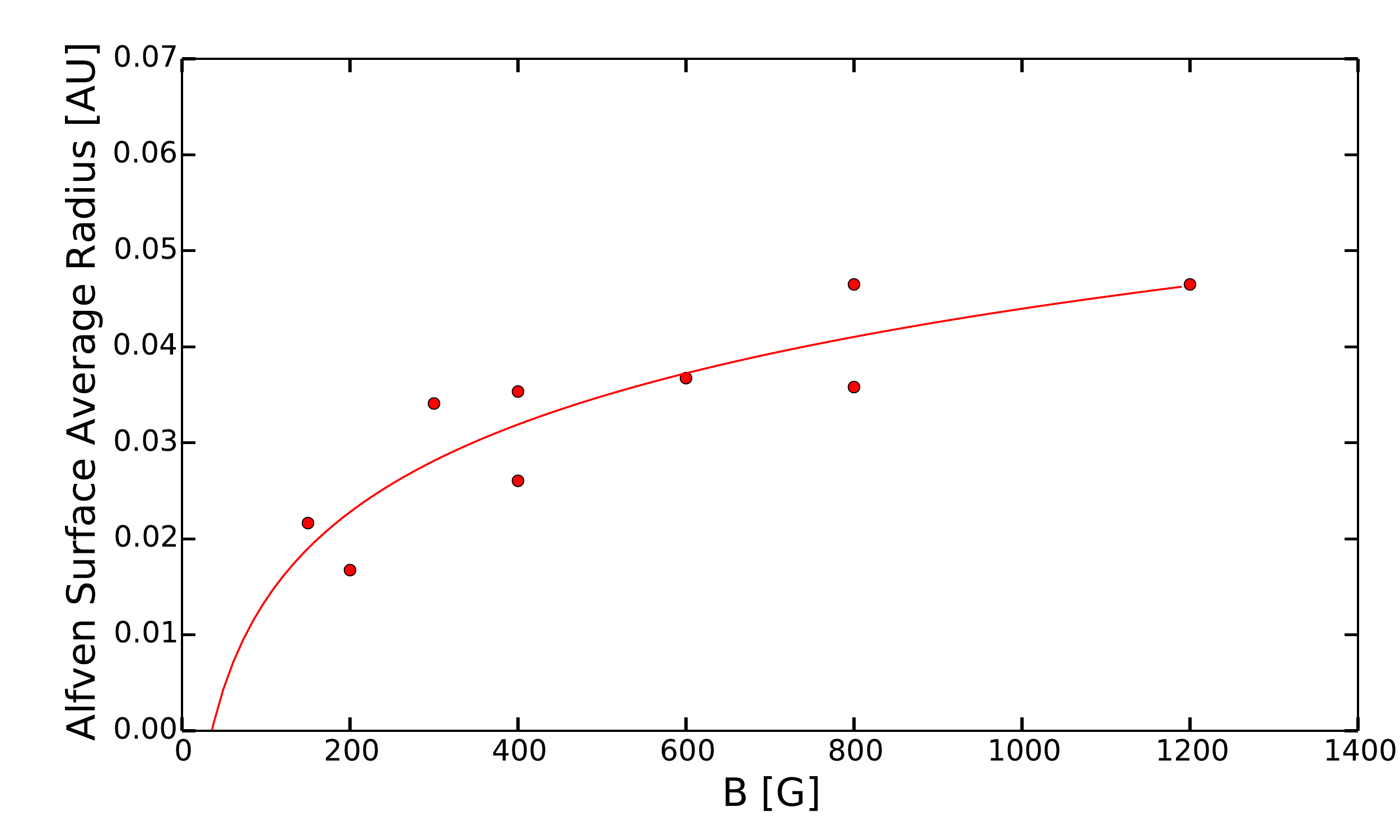}
 \caption{Dots represent the average Alfv\'en Surface size as a function of average magnetic field strength from MHD simulations. The line represents the trend derived from them. }
 \label{alfven_surface}
\end{figure}

Since the Alfv\'en surface can be a measure to the boundary between the stellar corona and the interplanetary space (filled with the fully developed stellar wind), planets residing within the Alfv\'en surface could be considered to be inside the corona. In addition to the extreme temperatures that can exceed a million degrees, planets in this regime will orbit in densities and magnetic fields that can reach 2-4 orders of magnitude higher than those at Earth \cite{Garaffo2016ApJ...833L...4G} and, as a result, experience extreme dynamic and magnetic pressures at the planetary orbit.  Such space environment conditions \cite{2017ApJ...843L..33G} may lead to an Alfv\'en wings (Io-like) topology of the planetary magnetosphere, at which a significant fraction of the planetary field is open to the stellar wind. As a result, the planetary atmosphere may be exposed to intense heating due to the incoming stellar wind energy in the form direct particle precipitation, and Alfv\'en wave energy that is transmitted by the stellar wind. Additionally, since the planets reside in the sub-Alfv\'enic corona, some of the Alfv\'en wave heating that is deposited in the corona may be transferred to the planet. While essentially no work has been done on these processes in exoplanets, the scenario described above may suggest that it is unlikely that these planets are habitable. \textbf{Therefore, the Alfv\'en surface might serve as an inner limit at which the HZ can be placed for a given stellar system.} The result of the simulations of the distance of the Alfv\'en surface from its parent star as a  function of the average magnetic field of that star is given in Figure \ref{alfven_surface}.  The spread in distance for a given magnetic field strength arises from possible differences in the geometric distribution of the magnetic field on the stellar surface. The Alfv\'en surface is smaller for complex field structures (i.e.\ higher order in the multipolar expansion) \cite{2016A&A...595A.110G}.  

\section{Developments Needed in Measurements and Modeling of Atmospheric Escape}


In the context of astrobiology, developments are needed in modeling, observations, and laboratory measurements for being able to observe and characterize rocky exoplanets' atmospheres and for reconstructing the history of the atmospheres of the planets in the Solar system. We review here some of the planned and suggested improvements. For a more comprehensive list of suggestion, white papers submitted to the \citeA{nataca} have a comprehensive list. 
The JWST, being the astrophysics flagship of the 2020s, is shaping the direction of astrobiology research and therefore many currently proposed studies and developments are linked to its targets of choice, the exoplanets around M-dwarfs.

\subsection{Measurements/Observations}

The detection of a technosignature, currently searched in radio-waves, would be the ultimate proof of a life that developed outside Earth \cite{Wright2019}. The development of observatories for such endeavor is outside the scope of this study since a mere detection would not provide data on how life developed there, under which conditions, and how it started. To answer these questions, better measurements are needed to understand the Sun and other Stars, laboratory measurements and sophisticated measurements are needed for better interpreting our planetary observations, and better instruments are needed for looking at exoplanets. 

Space missions such as CHEOPS \cite{Broeg} and Transiting Exoplanet Survey Satellite (TESS) \cite{Ricker} are expected to find thousands of transiting planets with many terrestrial-like planets including hundreds of super-Earths over the next few years \cite{Fridlund}. With the ability to discover transiting exoplanets, efforts are being pursued for spectroscopic observations of exoplanetary atmospheres. The National Academy of Sciences (NAS) ``Exoplanet Science Strategy'' report mandated by the US Congress recommends a direct-imaging telescope as a follow-on to the Wide Field InfraRed Survey Telescope (WFIRST) mission scheduled to fly in the mid-2020’s after the launch of the James Webb Space Telescope (JWST). Large amounts of time are being dedicated on the Hubble and Spitzer space telescopes, as well as major ground-based telescopes (e.g. VLT, Keck, Gemini, Magellan, CFHT, etc.) \cite{Madhusudhan}. JWST and European-Extremely Large Telescope (E-ELT) will revolutionize exoplanetary spectroscopy. M dwarfs are prime targets for the detection and characterization of terrestrial exoplanets by the JWST, as they are abundant in the solar neighborhood and their small radii allow for greater transit signals from Earth-sized exoplanets (e.g., \cite{Quintana}). Future spectra with JWST would be of unprecedented precision and resolution which will enable us to derive precise chemical abundances for transiting exoplanets. Our modeling can help constrain future observations in making precise determinations of detectable key species abundances, distributions and understanding of their processes. 
The NAS report also recommends that the US National Science Foundation (NSF) invest more in the future Giant Magellan telescope (GMT) and proposed Thirty Metre Telescope (TMT) now being built in Hawaii and Chile, respectively. These telescopes would provide more focused study of exoplanets by using spectroscopy to seek out signs of free oxygen in their atmospheres. Such a study would be perfect suited for the goals of the astrobiology community.

For F and G stars, much of the observing is likely to initially be in the IR so the lower bound to some of the observations are likely to be near the tropopause (outside of H$_2$O bands). For Venus-like exoplanets, this does not present a problem, but for Earth-like exoplanets it may be for F type stars.  However, for all K and M type stars, processes of interest governing distribution of key species in atmospheres for these types of detections may be directly observed in the mesosphere and thermosphere. Additionally, there are spectral regions in the UV (and possibly in the visible to IR) where O$_2$ and O$_3$ in the mesosphere/upper stratosphere would dominate remote measurements. Those spectral regions would be an ideal target for future observations. 


\subsubsection{Solar and Stellar measurements}


The capabilities of the JWST mean that the search for biosignature with that observatory is mainly limited to M dwarfs. The increased risk of atmospheric escape due to joule heating in the classical HZ around these stars\cite{Garaffo2016ApJ...833L...4G} means that it is extremely important to study the activity of these stars. The best way to perform this is with UV instruments such as STIS on the HST \cite{France2013}. The HST being on its end of life, with no repair mission planned, a mission dedicated to the EUV-XUV measurements on close stars (such measurements would be less affected by interstellar H absorption) would be critical to support the modeling and observations of planetary atmospheres that could harbor life.

The recent observation of a CME around another star \cite{argiroffi2019stellar} shows the possibilities of such measurements. However, they should be improved to have a better idea of the fluxes of particles at other stars and to validate the semi-empirical laws linking flares and CME \cite{Moschou_2019}. See also Section~\ref{s:stellar_forcing}.



\subsubsection{For planets}

%
As already emphasized, the knowledge of accurate cross sections is of critical importance for the precise evaluation of escape processes. For charge transfer, including double charge exchange, electron capture and stripping, it has been customary to study these processes in the laboratory in two main directions: (i) under the wide umbrella of nuclear research, radiation dosimetry, and the effect of radiation on living tissues \cite{Nikjoo2012}, especially in water and carbon, and (ii) in astrophysics and heliophysics studies, especially rather recently with respect to X-ray production \cite{Wargelin2008,Dennerl2010}. In (i), one of the goals is to calculate the stopping power of particles in matter using so-called track-structure Monte Carlo models. In (ii), the applications are numerous, from interstellar medium to cometary X-ray emissions. This has resulted in a rather well-understood behaviour of charge-transfer cross sections at energies typically above 10 keV/amu impactor energy and peaking in the MeV range \cite{Uehara2002}. At low energies, from a few tens of eV/amu to 10 keV/amu, which are the typical energies for solar wind charge exchange and in planetary ionospheres, the information is usually fragmented and one is often forced to extrapolate, more or less arbitrarily, the shape of the cross sections, leading to high uncertainties \cite{wedlund_solar_2019}.

Over the two last decades, experimental physicists have punctually studied aspects of solar wind charge exchange. Several international groups have specialised on different aspects \cite{Dennerl2010}, for example the UV spectroscopy group at the University of Groningen (Netherlands) for cometary environments \cite{Juhasz2004,Bodewits2004,Bodewits2007}, motivating studies of impacts of fast solar-wind like ions with several neutral species of planetary atmosphere relevance. The examples of H$_2$O, CH$_4$, CO and CO$_2$ are particularly relevant: \cite{Greenwood2000,Greenwood2004}  and \cite{Bodewits2006} have recently measured with good accuracy charge-transfer cross sections of protons and helium ions on H$_2$O, CH$_4$, CO and CO$_2$ for astrophysics applications. However, these cross sections were not measured in the very low-energy range (below 50 eV for helium ions, below 1.5 keV for protons). Moreover, certain electron capture and stripping reaction cross sections have yet to be altogether measured by any group. For example, the stripping reaction involving hydrogen or helium fast atoms and H$_2$O ((H,He) + H$_2$O $\rightarrow$ (H$^+$,He$^+$) + H$_2$O + $e^-$) has never been measured below 20 keV/amu energy; it may prove an important sink for the produced ENAs, and hence may play a role in the escape of such particles into space. 

A few online databases exist for several charge transfer cross-sections. Despite ongoing work made to create online database and recommended sets of cross sections (e.g. \citeA{lindsay2005charge}),  it is left for a supplementary critical review of charge-transfer cross sections in planetary and cometary atmospheres to list all of the available cross sections, their energy range, evaluate their uncertainties and the gaps in our present knowledge and provide a final recommendation that can be used in models and data analysis. Such a specific review is outside the scope of the present article, and we will here only point out one direction that experimental physics teams are encouraged to further study, that is, the solar wind charge transfer with a neutral atmosphere. 

Further studies will have to choose colliding species, such as:
\begin{itemize}
	\item Neutrals of interest (non-exhaustive list by increasing atomic/molecular weight): H, H$_2$, He, C, N, O, CH$_4$, OH, H$_2$O, Ne, N$_2$, CO, O$_2$, S, Ar, CO$_2$.
    \item Impactor of interest: H, H$^+$, He, He$^+$, He$^{2+}$, and high ion states of O and Fe.
\end{itemize}
and to consider the two following points:
\begin{itemize}
 	\item Study more systematically all sources and sinks for the ion-ENA system: single and multiple electron capture, single and multiple electron stripping, impact ionisation by fast atoms and ions.
	\item Measure new energy-dependent cross sections, and uniformly extend current cross section measurements to energies ranging from threshold to 20 keV/amu energy, most relevant for solar wind studies.
\end{itemize}

Other developments are needed for planets, such as Cassini-like missions to the Ice Giants, for a better understanding of the development of  these atmosphere and their satellite. Such missions would give more insight into the evolution of our Solar system. Improved instrumentation could be of use around the Earth to better discriminate the escaping species: it is currently extremely difficult to know if it is an O or a N that is leaving the atmosphere. 

From that point of view, both for planets and exoplanets, it is interesting to look at the X-Ray emission of the planets: the X-ray halo, created by charge exchange between the exosphere and the incoming solar wind, gives some insight to the composition of escaping species \cite{1997Sci...277.1488K,Dennerl2007,Dennerl2010}. This is of interest both for planets and exoplanets since detecting such an halo in another stellar system would give some direct insight in the composition of the exoplanetary exosphere.

\subsubsection{For exoplanets}


Atmospheric escape from exoplanets can be constrained by observations of components that affect the escape. This includes compositional observations (i.e., transmission spectra), direct observations of escaping material. These observations will pose challenges in the upcoming decades. 

It is also necessary to observe magnetic fields of exoplanets as they may play a key role in the atmospheric escape.  One promising option is to obtain information about exoplanets magnetic fields via observed signature star-planet interaction. These signatures include induced chromospheric activity(e.g.  \citeA{Shkolnik08,Cauley19}), or modulation of coronal radio emissions \cite{2018AJ....156..202C}. The direct detection of exoplanets magnetic fields (via radio observations of auroral emissions \citeA{Zarka07}) has recently been reported \cite{2020NatAs.tmp...34V}. This has only confirmed the existence of the magnetosphere: more work is needed to be able to estimate (e.g.) the magnetic moment from these observations. The modeling of the interaction of the stellar wind with the planetary magnetosphere of HD 209458b led to the estimation of its magnetic moment \cite{Kislyakova2014b} from the observation of Ly-$\alpha$.  Giant space UV-telescopes would be choice instruments to study the upper atmosphere of rocky exoplanets. \citeA{Gronoff2014} proposed a technique to detect hydrodynamic escape of CO$_2$ or O$_2$ rich planets using such laboratories. For the detection of biosignatures, a review of techniques and developments needed can be found in \citeA{Fujii2018}.
The generalization of the detection technique used by \citeA{kreidberg_absence_2019} to detect the absence of an atmosphere around LHS 3844b is also needed to look at the best target for future telescopes.

\subsection{Modeling}

To get a comprehensive view of the escape of planetary atmospheres, models have to be developed to take into account all the energetic inputs and all the processes leading to the escape. The outputs of such models have to be compared with observations. Problems lies with inputs parameters for the model (cross sections, observation of e.g. solar flux), the estimation of the uncertainties, but also with the neglected parameters. It is often the case that our instrumentation gives a very detailed view of the conditions on a planet; however, the uncertainties in the input parameters of the models make it challenging to interpret \cite{sanchez2018}. The estimation of model uncertainties from the different input parameters can be an arduous task \cite{gronoff2012_1,gronoff2012_2} and becomes problematic once free parameters are needed, which is often the case in our models of atmospheres, stellar wind, stellar wind interactions, etc. It is possible to begin solving the problem by careful comparison with solar system observations, then with extrapolation.
On the other hand, the instrumentation may not be sensitive enough to observe interesting phenomenon in exoplanetary atmospheres, or to provide significant model constrains.

Future modeling should also include the dynamical response of the planet's atmosphere to dynamic drivers, stellar evolution scale changes of atmospheric escape, as well as self-consistent coupling between the external drivers and the different regions of the atmosphere. 


\subsubsection{Modeling of Solar and Stellar Environments}

Global models for the solar corona have been developed since the late 1960s by solving the MHD equations. The models are driven by data of the photospheric radial magnetic field in combination with the potential field method \cite{1969SoPh....9..131A}. In recent years, more self-consistent models have been developed for the solar corona and solar wind (e.g. \citeA{2014ApJ...796..111L,2014ApJ...782...81V,2016ApJ...832..180D}). These models incorporate coronal heating and wind acceleration in the form of large-scale heating and momentum terms. These large-scale terms are parameterized and tuned to match solar observations, and the models have been successful in reproducing the observed density and temperature structure of the solar corona, and the observed structure of the solar wind. 

The limited availability of observations of photospheric magnetic field of selected stars using the Zeeman-Doppler Imaging technique \cite{Semel:80} has led to a growing global modeling in stellar coronae and stellar winds of Sun-like stars (e.g. \citeA{Cohen2010,2011MNRAS.412..351V,Garaffo2016ApJ...833L...4G}). However, since the stellar winds of solar analogs cannot be directly measured, the results of these studies are poorly constraints. Therefore, A better modeling work is needed to constrain the magnitude of the stellar wind, and the coronal structure and temperature for different stars as these parameters define the stellar environments at which exoplanets reside in. In particular, the scaling of the global heating and acceleration parameters needs further investigation and quantification to better understand how these processes scaled with stellar type.

\subsubsection{Modeling Atmospheric Escape from Exoplanets}

The current modeling tools for planetary atmospheric escape are built on and tune to known, measurable atmospheres within the Solar system. These tools have already been used to study escape from exoplanetary atmospheres with no significant constraints of the results. A number of features, which are different from Solar system bodies, has already been identified to be crucial for exoplanetary atmospheric escape, especially in the case of close-orbit planets. However, these features need more self-consistent modeling in order to be better defined and quantified.  

The first notable feature is that atmospheric escape from close-orbit planets may be extremely high, to the point that atmosphere could be completely lost. This is due to extremely high dynamic pressure of the stellar wind near these planets (e.g. \citeA{Garaffo2016ApJ...833L...4G,GarciaSage2017, Dong2017ApJ...837L..26D}), the strong orbital variations of the stellar wind conditions, and potential strong heating of the upper atmosphere (e.g. \citeA{2014ApJ...790...57C,2018ApJ...856L..11C}). A more detailed model is required to quantify the exact energy deposition between the wind and the planetary atmosphere, as current models focus on the stellar wind - magnetosphere interaction, without detailed modeling of the energy and mass transfer to and from the atmosphere itself.

The second notable feature is the impact on the planetary upper atmosphere and ionosphere. Current models provided estimation about the Joule Heating assuming specific, constant atmospheric conductance. Since the conductance is the key to determine the heating, further self-consistent modeling is needed to estimate the ionospheric conductance. In particular, these calculations are needed for the case where the EUV and X-ray stellar radiation are much higher than the Earth case, and for different atmospheric composition. 

Finally, close-orbit exoplanets may reside within the Alfv\'enic point inside the stellar corona. Therefore, a direct star-planet interaction is expected to occur. In order to investigate the impact of such a direct interaction between the stellar corona and the planet, a self-consistent modeling that couples the corona and the planetary atmosphere domains is needed. 

\begin{figure}
	\centering
	\noindent\includegraphics[width=40pc]{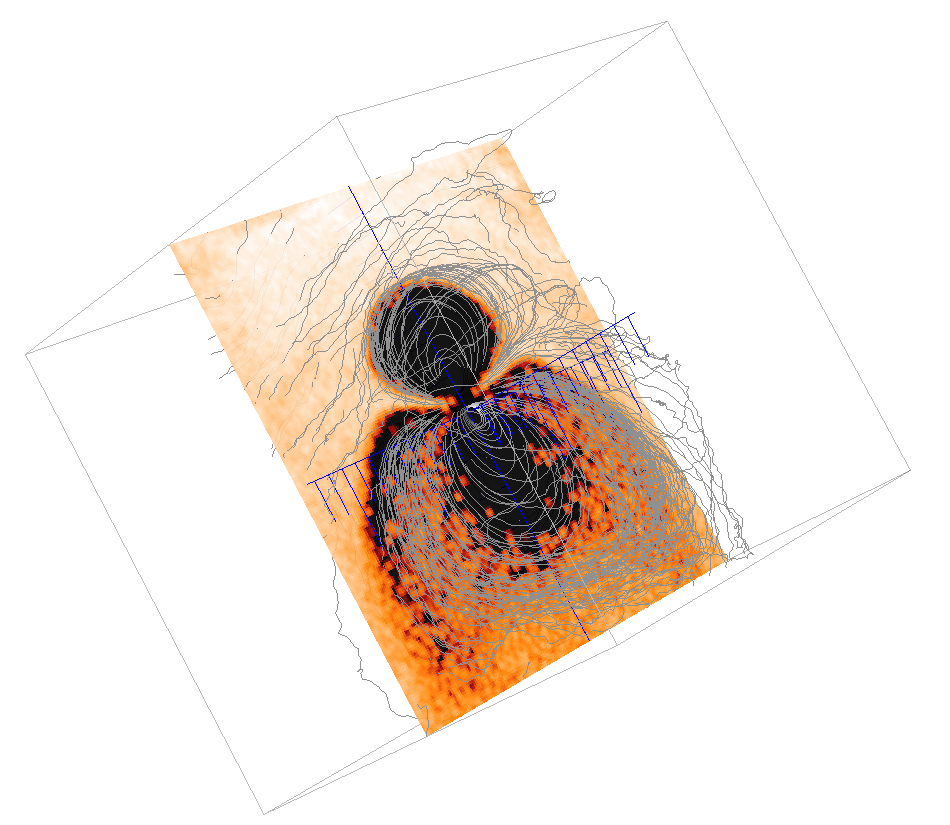}
	\caption{Field aligned current and prototype  kinetic simulation of exoplanet. These simulations are leading to a better understanding of the dynamic within an exoplanetary magnetosphere, which is needed for computing both energetic inputs in the atmosphere and loss of plasma from the magnetosphere. }
	\label{fig:largefieldsw}
\end{figure}

An example of a code in development to address some of these problems could be IAPIC, a particles-in-cell electromagnetic 3D global code, used  \cite{Baraka2010,Ben-Jaffel2013,Ben-Jaffel2014b,Baraka2016} to produce the magnetosphere (XZ plane) of an earth-like planet. Both plasma density and field lines are shown in Figure~\ref{fig:largefieldsw}. It is interesting to see that the PIC simulations naturally recover the field aligned currents (streams of particles appearing between cusps and current sheet in the figure) that drive particles precipitation from the magnetosphere into the polar regions, producing auroral emissions. IAPIC can provide both the angular and energy distributions of the impinging magnetospheric particles into the ionosphere. Charge separation is obtained in the code so that kinetic effects could be obtained while conserving charge \cite{Villasenor1992}.  These electrons and ions enter the upper atmosphere to trigger ion-chemistry, heating, and winds.  Their fluxes should be used as input in existing ionospheric models to evaluate new species produced and atmospheric inflation due to the extra heating deposit in the auroral region of any exoplanet. The simulation, shown in Figure~\ref{fig:largefieldsw}, was carried out with these code parameters for a grid size of  $ 0.1 R_{E} $ and an ion-electron mass ratio of $\frac{m_{i}}{m_{e}} =100$. 

\subsubsection{Modeling Exoplanetary Magnetic Field Observations}

Following the previous sections, it seems like exoplanetary Magnetic fields may play a crucial role in the evolution and sustainability of exoplanets atmospheres. However, these planetary field currently cannot be detected and observed. 

Modeling of star-planet interaction suggest that this interaction can potentially generate observable signatures that can help to quantify the planetary magnetic field (e.g. the broadening of Ly-$\alpha$ in \citeA{Kislyakova2014} and the soft X-ray emission in \citeA{Kislyakova2015}). However, it is clear that a deep understanding of the stellar background field is need for this purpose \cite{Shkolnik08,Cohen2011,Strugarek14,Matsakos15}. In that context, more detailed modeling work on the background stellar environment can support the interpretation of star-planet interaction observations.  

A  number of attempts were made to estimate auroral radio emissions from exoplanets \cite{Zarka07,Lazio07,Grismeier07,Vidotto15,See15,Nichols16,BurkhartLoeb17,Turnpenney18,Lynch18}. Most of these studies have concluded that such auroral emissions are not detectable with the current radio telescopes. An alternative approach has been recently proposed by \cite{2018AJ....156..202C}, who proposed to look for planetary modulations of the ambient coronal radio emission (exoplanet radio transit), instead of looking for the planet as a radio source. Using idealized models for the planetary and stellar fields, \cite{2018AJ....156..202C} have shown that observing the ambient coronal radio intensity, as well as the planetary modulation are more feasible. To better quantify and estimate the latter method, further, more detailed modeling work is needed, targeting specific planetary systems.

\section{Discussion and  conclusion}
The problem of atmosphere escape is more complex than just the estimation of Jeans' escape or energy-limited escape. Several outstanding questions in planetary science can be linked to atmosphere escape, from the problem of the faint young Sun to the question of detection of astrobiological signatures. These problems are  leading to a roadmap of future investigations.


\subsection{Problems still to be resolved}

\subsubsection{The faint young sun paradox}
\label{FYSparadox}
The Faint Young Sun paradox was introduced by \cite{SaganMullen1972}. The paradox states that based on stellar evolution models, during its earlier stages, the Sun's luminosity was about 30\% lower than its current luminosity. As a result, the surface equilibrium temperature of the Earth would be below the freezing point of water. However, many types of geological evidence for the existence of liquid water were found both at Earth and Mars. Therefore, we need to introduce some heating process which increased the average surface temperature of the Earth above zero degrees Celsius. 

The most prominent solution to the paradox is the existence of greenhouse gases in the atmosphere, which lock the infrared radiation and lead to a global warming of the Earth's surface (see e.g. \citeA{Kasting1993}).An enormous amount of work has been done on this topic in what came to be the science field of ``Global Climate Change'' \cite{feulner_faint_2012}. It has been suggested tropical cirrus clouds could also enhance the greenhouse effect, being either the main explanation to the FYS paradox or an complementary source to a greenhouse gas \cite{Rondanelli2010}. \citeA{Goldblatt2011} further develop the more general discussion on the effect of clouds on the climate, noting that \citeA{Rondanelli2010} base the cirrus solution to the FYS on the ``iris theory'' which stipulates that the cirrus coverage should increase if the surface temperatures decrease, which is quite controversial for the current Earth (see e.g. the comparison of the theory with observations by \citeA{chambers2002examination}), but may be applicable on other atmospheres. \citeA{2013Icar..226..229U} applied the theory to the Early Mars, but the work of \citeA{Ramirez2017} show that there is no room for error when considering cirrus clouds for warming: a large cirrus cloud coverage, greater than 70\%, should be present. This explanation is therefore unlikely to be applicable to Mars since such a cloud coverage is not realistic: cirrus cloud formation is limited by the parts of the atmosphere that are under-saturated in water. Overall, cloud warming has not been proven to be the solution to the FYS paradox, but it shows the importance of addressing the problem of cloud formation and humidity transport, and therefore shows the importance of looking at the climate using 3D GCMs.  On the other hand, solutions to the FYS paradox that involve external factors, i.e. particle precipitations or a heavier Sun, are attractive since they could solve the problem both at Mars and the Earth. Some theories suggested that cosmic-rays may affect the cloud condensation in the Earth's atmosphere, with an overall cooling effect when there is fewer GCR (i.e. when the solar activity is higher) \cite{Svensmark1997,Shaviv2005}. Thus, a significant reduction in the cosmic-ray flux may increase the surface temperature of the Earth. These models are controversial both in the cosmic-rays ability to affect the cloud condensation \cite{Kirkby2011} and the heating efficiency of the process. The results of the CLOUD experiment at CERN tend to indicate that the present day cloud formation cannot be effectively affected by GCR flux \cite{Dunne2016,Pierce2017}: the aerosols responsible for cloud condensation come mainly from the ground, and setting the GCR flux at 0 would only reduce the cloud coverage by $\approx$10\%. It is to be noted that this experiment does indicate that GCR can affect aerosol/particle formation \cite{Kirkby2016,Trostl2016}, which is something that is also observed on other atmospheres such as Titan \cite{DOBRIJEVIC2016313,LOISON2015218}. The contentious point is whether or not GCR variation has an influence on the current climate or not. Studies such as  \citeA{LANCI2020103095} do not find historical evidence of GCR control of climate while \citeA{2017NatCo...8.2199S} advocates for a strong control of the climate by GCR in the past, with increased GCR leading to aerosols that act as cloud condensation nuclei (CCN), that lead to the formation of clouds, and ultimately heating. It has to be noted that aerosols do not always act as CCN, but can act as coolant (e.g. \citeA{2006PNAS..10318035T}).
The work of \citeA{Airapetian2016} suggests, on the contrary, that the increased particle flux from the Solar Energetic Particle events, more frequent for the Young Sun, led to the creation of greenhouse gases, that could help resolve the paradox. This interaction between ionizing radiation and climate is fundamentally different since it requires extreme radiation rates (compared to present values) to get a significant increase in greenhouse gases. In addition, it does not suppose that CCN is the limiting factor in the creation of clouds: the atmosphere can be under-saturated \cite{Ramirez2017}.

In the context of Astrophysics, a solution for the paradox can be found if one can show that the solar mass was about 10\% larger than its current 
mass. This requires the young Sun's mass loss rate to be very high, with the ability to keep this high mass loss rate for rather long time. As mentioned in Section \ref{solarforcingintime}, it is unlikely that the mass loss to the ambient solar wind can be sufficient. However, it is possible that due to high activity levels at early stellar stages, the Sun lost large amount of its mass via CMEs, although present estimates also indicate this mechanism is insufficient \cite{Drake2013}.

\subsubsection{Impact of planetary magnetic field}
Whether a planet is magnetized or not impacts ionospheric ion outflow processes and the return rate of ions outflowing from the ionosphere. 
It was believed that a planetary magnetic field was shielding its atmosphere from escape, until observations, reported by \citeA{barabash2010venus}, was showing that the escape rate at Earth is higher than the one at Venus and Mars. This was further discussed by \citeA{strangeway2010does}, and explored more in \citeA{Brain2013}. This later work also discussed the influence of the magnetic field on incoming gases that could also have an effect on climate. \citeA{Tarduno2014} and \citeA{Ehlmann2016} looked at the effect of magnetospheric escape respectively for the Early Earth and exoplanets.

\citeA{gunell_why_2018} compiled the effect of the planetary magnetic moment on ionospheric ions outflow rate in the current solar system, and compared it to other sources of escape. Considering that the observed ionospheric ion outflow rate on Earth, Mars and Venus is of the same order of magnitude (10$^{25}$ s$^{-1}$) while only Earth has a strong magnetic field, the authors made empirical models of ion outflow for three hypothetical planets with atmospheric conditions similar to the Earth, Mars and Venus but with a variable magnetic moment. They show that for each of those planets the mass escape rate, including both oxygen and hydrogen is similar in the unmagnetized range and for high magnetizations. In-between, they identify two maxima where outflow is enhanced by a factor of 2-5, one corresponding to polar cap escape and dominant for hydrogen and another corresponding to cusp escape. The presence of a large magnetosphere around a planet actually diverts part of the stellar wind energy and protects the atmosphere from sputtering and ion pickup. The induced magnetospheres of the unmagnetized planets also provide protection from sputtering and ion pickup but to a lesser extent. However, magnetospheres are much bigger objects than the planets themselves. The presence of a magnetosphere increases the size of the interaction region between the stellar wind and the planet and thus the amount of energy which can potentially be transferred into the ionosphere. For instance the cross section of the Earth magnetosphere with the solar wind is about 100 times higher than the cross section of the Earth itself with the solar wind. Consequently, the amount of energy transferred from the stellar wind to the ionosphere of magnetized planet is not necessarily lower than for unmagnetized planets \cite{Brain2013}.

Large-scale magnetospheres enable polar cap and cusp escape, which increases the escape rate. Two outflow processes are enhanced by the presence of a magnetosphere. The first is the polar wind which corresponds to a thermal ion outflow on the open magnetic field lines at high latitudes, above the polar caps. It maximizes for moderate magnetic moments when the size of the polar cap is maximum. The second corresponds to outflow from the cusp region where a significant amount of the stellar wind energy is deposited. This energy deposition increases with the size of the magnetosphere (i.e. with its cross section with the stellar wind) but is limited by the amount of ions available at the ionospheric level. \\
The effect of magnetospheres on the ion return rate is discussed in Section 2.5. In that case the protective effect of the magnetosphere is not related to the outflow itself but to the trapping of outflowing ionospheric ions, even those with high energies well above the gravitational binding energy, which was thought to result in a significant return rate into the atmosphere. However, recent observations in the Earth magnetosphere question the validity of this protective effect. Indeed, that the measured flux of precipitating ions in the ionosphere is well below the measured flux of outflowing ionospheric ions and the escape route above the polar ionosphere, where polar cap and cusp outflows occur, seems to preferentially lead to a direct ion loss to interplanetary space rather than to a return in the atmosphere. 

On the other hand, the thermospheres of Mars and Venus are called cryospheres because of the cooling effect of CO$_2$: their thermospheric temperature is extremely low, effectively shielding the atmospheres from several escape processes: the Earth's atmosphere without a magnetic field is believed to be escaping more efficiently. In addition, the main species in the ionosphere of Venus is O$_2^+$, while it is O$^+$ at Earth \cite{10.1093/astrogeo/atz047}, and this may affect the amount of ions able to escape via non-thermal processes (Mars is a special case since above $\approx$200~km O$^+$ is in majority while it is O$_2^+$ below). A recent study by \citeA{WEI201494} shows that using an escape model developed for Mars, but with Earth's upper atmosphere, increases greatly the O escape. This study also suggests that there are some correlations between lower content of O$_2$ at Earth and magnetic field inversions. Such a study could be criticized on the basis that the magnetic field does not seem to disappear during inversions (e.g. \citeA{NOWACZYK201254}), or on the basis that the fluxes of O$_2$ at Earth are dominated by the biosphere and the oxidation of the crust. The fluxes of oxygen in \citeA{WEI201494} are indeed of the same order of magnitude as the current oxygen losses in the crust \cite{catling2014great}.  A reduction of carbon burial  --which is a life-controlled process leading to net O$_2$ fluxes -- could explain the loss of oxygen without the need for a fast process. It is also in disagreement with the observation of higher ion escape near magnetic anomalies at Mars \cite{Sakai2018,Inui2019}.

To summarize, while the presence of a magnetosphere has a clear impact on ionospheric outflow, recent developments in the study of the coupling between stellar wind, magnetospheres and ionospheres challenge the idea of a protective effect of magnetospheres on atmospheric erosion. It could simply be that the question is poorly asked and that a better question is ``what kind of atmospheres require a magnetic field to be sustainable in a given set of solar/stellar activity conditions''. In any case, recent studies such as \citeA{Brain2013,gunell_why_2018,airapetian2017hospitable,GarciaSage2017}, as well as the case of Mercury, show that an intrinsic magnetic field \underline{does not} totally protect an atmosphere. A contrario, the case of Venus shows that a magnetic field absence \underline{does not} prevent sustaining a dense atmosphere.

\subsubsection{Impacts of stellar dynamics}
In a large fraction of the studies of escape through time, the stellar parameters, i.e. the EUV-XUV fluxes and the stellar wind, are considered as slowly varying with the epoch. The impact of the frequency of intermittent stellar events such as flares, CMEs, and SEPs on the escape is seldom taken into account. This is a major problem for studies of close-in exoplanets since these events can extremely affect the atmosphere as the observations of the variations of escape rate at Mars due to a CME has shown \cite{jakosky2015}.  MAVEN is currently showing that extreme solar events have a very important role in the loss of atmosphere at Mars \cite{jakosky2018loss,Mayyasi2018}. As an example, the increase in the exospheric temperature due to a flare has been observed \cite{elrod2018september}, along with change in the upper atmospheric ion and neutral composition \cite{thiemann2018mars}, and an increase of $\approx$20\% of the escape \cite{lee2018observations}.

The work of \citeA{GarciaSage2017} has shown that the EUX-XUV flux can lead to extreme absorption at rocky exoplanets around M dwarfs, however, it does not answer the question of how much active a G-star an Earth-like planet could survive. 

\subsection{The role of non-atmospheric/stellar processes}
While it is not generally explicit in the discussion above, the mass of the planet that is experiencing escape is a critical factor. Closely related is its radius, and therefore its density. The planets of the Solar system are there to show that the mass is the first factor to consider when estimating if a body will have an atmosphere or not; the energy received/distance to the Sun being the second factor. Mass is still challenging to retrieve, especially the mass of small planets, whose signal in radial velocities can be hidden by the natural variations of the star. Once mass and distances are considered, it may be possible that interesting effects come from close-in exoplanets, such as the roche-limit of the star reaching for the planet's atmosphere. 
\textbf{Overall it should not be forgotten that the inventory of volatiles, which has be estimated from the density of the exoplanet, will define the lifetime of an habitable world with large escape rates.} 

\subsection{The future of research on escape processes}
The study of planetary atmosphere habitability and evolution has, as shown here, three main directions.
\begin{enumerate}
\item Escape modeling efforts. The approach of this review has been reductionist; we have sought to isolate the individual escape processes and identify simple ad hoc models that can determine whether or not a specific escape process is important. Yet a better approach to escape would be to create so-called ``grand-ensemble models'' that are able to examine the interactions between the different processes without a priori exclusion of processes. An already invoked example comes from \citeA{Chassefiere1997sw}, in which comprehensive treatment of multiple types of charge exchange predicted an increased exospheric temperature and therefore, indirectly, higher thermal escape.
Improved models will allow the evaluation of critical parameters to help people work on the deeper parts of the atmosphere to estimate which species are escaping and at which rate, in order to prevent poor estimates based on energy-limited escape (that do not take diffusion limitation into account). An additional consideration in modeling is to devise a standard procedure for asynchronously coupling climate, chemistry, and escape models at exoplanets to study the evolution of climate and composition in tandem with stellar evolution.
\item  Laboratory work. A major limiting factor of escape models is the quality of the input parameters, such as chemical reaction rates, cross sections, etc. Laboratory experiments and model-laboratory data comparisons such as that of \citeA{Wedlund2011} are needed to refine the accuracy of the physico-chemistry parameters, and, in turn, may help identify the needs of the community for the study of specific processes. Laboratory data are also crucial to retrieve parameters from observational data.
\item Observation work. Observations efforts are limited and currently concentrated to what is believed to be the ``best known targets'' for habitability. Unfortunately, our instrumentation is not optimized for detecting habitability signatures on these targets. 
Future observations should be designed not just to
 characterize the bulk properties of the atmosphere but also to consider known or potential observables affecting atmospheric stability. One advantage of doing so is that processes like thermal escape mostly take place above cloud and haze layers and so may not be as challenging as observing lower in the atmosphere. These targets have broader characteristics than current one, and could benefit from the whole range of existing instrumentation to answer questions leading to constrain the conditions required for habitability.   Techniques should be improved to better understand the stellar environment of exoplanets, such as the observation of the stellar winds potentially impacting exoplanets as well as stellar variability in general, which has a strong potential impact on transit observations of planetary atmospheres \cite{Wakeford_2018}. As characterizing the variability of a host star typically requires less sensitive instrumentation than detecting a telluric planet orbiting it (particularly for warmer stars), it may be worth prioritizing observations of variability of types other than those around which telluric planets currently can be observed.
 \end{enumerate}

Overall, the challenge is thus to couple a grand-ensemble escape model with a complex planetary atmosphere model, itself coupled with a planetary interior model. From there it would be possible to obtain the whole story of an empirical planetary atmosphere in time. The uncertainties in each of these sub-models have to be correctly evaluated, so that it is possible to address the overall model validity. This is why a validation strategy is also of utmost importance for this kind of work.

On the stellar part, the challenge will be to determine the activity history of a star, both from the slowly evolving parameters, such as luminosity, and the discrete events such as flares. From there, it would be possible to evaluate how a given star stresses an atmosphere over time. Finally, it will be necessary to develop observation missions dedicated to study the UV flux of stars to validate the model of activity in time.




\subsection{Effects of escape on biosignatures}
The escape processes reviewed above have significant influence on the composition of the upper atmosphere, and acting over geologic time can affect the bulk composition of the atmosphere, surface and interior. The consequences of atmospheric escape for our search for life via chemical biosignatures in the atmosphere and on the surface must therefore be considered \cite{DesMaHarwiJucks-etal:AB2002,DomagSegurClair-etal:ApJ2014,Airapetian2016}. The alteration of planetary chemistry by escape can result in both false positive and false negative biosignature errors if it is not accounted for \cite{0004-637X-806-2-249}.

False positive biosignatures that can be produced by interactions at the top of the atmosphere include oxygen and oxidized species such as NO as well as organics such as the haze materials produced through UV photolysis at Titan and elsewhere. The preferential loss of hydrogen from water is one way for oxygen to arise from escape-related abiogenic processes. The processing of sufficient water to influence the bulk oxidation state of the surface materials is likely the cause of high oxidation in the Martian surface \cite{LammeLichtKolb-etal:I2003}. False negative biosignatures would result from the masking of true biogenic molecules by escape processes, either through rapid modification by particle or photon radiation or through chemical interaction with, for example, photolysis-produced oxygen. 

Biosignatures related to disequilibrium chemistry \cite{KrissOlsonCatli:SA2018} must contend with non-LTE behavior in the upper atmosphere and the potential for disequilibrium signatures to be transferred from the upper to the lower atmosphere. Even biosignatures that are based on time variations \cite{OlsonSchwiReinh-etal:ApJ2018} need to contend with seasonal variations in star-planet interaction caused by a tilted magnetic field axis which could produce either false positive or false negative results.

This discussion is not intended to be exhaustive or definitive, but instead we seek to highlight the importance of understanding the impact of the stellar environment on the production, destruction, or masking of putative chemical biosignatures. In general, although space weather processes involve small fractions of the planetary mass they can, like biological reactions, be quite selective in their reactants and products and over time can yield significant signals that must be differentiated from biological ones.

\subsection{Final thoughts}
\begin{figure}
\vspace{-2cm}
 \noindent\includegraphics[width=30pc]{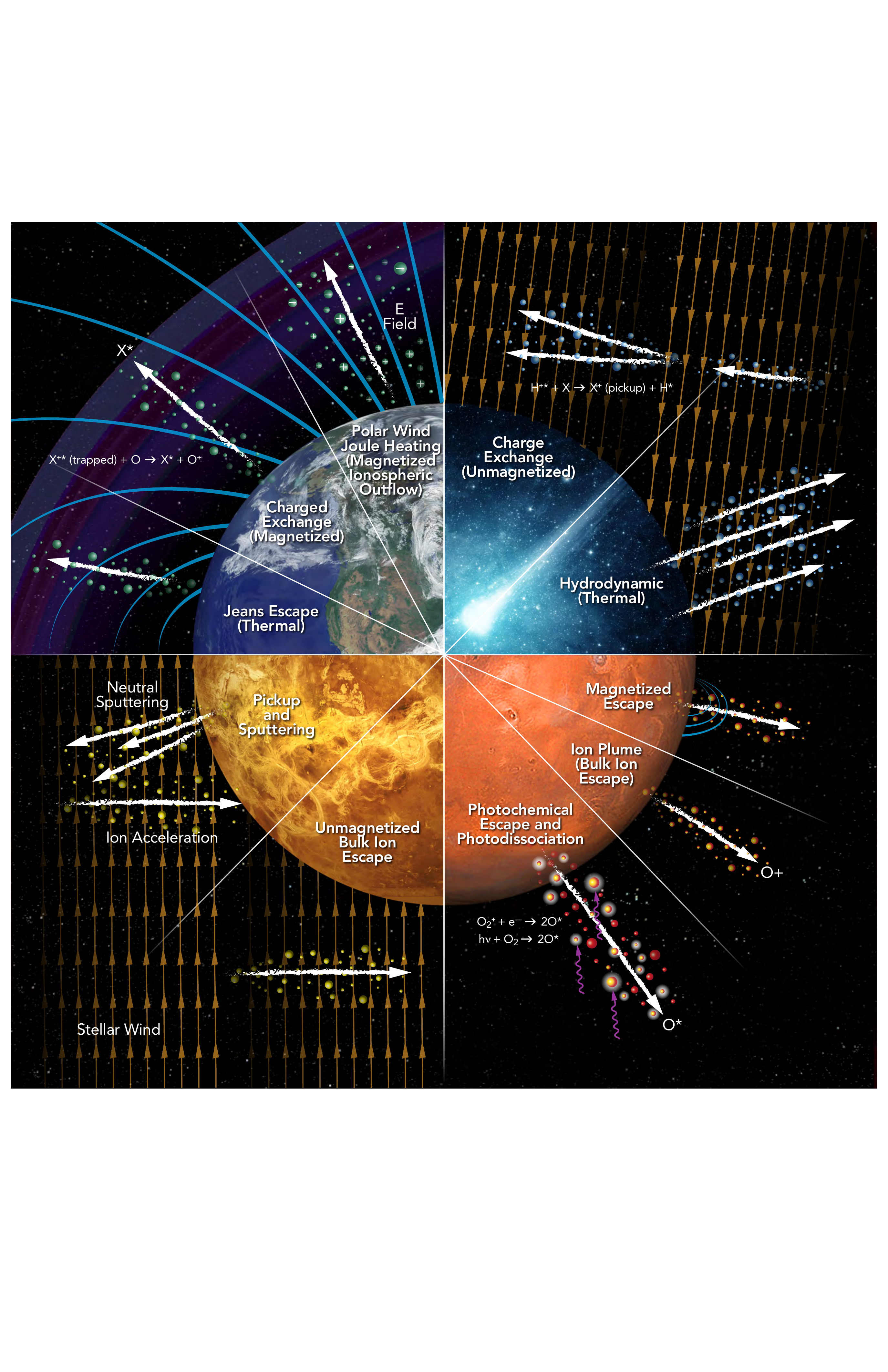}
 \vspace{-2cm}
 \caption{Overview of the escape processes, along with an example of a planet where they are major/important. The Earth's main escape processes are  Jeans', charge-exchange, and polar wind. At Venus, interaction with the solar wind; at Mars, photochemical escape and ion escape; at some exoplanets/comets hydrodynamic escape.}
 \label{escapeconclusion}
\end{figure}

We have reviewed the different escape processes considered so far in the literature, and summarized in Figure \ref{escapeconclusion}. Understanding these processes, and also ones that are still to be discovered, makes it possible to understand how a planetary atmosphere evolves. This is however not enough to understand the whole history of an atmosphere: change in the atmosphere composition, change in the stellar activity, and change of the outgassing from the planetary interior are examples of processes that affect the development of an atmosphere, and can lead to very different pathways. To that extent, life is one of the major modifiers of Earth's atmosphere. It would be easy to consider an atmosphere  that is out of equilibrium, or  that is very difficult to model/undertand by our current means, as harboring life; this is the idea behind the concept of biosignatures. However, the detection of a Titan-like atmosphere outside our Solar system may lead to life detection claims that are not (at least to date) consistent with Titan's observations. On the other hand, the atmosphere of an Early Earth may be detected, but considered as sterile. 

Since this paper has been showing that habitability is a dynamic process, and that the habitability of a planet is the result of its history, and not just of its location with respect to its star, it is important to extend that notion to biosignatures. In this respect, it would be preferable to announce the detection of molecules relevant to pre-biotic chemistry instead of directly announce biosignatures, so that no extraordinary claim is made without extraordinary evidence.

   \begin{acronyms}
   \acro{CME}
   Coronal Mass Ejection
   \acro{DSMC}
   Direact Simulation Monte Carlo
   \acro{ENA}
   Energetic Neutral Atom
   \acro{ESA}
   European Space Agency
   \acro{EUV}
   Extreme Ultraviolet
   \acro{FAC}
   Field Aligned Currents
       \acro{FYS} 
    Faint Young Sun (typically for speaking about the Faint Young Sun paradox)
    \acro{GCM}
    Global Circulation Model
   \acro{HZ}
    Habitable Zone
    \acro{KHI}
    Kelvin-Helmholtz instability
    \acro{MAVEN}
    NASA / Mars Atmosphere and Volatile and EvolutioN mission
    \acro{MEX}
    ESA / Mars Express
    \acro{MHD}
    Magneto-Hydro Dynamic
    
    \acro{NASA}
    National Aeronautics and Space Administration
    \acro{SEP}
    Solar Energetic Particle
    \acro{UV}
    Ultraviolet
    
    \acro{VEX}
    ESA / Venus Express
    \acro{XUV}
    X-Ultraviolet
   \end{acronyms}

%

 \begin{notation}
	 \notation{$k$} the Boltzman constant
	 \notation{$m$} the average molecular mass
	 \notation{$g$} the gravitational acceleration (typically dependent upon the altitude)
	 \notation{$z$} the altitude
	 \notation{$\vec{x}$} the location in space
	 \notation{$\Theta$, $\Phi$} angles in spherical coordinates
	  \notation{$T$} the temperature (of neutral constituents, the subscript can show if it is of electrons or ions, and it is generally dependent upon the altitude)
	 \notation{$T_{exo}$} the exospheric temperature
	 \notation{$n$} the density of the considered species or of the gas (typically dependent upon the altitude). $n_a$ is usually used to note the total density. A typical unit is species.$cm^{-3}$.
    \notation{$X_i = \frac{n_i}{n_a}$} mole fraction of the gas i.
	 \notation{$H = \frac{k T}{m g}$} The scale height. 
	 \notation{$H_{exo}$} is the scale height at the exosphere, so when $T= T_{exo}$ 
	\notation{$R$} radius of the planet. Sometimes the radius of the exosphere $R_{exo} = R + z_{exo}$
	\notation{$\lambda$} photon wavelength
    \notation{$l = \frac{1}{\sqrt{2} n \sigma}$} the characteristic length between collisions.
    \notation{$v_{esc} = \sqrt{2GM/R}$} the escape speed.
    \notation{$K_n = l / H$} the Knudsen number: characteristic parameter for the transition between collisionless and fluid regimes. If $K_n\rightarrow 0$, the collisions are dominant, we are in a fluid regime. If $K_n>1$,  we are in a collisionless regime. 
    \notation{$\lambda_{ex} = R_{exo} / H_{exo}$} is the characteristic number for the thermal escape. In other work, such as \citeA{Selsis2006}, this parameter is designed by $\chi$.
    \notation{$\gamma = \frac{C_p}{C_V}$} heat capacity ratio, or adiabatic index. directly linked to the degree of freedom $f$ of the molecule/atom by the equation $\gamma = 1 + \frac{2}{f}$.
    
 \end{notation}

%
%
%
%
%
%
%
%

\acknowledgments
The Living Breathing Planet team is funded by the NASA Nexus for Exoplanet System Science under grant NNX15AE05G.
Work at the Royal Belgian Institute for Space Aeronomy was supported by PRODEX/Cluster contract 13127/98/NL/VJ(IC)-PEA90316.  
The work of C.S.W. has been partially funded by the Austrian Science Fund under project number P\,32035-N36.
We thank Mary Pat Hrybyk-Keith at NASA/GSFC for her graphics work on the summary figures. We would like to thank N. Wright (Keele University) for his assistance in providing additional figures. The authors would like to thank the Institut d'Astrophysique de Paris (IAP), France and Ben Jaffel for the IAPIC continuous development.


%
%

%
%
%
%
%

\end{document}